\documentclass[fleqn,usenatbib]{mnras}

\usepackage[T1]{fontenc}

\DeclareRobustCommand{\VAN}[3]{#2}
\let\VANthebibliography\thebibliography
\def\thebibliography{\DeclareRobustCommand{\VAN}[3]{##3}\VANthebibliography}

\usepackage{graphicx}	
\usepackage{amsmath}	
\usepackage{amssymb}	

\usepackage{siunitx}
\usepackage{filecontents}
\usepackage{subfigure}
\usepackage{tikz}
\usepackage{multirow}
\usepackage{xcolor}
\usepackage{multicol}

\usepackage{caption}
\usepackage{float}

\newcommand{\Teff}{\mbox{$T_\mathrm{eff}$}}
\newcommand{\kms}{\mbox{km\,s$^{-1}$}}

\title[Cooler white dwarfs as IR standards]{A network of cooler white dwarfs as infrared standards for flux calibration}

\author[A. K. Elms]{Abbigail K. Elms,$^{1}$\thanks{Contact e-mail: \href{mailto:Abbigail.Elms@warwick.ac.uk}{Abbigail.Elms@warwick.ac.uk}}
{Nicola Pietro Gentile Fusillo,$^{2}$}
{Pier-Emmanuel Tremblay,$^{1}$}
{Ralph C. Bohlin,$^{3}$}
\newauthor
{Mark A. Hollands,$^{1}$}
{Snehalata Sahu,$^{1}$}
{Mairi W. O’Brien,$^{1}$}
{Susana Deustua$^{3,4}$ and}
{Tim Cunningham$^{5}$}
\\
$^{1}$Department of Physics, University of Warwick, Coventry, CV4 7AL, UK\\
$^{2}$Department of Physics, Universit\`a degli Studi di Trieste, Via A. Valerio 2, 34127, Trieste, Italy\\
$^{3}$Space Telescope Science Institute, 3700 San Martin Drive, Baltimore, MD 21218, USA\\
$^{4}$Sensor Science Division, National Institute of Standards and Technology, Gaithersburg, MD 20899-8441, USA\\
$^{5}$Center for Astrophysics | Harvard \& Smithsonian, 60 Garden St., Cambridge, MA 02138, USA\\
}

\date{Accepted XXX. Received YYY; in original form ZZZ}

\pubyear{2024}

\usepackage{newtxtext,newtxmath}

\begin{document}
\label{firstpage}
\pagerange{\pageref{firstpage}--\pageref{lastpage}}
\maketitle

\begin{abstract}
The accurate flux calibration of observational data is vital for astrophysics and cosmology because absolute flux uncertainties of stellar standards propagate into scientific results. With the ever higher precision achieved by telescopic missions (e.g. \textit{JWST}) in the infrared (IR), suitable calibrators are required for this regime. The basis of the \textit{Hubble Space Telescope} (\textit{HST}) flux scale is defined by model fits of three hot ($\Teff > 30\,000$\,K) hydrogen-atmosphere (DA) white dwarfs, which achieve an accuracy better than 1 per cent at optical wavelengths but falls below this level in the IR range. We present a network of 17 cooler DA white dwarfs with $\Teff < 20\,000$\,K as spectrophotometric flux standards that are equally, if not more, accurate at IR wavelengths. Cooler white dwarfs do not suffer from non-local thermal equilibrium (NLTE) effects in continuum flux or from UV metal line blanketing, have a larger sky density, are generally closer to Earth with little or negligible interstellar reddening, and have energy distributions peaking in the optical or near-IR. Using the latest grid of DA LTE atmosphere models with three-dimensional (3D) convection, the observed Space Telescope Imaging Spectrometer (STIS) and Wide Field Camera three (WFC3) fluxes of our network are accurate to 3 per cent over most of the range $1450 - 16\,000$\,\AA, with a median standard deviation of 1.41 per cent. Fitting the \textit{HST} STIS and WFC3 white dwarf SEDs and Balmer lines independently yields SEDs that agree within $3\sigma$, which demonstrates the precision of the models for our network.
\end{abstract}




\section{Introduction}

Now, more than ever, the creation and use of a reliable and self-consistent flux scale is of paramount importance. We have entered a pioneering era of space telescopes and instruments, which are providing previously unattainable data for objects in our Galaxy and beyond, especially in the near-IR (NIR; $0.8 - 5$\,$\mu$m) and mid-IR (MIR; $5 - 32$\,$\mu$m). 

The \textit{Hubble Space Telescope} (\textit{HST}) provides high-resolution imaging and broad wavelength coverage from the far-ultraviolet (FUV; $0.0912 - 0.2$\,$\mu$m) to NIR. New generation observatories will compliment and further the scientific reach of \textit{HST}, for instance the \textit{James Webb Space Telescope} (\textit{JWST}), \textit{Euclid} spacecraft and \textit{Nancy Grace Roman Space Telescope} (\textit{RST}) have an array of high-resolution and high-sensitivity instruments which facilitate photometric and spectroscopic observations from the FUV to MIR regimes. Similar capabilities will be achieved with first generation high-resolution instrumentation on ground-based telescopes, such as the Multi-AO Imaging Camera for Deep Observations (MICADO) on the Extremely Large Telescope (ELT). Data from these instruments will revolutionise fields in cosmology, such as dark energy and dark matter, and astrophysics, such as stellar evolution and exoplanet studies. However, observational data calibration is a major source of uncertainty for many instruments and scientific fields \citep{Hounsell2018, Tayar2022, Wilson2023}. For example, the accurate comparison between the ﬂuxes of distant redshifted and nearby supernovae (SNe) in the rest frame has a crucial impact on our understanding of the nature of the dark energy that is driving the observed accelerating cosmic expansion \citep{Stubbs2015, Scolnic2015, Scolnic2022, Brout2022}. To reach this full scientific potential, photometric and spectroscopic instruments must be accurately calibrated on a single consistent flux scale, as uncertainties from flux standards propagate into scientific results \citep{Hounsell2018}.

To date, the most internally consistent set of primary flux standards are three hot ($\Teff > 30\,000$\,K) hydrogen-atmosphere (DA) white dwarfs: GD\,71, GD\,153 and G191-B2B \citep{BohlinGordonTremblay2014}. The spectral energy distributions (SEDs), which are flux as a function of wavelength, of these primary standards provide the basis for calibration of the \textit{HST} spectrophotometry, primarily from the Space Telescope Imaging Spectrometer (STIS), the Wide Field Camera three (WFC3) IR grisms, and the Near Infrared Camera and Multi-Object Spectrometer (NICMOS) \citep{Bohlin2014, Bohlin2020}. While fits to Balmer lines determine $\Teff$ and $\log g$, i.e. the shape of the SEDs, the absolute flux level is set by a reconciliation of the original Vega flux at 5557.5\,\AA\ \citep[vacuum; ][]{Megessier1995} with the \textit{Midcourse Space Experiment} (\textit{MSX}) IR absolute flux to define the absolute flux at all wavelengths over the ultraviolet (UV), visible and NIR \citep{Bohlin2014, Bohlin2020}. These SEDs provide the basis for the \textit{HST} absolute flux scale and are publicly available on CALSPEC\footnote{\href{https://www.stsci.edu/hst/instrumentation/reference-data-for-calibration-and-tools/astronomical-catalogs/calspec}{https://www.stsci.edu/hst/instrumentation/reference-data-for-calibration-and-tools/astronomical-catalogs/calspec}}, which is a database of primary and secondary spectrophotometric absolute flux standard stars on the \textit{HST} flux scale containing measured stellar spectra and modelled SEDs. CALSPEC standards are used for the calibration of \textit{HST}, \textit{JWST}, \textit{Gaia} and other ground- and space-based instrumentation. Therefore, the precision and accuracy of observational astrophysical and cosmological studies which use CALSPEC standards are entirely dependent on the CALSPEC precision of measured and modelled SEDs. 

Hot DA white dwarfs have fully radiative atmospheres, therefore, their optical spectra are relatively simple to model. Their effective temperature (\Teff) and surface gravity ($\log g$) determine accurate model SEDs \citep{Holberg1985}. SED models of the current three hot standards are consistent within 1 per cent at optical wavelengths ($0.3 - 1$\,$\mu$m) with \textit{HST} spectrophotometric observations \citep{Bohlin2014, Narayan2019}. However, outside the optical, discrepancies of $>$\,1 per cent are evident \citep{Bohlin2011, Bohlin2014, Bohlin2022}, because uncertainties arise in hot white dwarf model atmospheres from complex non-local thermal equilibrium (NLTE) effects, trace metal contamination and UV metal line blanketing \citep{Werner1996,Gianninas2010, Rauch2013}. Hot white dwarfs have a lower sky density ($\simeq 0.75$\,deg$^{-2}$ at $G < 20$) than cooler white dwarfs, forcing the selection of more distant hot stars with larger reddening effects and more uncertain \textit{Gaia} parallaxes. Also, hot white dwarf SEDs drop rapidly in flux at longer wavelengths. These sources of uncertainty are non-negligible and propagate into the results of scientific studies. 

The need for faint standards, e.g. below the saturation limits of several instruments, has motivated the expansion of hot white dwarf standards in recent years using \textit{HST} spectrophotometry from faint ($16.5 < V < 19.5$) and photometrically stable DA white dwarfs spanning the whole sky \citep{Narayan2016, Narayan2019, Calamida2022, Axelrod2023}. These studies have achieved consistency within 1 per cent of the \textit{HST} flux scale at optical wavelengths, thus establishing all-sky spectrophotometric secondary standards for wide-field optical surveys. Good agreement was also found in the NIR regime with WFC3 observations. 

We propose to add a network of 17 cooler DA white dwarfs with $\Teff < 20\,000$\,K as secondary spectrophotometric flux standards on the \textit{HST} flux scale. Our network has \textit{HST}/STIS flux calibrated spectra, with 16/17 supplemented by \textit{HST}/WFC3 spectrophotometry, thus they are tied to the three CALSPEC primary standards. Cooler white dwarfs than the current primary hot standards have been proposed in recent years to have a comparable, or even higher, accuracy at IR wavelengths compared to hot white dwarfs \citep{Tremblay2017, GF2020}. DA white dwarfs with $\Teff < 20\,000$\,K suffer much less from NLTE effects or UV metal line blanketing, have a larger sky density than hot white dwarfs, their energy distributions peak in the optical/NIR, and are still relatively bright in the IR compared to their hotter counterparts. Cooler white dwarfs at distances $\lesssim 100$\,pc have the advantage of having less reddening than further away hot white dwarfs, while still being reliable calibrators for IR observing modes. The latest DA model atmospheres can effectively account for convective effects which are important for white dwarfs with $\Teff \lesssim 13\,000$\,K \citep{Tremblay2013, Tremblay2019b} so there are no reservations about modelling cooler white dwarfs which might have arisen decades ago, even though dense atmospheres below 6000\,K are still subject to a low-mass problem \citep{OBrien2024}.

The exclusive use of hot white dwarfs as flux calibrators may not be the optimal choice for achieving the highest accuracy possible, in particular with the dawn of next-generation IR missions. Our proposed network of 17 cooler calibrators can be used in parallel with hot white dwarf calibrators depending on the science case. This network will provide an independent estimate for secondary SEDs compared to the current primary SEDs; and the difference between SEDs based on cool or hot white dwarfs sets limits on the precision of flux calibrations. We split our network into 13 warm (13\,000\,K $< \Teff < 20\,000$\,K) flux standard candidates and four extra flux standard candidates, which consist of the three coolest ($\Teff < 13\,000$\,K) white dwarfs in the network and one which has no \textit{HST}/WFC3 spectrophotometry. There is an on-sky separation of $<4$\,h between the white dwarfs across the whole sky, which ensures adequate sky coverage for the vast majority of science cases. 

Section~\ref{sec:observations} presents \textit{HST} spectrophotometric observations of the 17 white dwarfs in our network, in addition to their \textit{Gaia} Data Release 3 (DR3) astrometry and optical and IR photometry. We describe the fitting procedures implemented on the observed \textit{HST} data with our model atmospheres in Section~\ref{sec:analysis} which allow us to derive two sets of independent \Teff\ and $\log g$ best-fitting parameters, in addition to synthetic photometry and \textit{Gaia} atmospheric parameters. In Section~\ref{sec:Discussion} we discuss our results and conclude in Section~\ref{sec:Conclusions}.

\section{Observations}
\label{sec:observations}

\subsection{Hubble Space Telescope spectrophotometry}
\label{sec:HST_spectrophotometry}

Our network of 17 flux standard candidates was built using spectrophotometric observations from three \textit{HST} programmes that made use of the STIS and WFC3 instruments. WD\,1327$-$083 and WD\,2341$+$322 were observed in a Cycle 23 (GP-14213) pilot programme on using cooler white dwarfs than the current hot primary standards as near-IR and IR flux calibrators. WD\,0352$+$096 has been observed intermittently over the past few decades, with the latest observations being from a Cycle 25 (GP-15485) programme for CALSPEC and \textit{JWST} standard stars. The remaining 14 white dwarfs were observed in a Cycle 28 (GP-16249) programme specifically targeting candidates for our flux standard network. All 17 stars were selected because they do not have unresolved stellar companions in \textit{HST} and \textit{Gaia}, have hydrogen-dominated atmospheres (spectral type DA), have $\Teff < 20\,000$\,K, and no observed photometric or spectroscopic variability due to pulsations or magnetic phenomena - see Section~\ref{sec:flags} for flags from the literature on variability, metal traces, observed magnetism and binarity for the individual white dwarfs in our network. 

The STIS low dispersion (L) spectra were obtained with a $52" \times 2.0"$ long slit on the G140L, G230L, G430L and G750L gratings which cover $1140 - 10\,200$\,\AA\, and the WFC3 spectra were obtained with the G102 and G141 grisms which cover $8000 - 17\,000$\,\AA. STIS and WFC3 spectrophotometry were merged at 10\,115\,\AA\ to form complete \textit{HST} SEDs spanning $1140 - 17\,000$\,\AA\ for all 17 candidates except WD\,0352$+$096 which does not have WFC3 data. A log of the \textit{HST} spectrophotometric observations are given in Table\,~\ref{tab:HST_log}.

\subsection{\textit{Gaia} DR3 astrometry}
\label{sec:astrometry}

\begin{figure}
    \centering
    \includegraphics[width=\columnwidth]{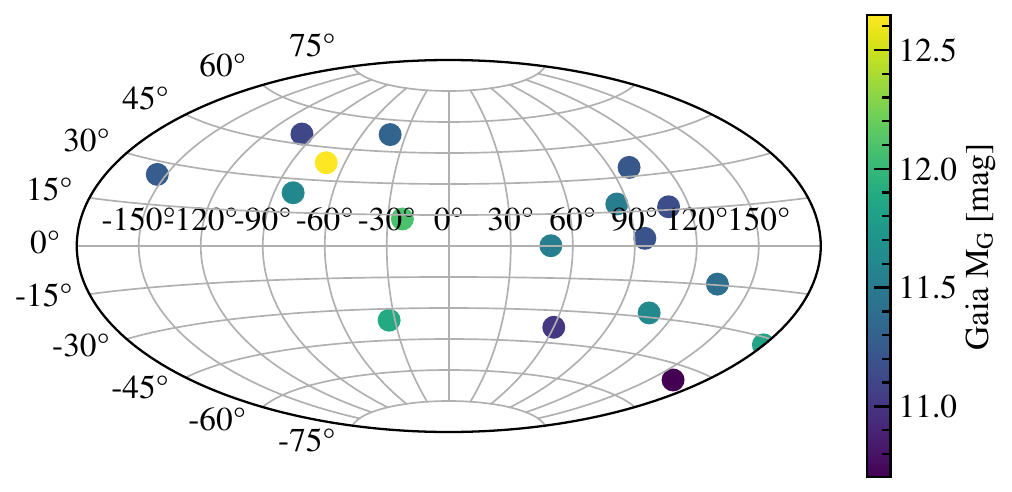}
\caption{Galactic coordinates (latitude and longitude in the J2016.0 epoch) of the 17 white dwarf flux standard candidates from \textit{Gaia} DR3, with their colour representing their absolute magnitudes, $M_{\rm G}$, computed using the \textit{Gaia} $G$ magnitude and the \textit{Gaia} parallax (Table~\,\ref{tab:astrometry}).}
\label{fig:Gaia_lb}
\end{figure}

The \textit{Gaia} DR3 astrometry for the 17 DA white dwarfs proposed as flux standards is displayed in Table~\,\ref{tab:astrometry} and the spatial distribution in Galactic coordinates of the network is shown in Fig.~\ref{fig:Gaia_lb}. The stars are all within 40\,pc and have absolute \textit{Gaia} $G$ magnitudes ($M_{\rm G}$) of $< 12.7$\,mag so they are relatively bright compared to white dwarfs in the same volume, which peak at absolute magnitudes of approximately 14.5\,mag \citep{OBrien2024}.

Data for two of the current hot primary standard white dwarfs, GD\,153 and GD\,71, are also given in Table~\,\ref{tab:astrometry} for comparison to the 17 flux standard candidates. The third hot primary standard, G191-B2B, is not included in this paper (see Section~\ref{sec:Spectrophotometric_fitting} for more details).

\begin{table*}
    \centering
	\caption{\textit{Gaia} DR3 astrometry for the 17 DA white dwarfs proposed as flux standards. Proper motions, $\mu$, are given in the right ascension ($\alpha$) and declination ($\delta$) directions. Absolute magnitudes, $M_{\rm G}$, are computed using the \textit{Gaia} $G$ magnitude and the \textit{Gaia} parallax. Values are given in the J2016.0 epoch.}
	\label{tab:astrometry}
	\resizebox{\textwidth}{!}{\begin{tabular}{ccccccccc}
		\hline
		\hline
		Object & \textit{Gaia} DR3 Designation & RA & Dec & $\mu_\alpha\,\textnormal{cos}(\delta)$ & $\mu_\delta$ & Parallax & Distance & $M_{\rm G}$ \\
		&  & [deg] & [deg] & [mas\,yr$^{-1}$] & [mas\,yr$^{-1}$] & [mas] & [pc] & [mag] \\
		\hline
		\hline
        \multicolumn{9}{c}{\textbf{Hot primary standards}}\\
        \hline
		GD\,153 & 3944400490365194368 & 194.259 & 22.030 & $-$38.402 $\pm$ 0.045 & $-$202.990 $\pm$ 0.051 & 14.593 $\pm$ 0.038 & 68.526 $\pm$ 0.177 & 9.132 $\pm$ 0.003\\
  		GD\,71 & 3348071631670500736 & 88.115 & 15.886 & 76.728 $\pm$ 0.053 & $-$172.960 $\pm$ 0.038 & 19.564 $\pm$ 0.055 & 51.115 $\pm$ 0.144 & 9.457 $\pm$ 0.003\\
        \hline
        \multicolumn{9}{c}{\textbf{Warm flux standard candidates}}\\
        \hline
		WD\,0148$+$467 & 356922880493142016 & 28.012 & 47.002 & 4.645 $\pm$ 0.031 & 122.025 $\pm$ 0.027 & 60.564 $\pm$ 0.031 & 16.511 $\pm$ 0.008 & 11.407 $\pm$ 0.003\\
		WD\,0227$+$050 & 2516322146457318144 & 37.570 & 5.264 & 76.958 $\pm$ 0.049 & $-$24.502 $\pm$ 0.039 & 37.711 $\pm$ 0.051 & 26.517 $\pm$ 0.036 & 10.703 $\pm$ 0.003\\
		WD\,0809$+$177 & 657056745624156416 & 123.158 & 17.617 & 73.482 $\pm$ 0.026 & $-$87.171 $\pm$ 0.020 & 36.977 $\pm$ 0.025 & 27.044 $\pm$ 0.018 & 11.269 $\pm$ 0.003\\
		WD\,1105$-$340 & 5401688425816913920 & 166.950 & $-$34.349 & 39.894 $\pm$ 0.021 & $-$263.429 $\pm$ 0.019 & 38.195 $\pm$ 0.021 & 26.181 $\pm$ 0.014 & 11.610 $\pm$ 0.003\\
		WD\,1105$-$048 & 3788194488314248832 & 167.000 & $-$5.160 & $-$55.547 $\pm$ 0.031 & $-$442.634 $\pm$ 0.028 & 40.293 $\pm$ 0.032 & 24.818 $\pm$ 0.020 & 11.117 $\pm$ 0.003\\
		WD\,1327$-$083 & 3630035787972473600 & 202.552 & $-$8.577 & $-$1111.205 $\pm$ 0.048 & $-$472.540 $\pm$ 0.028 & 62.148 $\pm$ 0.044 & 16.091 $\pm$ 0.011 & 11.322 $\pm$ 0.003\\
        WD\,1713$+$695 & 1638563322306634368 & 258.275 & 69.522 & $-$55.513 $\pm$ 0.020 & $-$343.042 $\pm$ 0.025 & 38.094 $\pm$ 0.018 & 26.251 $\pm$ 0.012 & 11.232 $\pm$ 0.003\\
        WD\,1911$+$536 & 2140481412496465152 & 288.203 & 53.721 & 144.450 $\pm$ 0.022 & 136.061 $\pm$ 0.019 & 45.109 $\pm$ 0.018 & 22.169 $\pm$ 0.009 & 11.518 $\pm$ 0.003\\
        WD\,1919$+$145 & 4319908862597055232 & 290.418 & 14.678 & $-$33.016 $\pm$ 0.027 & $-$75.933 $\pm$ 0.023 & 50.308 $\pm$ 0.030 & 19.878 $\pm$ 0.012 & 11.527 $\pm$ 0.003\\
        WD\,2039$-$682 & 6424566979354709248 & 311.092 & $-$68.090 & 182.095 $\pm$ 0.018 & $-$228.167 $\pm$ 0.023 & 51.099 $\pm$ 0.025 & 19.570 $\pm$ 0.010 & 11.882 $\pm$ 0.003\\
        WD\,2117$+$539 & 2176116580055936512 & 319.734 & 54.212 & $-$85.450 $\pm$ 0.026 & 193.193 $\pm$ 0.026 & 57.764 $\pm$ 0.022 & 17.312 $\pm$ 0.006 & 11.201 $\pm$ 0.003\\
        WD\,2126$+$734 & 2274076297221555968 & 321.741 & 73.644 & 55.338 $\pm$ 0.030 & $-$314.200 $\pm$ 0.030 & 44.992 $\pm$ 0.026 & 22.226 $\pm$ 0.013 & 11.153 $\pm$ 0.003\\
        WD\,2149$+$021 & 2693940725141960192 & 328.106 & 2.387 & 15.323 $\pm$ 0.042 & $-$300.533 $\pm$ 0.044 & 44.326 $\pm$ 0.043 & 22.560 $\pm$ 0.022 & 11.010 $\pm$ 0.003\\
		\hline
        \multicolumn{9}{c}{\textbf{Extra flux standard candidates}}\\
        \hline
        WD\,0352$+$096 & 3302846072717868416 & 58.842 & 9.788 & 173.274 $\pm$ 0.035 & $-$5.569 $\pm$ 0.025 & 28.587 $\pm$ 0.037 & 34.981 $\pm$ 0.045 & 11.829 $\pm$ 0.003\\
        WD\,1202$-$232 & 3489719481290397696 & 181.361 & $-$23.552 & 41.819 $\pm$ 0.021 & 226.558 $\pm$ 0.020 & 95.902 $\pm$ 0.018 & 10.427 $\pm$ 0.002 & 12.647 $\pm$ 0.003\\
        WD\,1544$-$377 & 6009537829925128064 & 236.873 & $-$37.920 & $-$423.692 $\pm$ 0.031 & $-$209.108 $\pm$ 0.025 & 65.689 $\pm$ 0.026 & 15.223 $\pm$ 0.006 & 12.088 $\pm$ 0.003\\
        WD\,2341$+$322 & 2871730307948650368 & 355.960 & 32.546 & $-$215.905 $\pm$ 0.033 & $-$59.871 $\pm$ 0.021 & 53.762 $\pm$ 0.027 & 18.601 $\pm$ 0.009 & 11.619 $\pm$ 0.003\\
	\end{tabular}}
\end{table*}

\subsection{Photometry}
\label{sec:photometry}

Most white dwarfs in our network have optical photometry from \textit{Gaia} DR3 $G$, $G_{\rm BP}$ and $G_{\rm RP}$ bands \citep{Gaia2023}, in addition to IR photometry from the Two Micron All Sky Survey $J$, $H$ and $K_S$ bands \citep[2MASS;][]{Skrutskie2006} and the Wide-field Infrared Survey Explorer \citep[WISE;][]{WISE2010} $W$1 and $W$2 bands from the AllWISE source catalogue. Observed photometry is displayed in Table~\ref{tab:obs_syn_photometry}.

\section{Analysis}
\label{sec:analysis}

\subsection{Spectrophotometric fitting}
\label{sec:Spectrophotometric_fitting}

The latest (March 2024) grid of 3D pure-hydrogen (DA) local thermal equilibrium (LTE) atmosphere models were used to fit the \textit{HST} spectrophotometry of the 17 DA white dwarfs proposed as flux standards. The grid is an updated version of the work in \citet{Tremblay2013} and \citet{Tremblay2015a}, which now has an extended resolution and wavelength coverage ($40 - 600\,000$\,\AA), is in vacuum wavelengths, and can be used with or without H$_2$ molecular lines in the UV. This work does not use high-resolution UV data so we employ the grid without H$_2$ molecular lines.  The grid includes models for $7.0 \leq \log g \leq 9.0$ in steps of 0.5\,dex (cgs units) and $1500 \leq \Teff \leq 40\,000$\,K. We have used LTE models because NLTE effects at $< 40\,000$\,K are only noticeable on the central, high spectral resolution cores of Balmer lines \citep{Napi2020,Munday2024}, which is inconsequential for continuum flux calibration. At $\Teff \lesssim 13\,000$\,K atmospheric convection becomes important to predict continuum fluxes from UV to IR, hence we use 3D models which have no free parameters unlike the previous 1D grids based on the mixing-length theory of convection \citep{Tremblay2013}. 

Ideally, atmospheric models should correctly predict the SED flux and Balmer line profiles of white dwarf flux standards independently to sub-percent precision. Therefore, we fit the network in two different ways: 1) SED fit of the continuum and \textit{Gaia} DR3 parallax, omitting the Balmer line regions; 2) Balmer line fit of H$\beta$ - H$\eta$. This method allows us to determine two independent \Teff\ and $\log g$ values for each star, thus defining model absolute fluxes with the help of a mass-radius ($M$-$R$) relation \citep[see e.g.][]{Bedard2020}. 

The same fitting methods were conducted on the current hot standards GD\,153 and GD\,71, except that we used a 1D DA NLTE grid of models due to these stars having $\Teff \approx 40\,000$\,K and being outside of the range of convection \citep{Tremblay2011}. This grid is in air wavelengths and ranges from $6.5 \leq \log g \leq 9.5$ in steps of 0.5\,dex and $1500 \leq \Teff \leq 140\,000$\,K. We did not include the current hot standard G191-B2B in our analysis because it has trace metals that strongly influence NLTE effects, thus a different model grid would have to be used to accurately fit its SED \citep{Gianninas2010,Bohlin2020}.

The input physics prescription and numerical methods for both DA 3D LTE and 1D NLTE model grids can be found on P.-E. Tremblay's Source Model Data webpage\footnote{\url{https://warwick.ac.uk/fac/sci/physics/research/astro/people/tremblay/modelgrids/}}.

\subsubsection{STIS and WFC3 SED fitting}
\label{sec:SED_fitting}
Each observed STIS and WFC3 \textit{HST} spectra were fit with model atmospheres by minimising the reduced $\chi^2$ using the \texttt{scipy optimize} trust region reflective (\texttt{trf}) algorithm, which is a non-linear least-squares method \citep{Byrd1987}. The fits neglect regions bluer than 1450\,\AA\ to avoid the broad Lyman-$\alpha$ H$^+_2$ satellite region at $1380 - 1410$\,\AA, which appears for stars $\Teff < 20\,000$\,K \citep{Koester1985} and increases in strength (i.e. extends to redder wavelengths) for decreasing temperatures - thus, the fits were started at 1700\,\AA\ for the two coolest stars in the network WD\,1202$-$232 and WD\,1544$-$377. WD\,2341$+$322 is the third coolest star in the network and has a Lyman-$\alpha$ satellite from H-H collisions at $\approx 1600$\,\AA, however our models fit this feature well and has been proven to be a good indicator of atmospheric parameters \citep{Xu2013, Greenstein1979, Wegner1982, Koester1985, Nelan1985} so we kept it in the fit by starting at 1450\,\AA\ like the warmer objects in the network. The reddest WFC3 grism extends to 17\,000\,\AA\ so to avoid the end of the grism where measurements can get unreliable we neglected all wavelengths redder than 16\,000\,\AA, except WD\,0352$+$096 which does not have WFC3 data so we end the fit at 10\,000\,\AA\ to avoid the edge of the STIS G750L grating (at $10\,238$\,\AA). The Balmer line regions $3815-4490$\,\AA, $4681-5041$\,\AA\ and $6385-6745$\,\AA\ were masked in all fits to ensure independent atmospheric parameters were obtained to the separate Balmer line fit parameters. 

The DA model Eddington fluxes, $H_{\nu}$, were converted to $H_{\lambda}(\lambda)$ for the fits, then compared to the observed fluxes, $f_{\lambda}(\lambda)$, using

\begin{equation}
 f_{\lambda}(\lambda) = 4\pi \left(\frac{R}{D}\right)^2 H_{\lambda}(\lambda) ,
 \label{flux_scaling}
\end{equation}

{\noindent}where $R$ is the white dwarf radius and $D$ is the distance to the white dwarf from Earth, constrained from the \textit{Gaia} DR3 parallax. The model fluxes were reddened (extinguished) using the G23 model from \citet{Gordon2023}, which is the latest extinction curve in the \texttt{python} package \texttt{dust$\_$extinction}\footnote{\url{https://github.com/karllark/dust_extinction}} for 912\,\AA\ to 32\,$\mu$m and $R(V) = A(V)/E(B-V)$. We used $R(V) = 3.1$ as this is the often used Milky Way average value \citep{Johnson1965, Schultz1975, Whittet1980, Fitzpatrick1999}. The \Teff, $\log g$, parallax and $E(B-V)$ were free parameters in the fits. The bounds for \Teff\ and $\log g$ were determined by the model grid limits, parallax was bounded by $\pm 3\sigma$ of the \textit{Gaia} DR3 parallax errors, and $E(B-V)$ had arbitrary bounds of $0.00 - 0.05$\,mag due to all white dwarfs in the network being $< 40$\,pc thus having minimal reddening effects. We used the DA $M$-$R$ relation with thick hydrogen layers and carbon-oxygen cores from \citet{Bedard2020}\footnote{\url{https://www.astro.umontreal.ca/~bergeron/CoolingModels/}} to determine $R$ (and white dwarf mass) using the best-fitting \Teff, $\log g$ and parallax. 

Statistical uncertainties were computed from the covariance matrix of the model atmosphere fit scaled by reduced $\chi^2$ to account for the goodness of fit. The systematic error on SED fits were computed by fitting the UV ($1450 - 3065$\,\AA) and combined optical+NIR ($3065 - 16\,000$\,\AA) regions separately to obtain two sets of \Teff\ and $\log g$ best-fitting parameters.

These pairs of \Teff\ and $\log g$ disagreed by far more than the statistical errors, indicating
some unmodelled systematic uncertainty. To quantify these, we assumed that each white dwarf
required some additional variance in $\ln \Teff$ and $\log g$ that is the same for all objects.
We then sought to determine these systematic uncertainties via a maximum-likelihood approach.
For a quantity $x$ (where $x$ is either $\ln \Teff$ or $\log g$) observed for a single
system $N$ times (in our case $N=2$), the $i$th
observation, $x_i$, can be considered to be drawn from a normal distribution with unknown
mean $\mu$ and variance $\sigma^2$, with additional measurement uncertainty, $e_i$.
The likelihood of all $N$ observations is therefore
\begin{equation}
    L(\mathbf{x}|\mu, \sigma, \mathbf{e}) = \prod_i^N \frac{1}{\sqrt{2\pi v_i}}
    \exp\left(-\frac{(x_i-\mu)^2}{2 v_i}\right),
\end{equation}
where $v_i = \sigma^2 + e_i^2$. Since we are not specifically interested in $\mu$, only $\sigma$,
we can integrate over $\mu$, reducing the dimensionality to a marginal likelihood
that depends only $\sigma$. This integral has an analytic solution which can be written as
\begin{equation}
    \ln L(\mathbf{x}|\sigma,\mathbf{e}) = k - \frac{1}{2}\left(\ln a -b^2/a + c + d\right),
\end{equation}
where $k$ is a normalising constant, and $a$--$d$ are sums over the observations given by
\begin{equation}
    a = \sum_i^N \frac{1}{v_i},\quad
    b = \sum_i^N \frac{x_i}{v_i},\quad
    c = \sum_i^N \frac{x_i^2}{v_i},\quad
    d = \sum_i^N \ln v_i,
\end{equation}
and are themselves functions of $\sigma$. Finally, for $M$ systems with index $j$,
the log-likelihood (dropping proportionality constants)
\begin{equation}
    \ln L(\mathbf{x}_1,\ldots,\mathbf{x}_M|\sigma,\mathbf{e}_1,\ldots,\mathbf{e}_M) = 
    \sum_j^M \ln L(\mathbf{x}_j|\sigma,\mathbf{e}_j).
\end{equation}
Because of the marginalisation step, the final likelihood depends only on $\sigma$.
Therefore determining the maximum-likelihood is a simple case of plotting as a function
of $\sigma$ and visually determining the location of the maximum to whatever precision
is desired.

For our 13 warm flux standard candidate white dwarfs, each with two observations, we determined systematic uncertainties of 1.1 per cent in \Teff\ (i.e. 0.011 in $\ln \Teff$), and $0.01$ dex in $\log g$. Data from the extra flux standard candidates is not included in the systematic uncertainty determination due to their SED fits spanning different wavelength ranges to the other 13 candidates. We apply the systematic uncertainties in \Teff\ and $\log g$ to all 17 flux standard candidates and the two primary standards. The combined statistical and systematic uncertainties are displayed in Table~\ref{tab:fit_results} with the best-fitting \Teff\ and $\log g$ parameters for our network and the two primary standards. The best-fitting extinction value is also quoted in Table~\ref{tab:fit_results} from each SED fit to three decimal places. 

\begin{figure*}
	\includegraphics[width=1.42\textwidth,height=0.36\textheight,keepaspectratio]{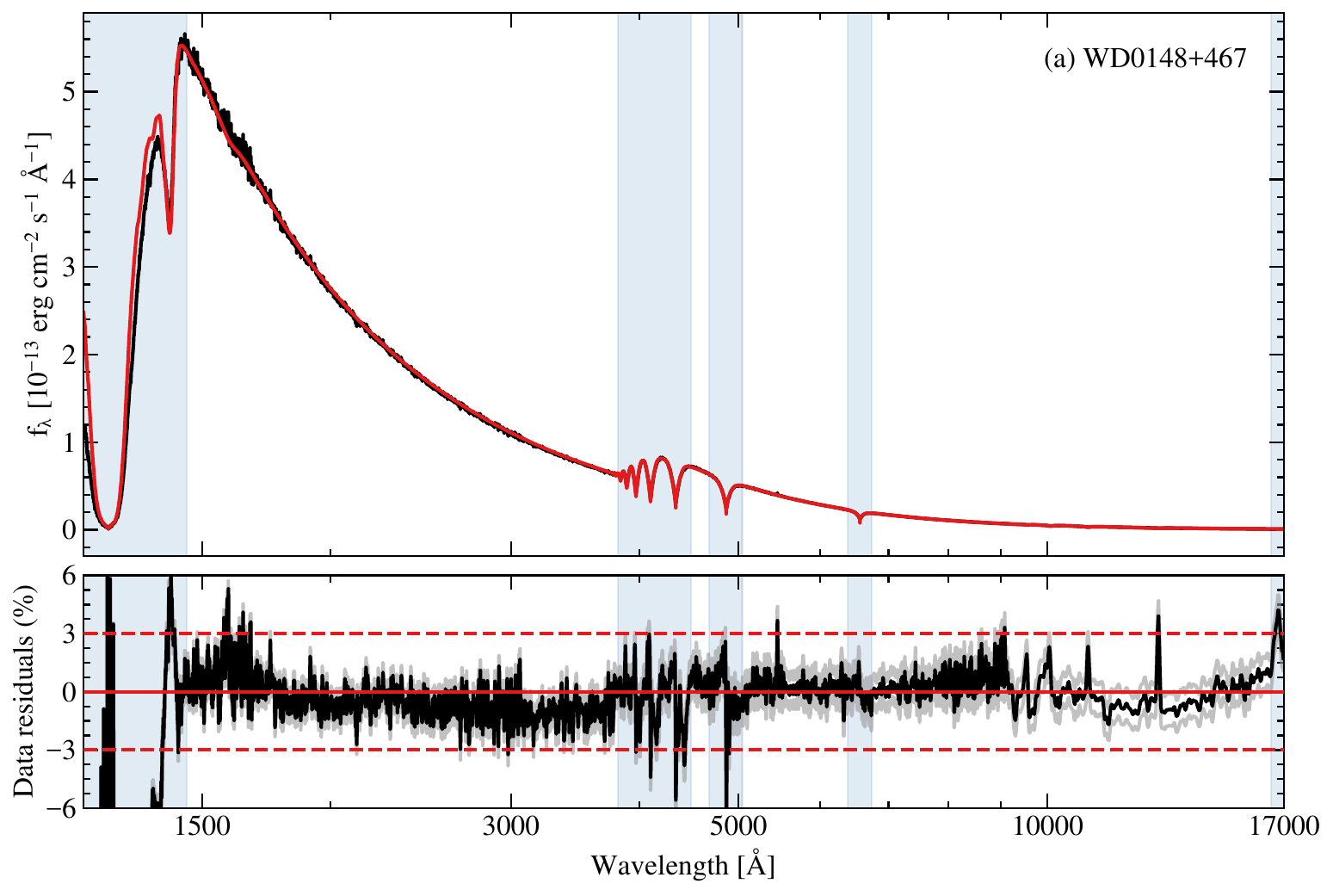}
	\includegraphics[width=0.45\textwidth,height=0.37\textheight,keepaspectratio]{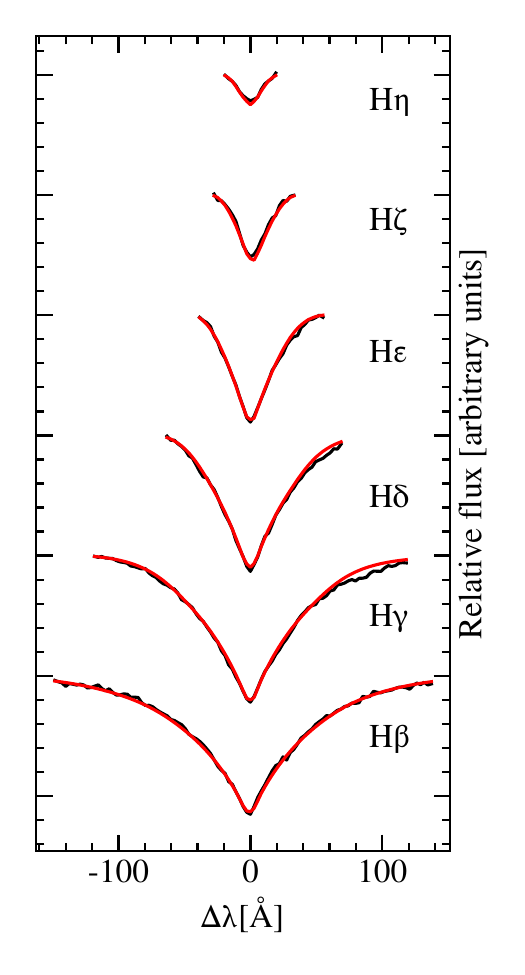}
 	\includegraphics[width=1.42\textwidth,height=0.36\textheight,keepaspectratio]{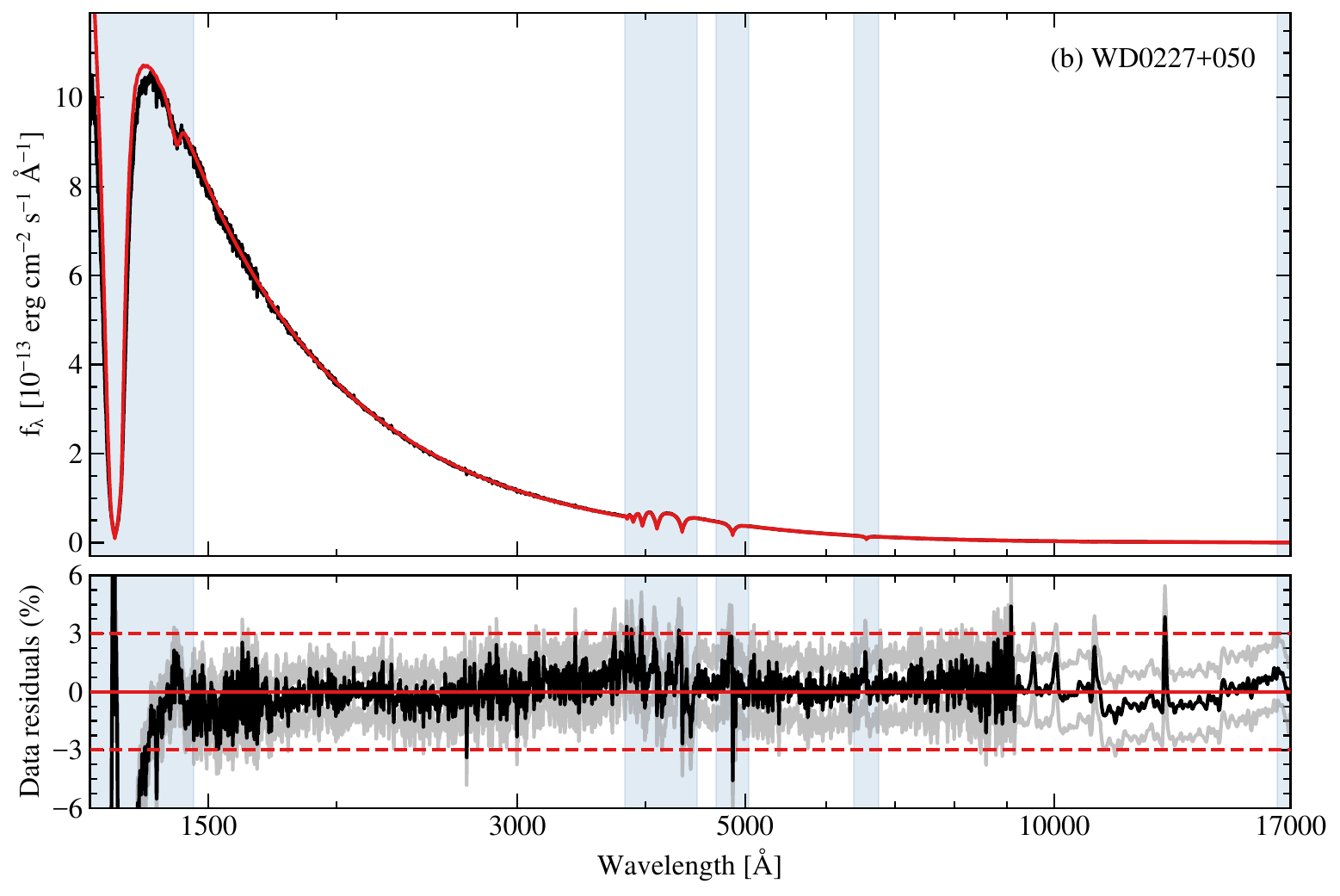}
	\includegraphics[width=0.45\textwidth,height=0.37\textheight,keepaspectratio]{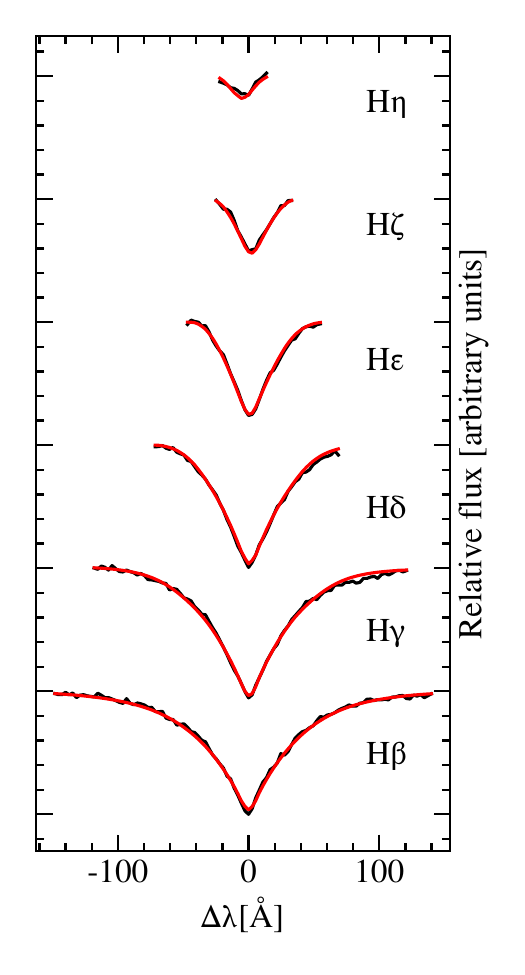}
    \caption{Spectrophotometric fits of the STIS and WFC3 data for two of the 13 warm white dwarfs proposed as flux standard candidates, with the WD name of each star given in the corner of the top left panels. The spectrophotometric fits of the remaining 11 warm white dwarfs are in Figures~\ref{fig:SED_Balmer_warm_appendix}. \textit{Top left}: SED fit between the observed spectrophotometry (black) and best-fitting model (red). \textit{Bottom left}: Flux residuals from the corresponding SED fit, where the black line is the calculated residual, grey lines indicate residuals $\pm 1\sigma$ using only the statistical errors from the fits, and red lines show residuals of 0 and $\pm 3$ per cent as a guide. The shaded blue regions in the left panels indicate wavelength ranges excluded from the fits. \textit{Right}: Balmer line fits for H$\beta$ to H$\eta$ between the observed spectrophotometry (black) and best-fitting model (red). The line profiles are vertically offset for clarity. The best-fitting parameters for the SED and Balmer line fits are found in Table~\ref{tab:fit_results}.}
    \label{fig:SED_Balmer_warm}
\end{figure*}

The SED fits between the observed spectrophotometry and best-fitting models for the 13 warm and four extra flux standard candidates are shown in the top left panels of Figures~\ref{fig:SED_Balmer_warm} (continued in Figure~\ref{fig:SED_Balmer_warm_appendix}) and \ref{fig:SED_Balmer_extra}, respectively. The SED flux residuals are shown in the bottom left panels of the same figures. The best-fitting SED models for the flux standard candidates include the parameters \Teff, $\log g$, $E(B-V)$, radius and mass. The best-fitting SED models and parameters are available with the electronic distribution of this paper and on CALSPEC, except the models for WD\,1202$-$232 and WD\,1544$-$377 which are not included in CALSPEC due to potential modelling issues and/or contamination in the IR (see Section~\ref{sec:fluxes}). The models for WD\,1202$-$232 and WD\,1544$-$377 included in the electronic distribution of this paper are cropped at $2300 - 12\,700$\,\AA\ and $2400 - 12\,500$\,\AA, respectively, as the flux residuals are majoritively within 5\,per cent in these regions.

We calculated the standard deviation of the flux residuals for all 17 candidates and the two hot primary standards over the entire wavelength range of each SED fit, in addition to the UV and combined optical+NIR ranges. The percentages are given in Table~\ref{tab:std_flux_residuals}. 

\begin{table*}
    \centering
    \caption{Atmospheric parameters of the 17 DA white dwarfs proposed as flux standards from two fitting methods: 1) SED fit of the continuum and \textit{Gaia} DR3 parallax, omitting the Balmer regions; 2) Balmer line fit of H$\beta$ - H$\eta$. The uncertainties on the SED fits correspond to $1\sigma$ and are a combination of the statistical errors from the fits and the systematic errors (1.1 per cent in \Teff\ and 0.01\,dex in $\log g$) obtained in Section~\ref{sec:SED_fitting}. The quoted uncertainties on the Balmer line fits correspond to $1\sigma$ and are purely of statistical nature as derived from the fits. These Balmer line fit uncertainties are found to be underestimated in Section~\ref{sec:atmospheric_parameters} hence we recommend the reader to apply a scale factor of $\times$3 for further analysis. $\sigma_{\Teff}$ and $\sigma_{\log g}$ correspond to how many $\sigma$ apart the best-fitting \Teff\ and $\log g$ parameters are from the SED and Balmer line fits. $E(B-V)$ values which are quoted as 0.000 may be non-zero (<0.0005\,mag) in the fit and analysis. }
    \label{tab:fit_results}
    \begin{tabular}{cccccccc}
        \hline
		\hline
         Object & \multicolumn{2}{c}{SED fit} & \multicolumn{2}{c}{Balmer line fit} & $E(B-V)$ & $\sigma_{\Teff}$ & $\sigma_{\log g}$ \\
         & \Teff &  $\log g$ &  \Teff &  $\log g$ &  ($R_V = 3.1$) &  &  \\
         &  [K] &  [dex] &  [K] &  [dex] &  [mag] &  &  \\
        \hline
		\hline
        \multicolumn{8}{c}{\textbf{Hot primary standards}}\\
        \hline
        GD\,153 & 40120 $\pm$ 443 & 7.818 $\pm$ 0.021& 39240 $\pm$ 251 & 7.750 $\pm$ 0.026 & 0.000 & 1.007 & 0.843\\
        GD\,71 & 33379 $\pm$ 367 & 7.822 $\pm$ 0.017& 33054 $\pm$ 107 & 7.780 $\pm$ 0.020 & 0.000 & 0.666 & 0.673\\
		\hline
        \multicolumn{8}{c}{\textbf{Warm flux standard candidates}}\\
        \hline
         WD\,0148$+$467 & 14483 $\pm$ 160 & 8.025 $\pm$ 0.012& 13784 $\pm$ 81 & 8.032 $\pm$ 0.014 & 0.000 & 2.403 & 0.182\\
         WD\,0227$+$050 & 19190 $\pm$ 213 & 7.903 $\pm$ 0.017& 19093 $\pm$ 55 & 7.894 $\pm$ 0.013 & 0.000 & 0.362 & 0.200\\
         WD\,0809$+$177 & 16371 $\pm$ 181 & 8.078 $\pm$ 0.015& 16152 $\pm$ 73 & 8.102 $\pm$ 0.015 & 0.000 & 0.767 & 0.509\\
         WD\,1105$-$340 & 14099 $\pm$ 156 & 8.102 $\pm$ 0.014& 13618 $\pm$ 190 & 8.120 $\pm$ 0.029 & 0.000 & 0.815 & 0.198\\
         WD\,1105$-$048 & 15826 $\pm$ 175 & 7.940 $\pm$ 0.015& 15602 $\pm$ 74 & 7.949 $\pm$ 0.014 & 0.000 & 2.795 & 0.199\\
         WD\,1327$-$083 & 15100 $\pm$ 167 & 8.007 $\pm$ 0.015& 14368 $\pm$ 103 & 7.984 $\pm$ 0.019 & 0.004 & 2.091 & 0.356\\
         WD\,1713$+$695 & 16063 $\pm$ 178 & 8.026 $\pm$ 0.015& 15629 $\pm$ 71 & 7.979 $\pm$ 0.014 & 0.000 & 1.563 & 1.072\\
         WD\,1911$+$536 & 17424 $\pm$ 192 & 8.316 $\pm$ 0.016& 17198 $\pm$ 71 & 8.335 $\pm$ 0.012 & 0.000 & 0.784 & 0.481\\
         WD\,1919$+$145 & 15364 $\pm$ 171 & 8.174 $\pm$ 0.017& 14750 $\pm$ 67 & 8.173 $\pm$ 0.015 & 0.000 & 2.335 & 0.003 \\
         WD\,2039$-$682 & 17076 $\pm$ 188 & 8.497 $\pm$ 0.016& 16726 $\pm$ 76 & 8.502 $\pm$ 0.016 & 0.002 & 1.182 & 0.088\\
         WD\,2117$+$539 & 15583 $\pm$ 172 & 7.949 $\pm$ 0.016& 14541 $\pm$ 72 & 7.921 $\pm$ 0.013 & 0.002 & 3.768 & 0.647\\
         WD\,2126$+$734 & 16050 $\pm$ 177 & 7.967 $\pm$ 0.015& 15535 $\pm$ 65 & 7.967 $\pm$ 0.013 & 0.003 & 1.953 & 0.005 \\
         WD\,2149$+$021 & 17692 $\pm$ 196 & 8.009 $\pm$ 0.016& 17695 $\pm$ 72 & 8.019 $\pm$ 0.015 & 0.000 & 0.011 & 0.221\\
        \hline
        \multicolumn{8}{c}{\textbf{Extra flux standard candidates}}\\
        \hline
        WD\,0352$+$096 & 14810 $\pm$ 164 & 8.320 $\pm$ 0.015& 14000 $\pm$ 260 & 8.341 $\pm$ 0.023 & 0.000 & 1.017 & 0.308\\
        WD\,1202$-$232 & 8854 $\pm$ 100 & 8.024 $\pm$ 0.023& 8736 $\pm$ 21 & 7.980 $\pm$ 0.019 & 0.008 & 0.999 & 0.716\\
        WD\,1544$-$377 & 10303 $\pm$ 121 & 7.999 $\pm$ 0.033& 10581 $\pm$ 46 & 8.012 $\pm$ 0.025 & 0.003 & 1.517 & 0.154\\
        WD\,2341$+$322 & 12852 $\pm$ 142 & 8.027 $\pm$ 0.013& 12775 $\pm$ 128 & 8.036 $\pm$ 0.021 & 0.000 & 0.190 & 0.143\\
    \end{tabular}
\end{table*}

\begin{figure*}
	\includegraphics[width=1.42\textwidth,height=0.36\textheight,keepaspectratio]{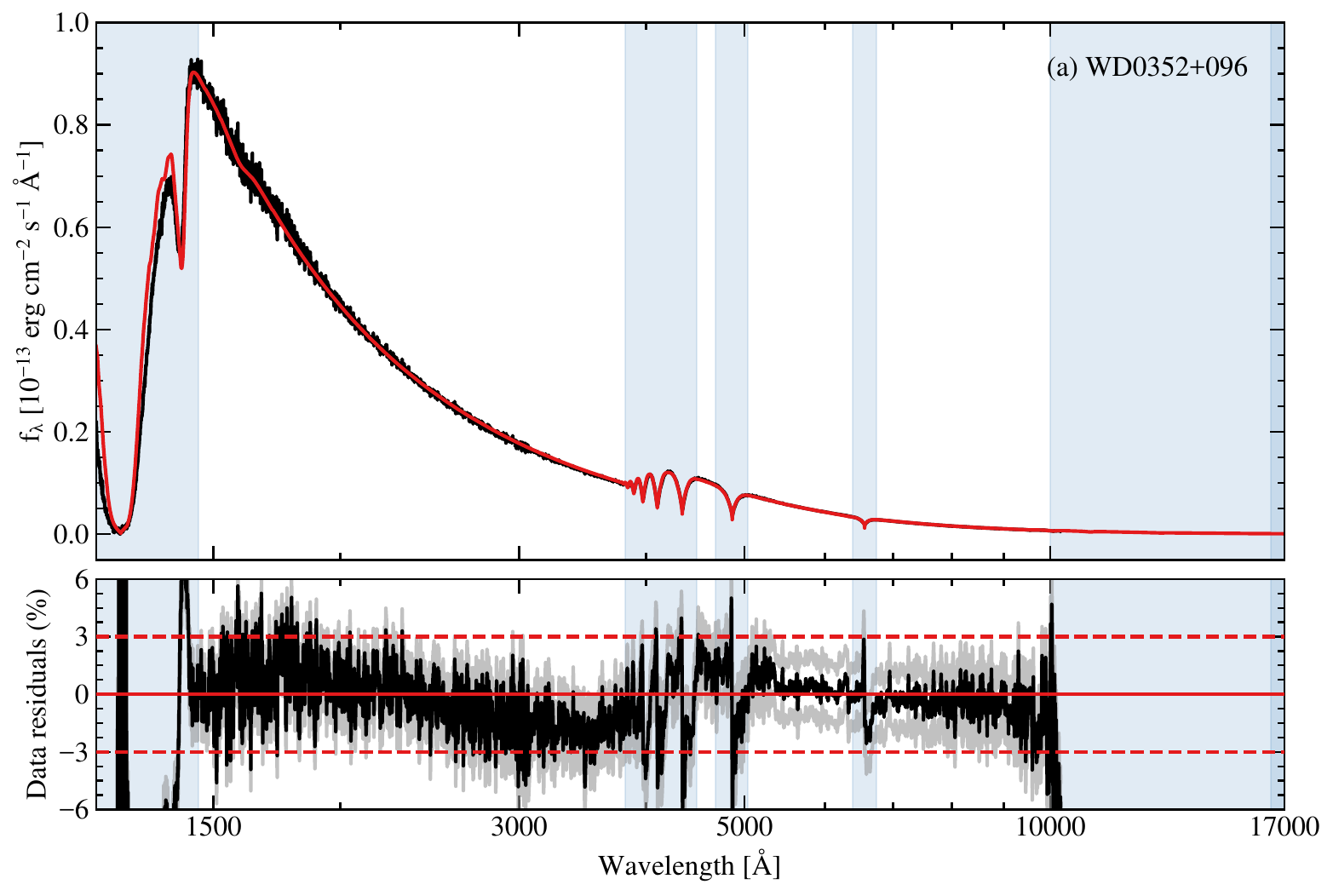}
	\includegraphics[width=0.45\textwidth,height=0.37\textheight,keepaspectratio]{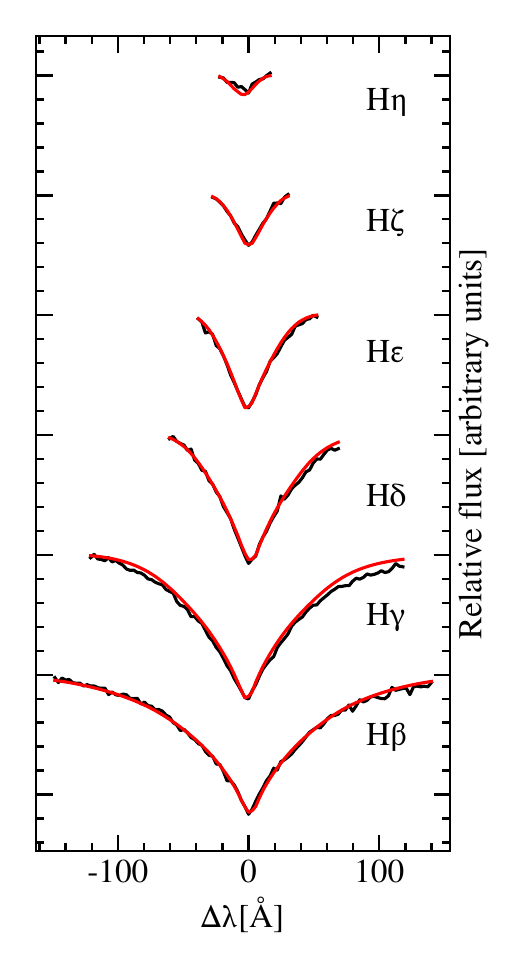}
 	\includegraphics[width=1.42\textwidth,height=0.36\textheight,keepaspectratio]{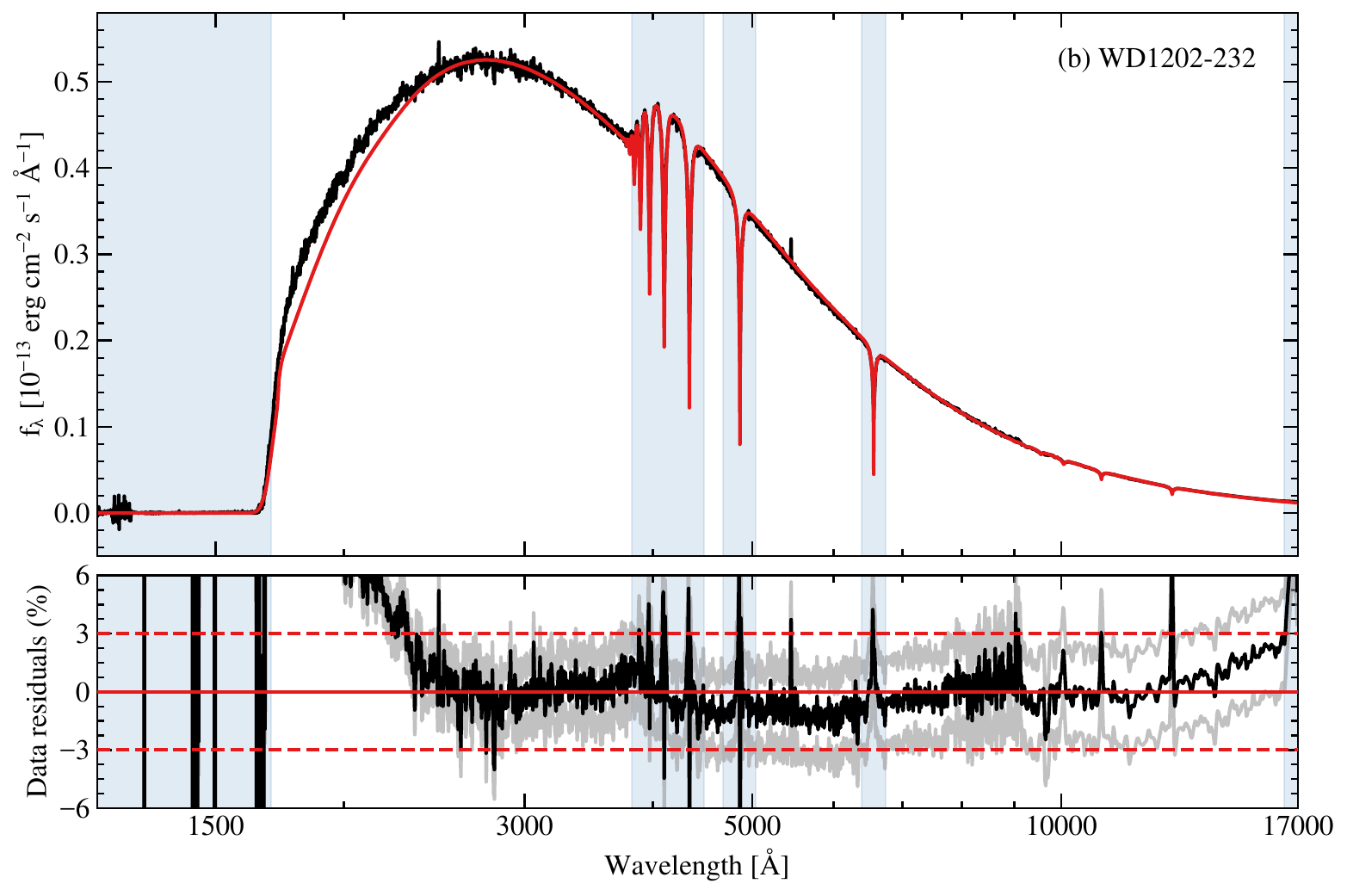}
	\includegraphics[width=0.45\textwidth,height=0.37\textheight,keepaspectratio]{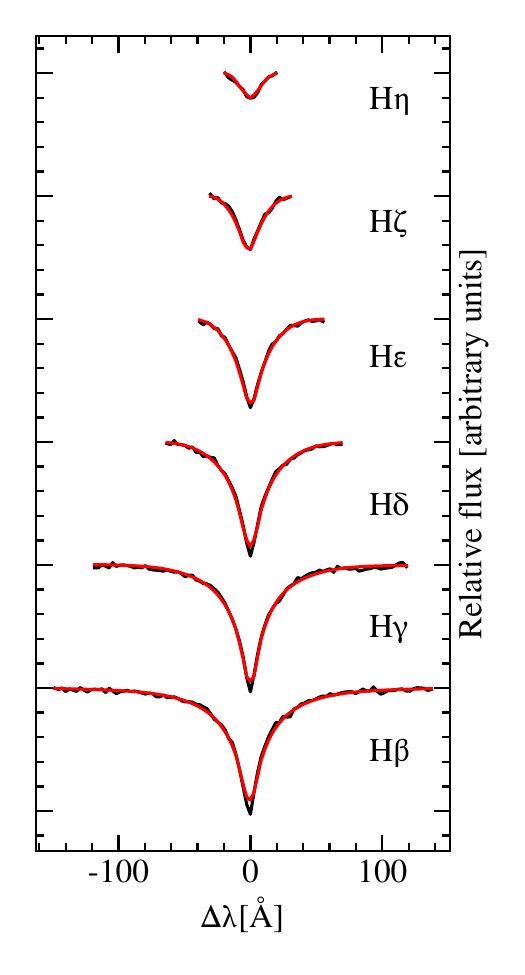}
    \caption{Similar to Figure~\ref{fig:SED_Balmer_warm} but for STIS and WFC3 spectrophotometry of the four extra white dwarfs proposed as flux standard candidates. (a) WD\,0352$+$096 does not have WFC3 spectrophotometry.}
    \label{fig:SED_Balmer_extra}
\end{figure*}

\begin{figure*}
\contcaption{}
	\includegraphics[width=1.42\textwidth,height=0.36\textheight,keepaspectratio]{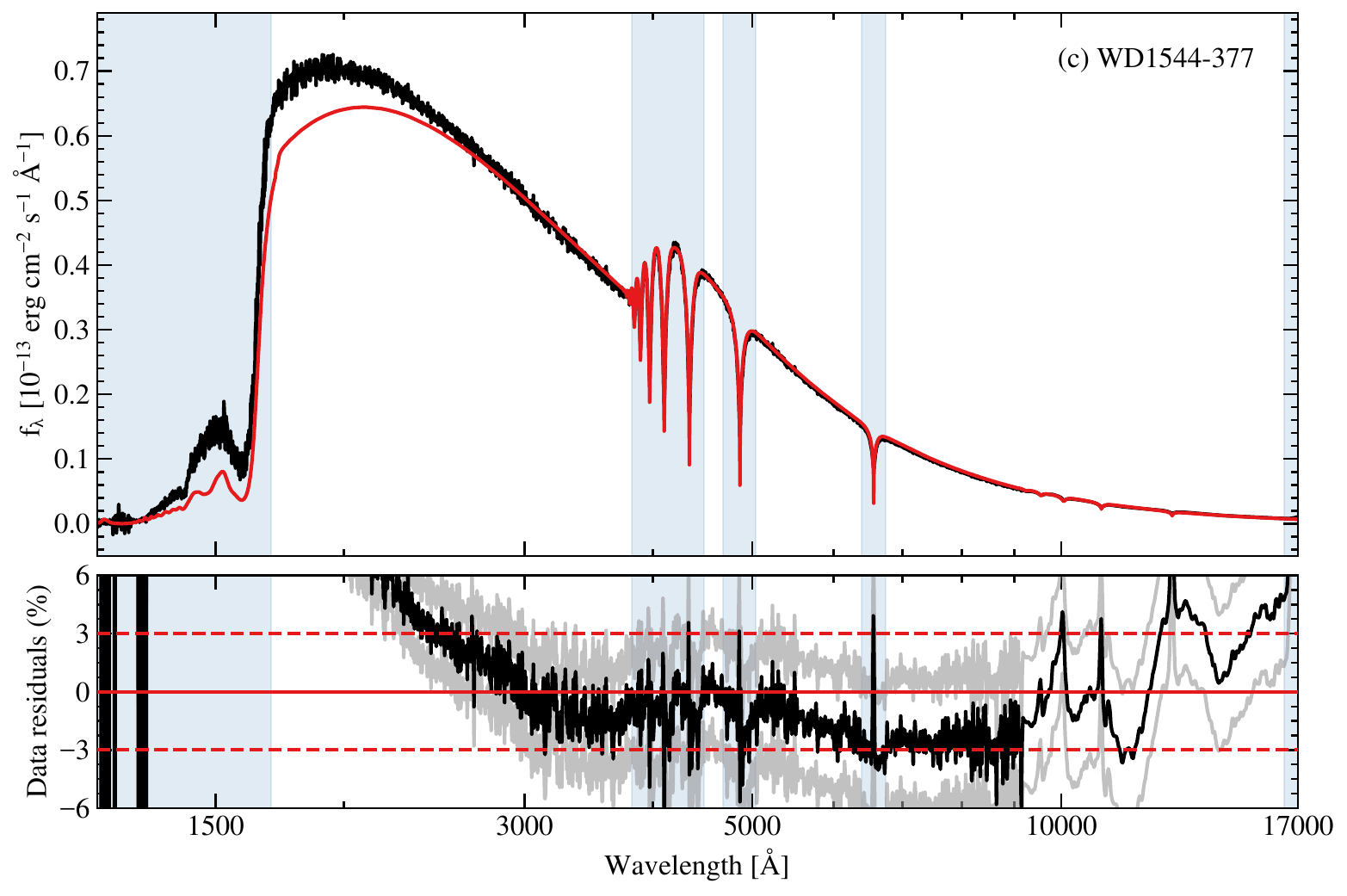}
	\includegraphics[width=0.45\textwidth,height=0.37\textheight,keepaspectratio]{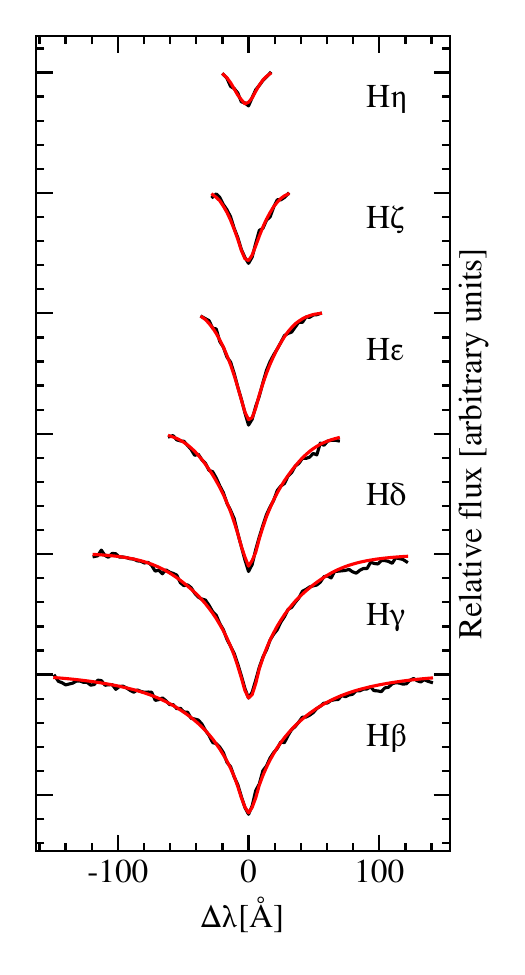}
 	\includegraphics[width=1.42\textwidth,height=0.355\textheight,keepaspectratio]{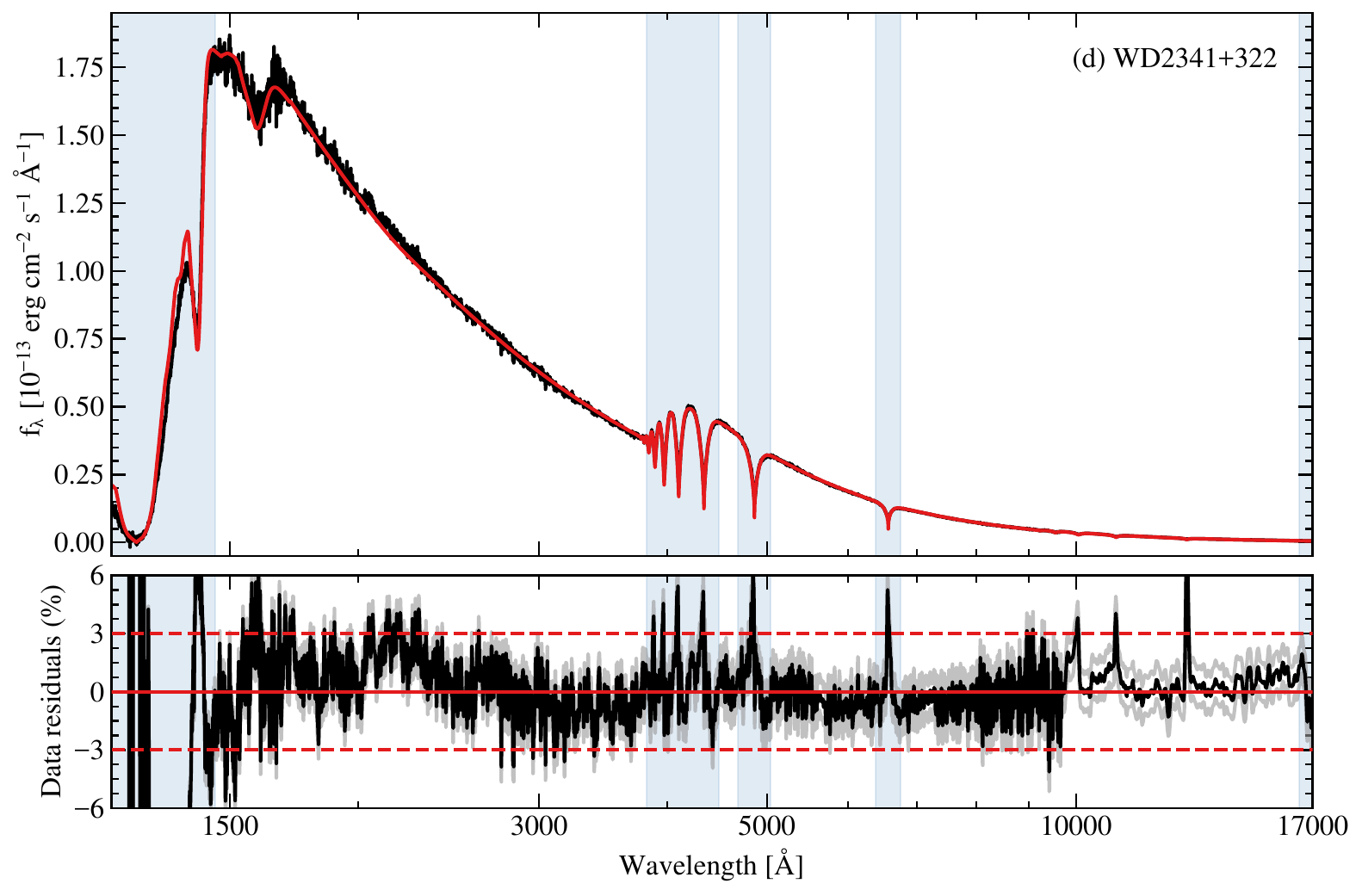}
	\includegraphics[width=0.45\textwidth,height=0.37\textheight,keepaspectratio]{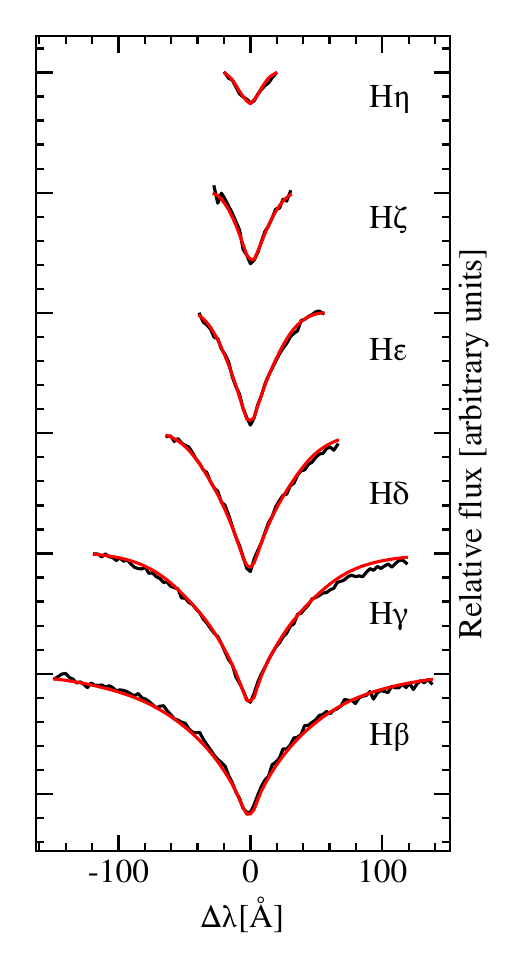}
\end{figure*}

\begin{table*}
    \centering
    \caption{The standard deviation flux residual across three different wavelength ranges of the SED fit: the entire wavelength range of each SED fit; the UV region; the optical+NIR region. Quantities are given as a percentage.}
    \begin{tabular}{cccc}
        \hline
		\hline
        Object & Entire wavelength range & UV & optical+NIR \\
         &  per cent &  per cent &  per cent \\
        \hline
		\hline
        \multicolumn{4}{c}{\textbf{Hot primary standards}}\\
        \hline
        GD\,153 & 0.36 & 0.31 & 0.36 \\
        GD\,71 & 0.33 & 0.30 & 0.35 \\
		\hline
         \multicolumn{4}{c}{\textbf{Warm flux standard candidates}}\\
         \hline
         WD\,0148$+$467 &  1.24 & 1.30 & 1.04 \\
         WD\,0227$+$050 &  1.12 & 1.17 & 0.99 \\
         WD\,0809$+$177 &  1.43 & 1.58 & 1.26 \\
         WD\,1105$-$340 &  2.58 & 2.68 & 1.59 \\
         WD\,1105$-$048 &  1.31 & 1.50 & 1.12 \\
         WD\,1327$-$083 &  1.49 & 1.43 & 1.19 \\
         WD\,1713$+$695 &  1.41 & 1.56 & 1.21 \\
         WD\,1911$+$536 &  1.31 & 1.35 & 1.18 \\
         WD\,1919$+$145 &  1.38 & 1.63 & 1.10 \\
         WD\,2039$-$682 & 1.33 & 1.47 & 1.15 \\
         WD\,2117$+$539 & 1.49 & 1.33 & 0.91 \\
         WD\,2126$+$734 & 1.23 & 1.30 & 0.98 \\
         WD\,2149$+$021 & 1.34 & 1.23 & 1.01 \\
        \hline
         \multicolumn{4}{c}{\textbf{Extra flux standard candidates}}\\
         \hline
         WD\,0352$+$096 & 2.08 & 2.22 & 1.52 \\
         WD\,1202$-$232 & 4.61 & 2.34 & 0.96 \\
         WD\,1544$-$377 & 4.42 & 1.27 & 2.44 \\
         WD\,2341$+$322 & 1.89 & 2.19 & 1.28 \\
    \end{tabular}
    \label{tab:std_flux_residuals}
\end{table*}

\subsubsection{Balmer line fitting}
\label{sec:Balmer_fitting}
For each of our 17 flux standard candidates, an independent set of \Teff\ and $\log g$ values can be obtained by directly comparing the Balmer line profiles with our sets of models. We do this by first continuum normalising both the models and the STIS spectra by fitting a spline, and then cropping out each Balmer line (e.g. Figure~\ref{fig:SED_Balmer_warm} right panels). We then fit the cropped H$\beta$ to H$\eta$ absorption features with our DA model atmospheres using the same \texttt{scipy optimize} \texttt{trf} reduced $\chi^2$ minimisation technique as our SED fits. \Teff, $\log g$ and radial velocity were free parameters in the fits. The bounds for \Teff\ and $\log g$ were determined by the model grid limits, whereas radial velocity was bounded by the range $-300 - 300$\,\kms. The same DA $M$-$R$ relation as in our SED fits (Section~\ref{sec:SED_fitting}) was used to determine $R$ (and white dwarf mass) using the best-fitting \Teff\ and $\log g$. The H$\eta$ Balmer line was excluded from the fits for GD\,153 and GD\,71 because it has negligible equivalent width for these hot stars.

Unlike the SED fits, for Balmer line modelling the arbitrarily high natural resolution of our DA models needs to be degraded to match that of STIS. This can be achieved by convolving the models with the appropriate STIS grating Line Spread Function (LSF) at the slit width matching that of our observations. Tabulated LSFs are available at the \textit{HST} STIS web pages\footnote{\url{https://www.stsci.edu/hst/instrumentation/stis/performance/spectral-resolution}}, which specifies the shape of a monochromatic line feature as observed with the specified instrumental resolution and provides an empirical description of the light distribution along the primary dispersion axis. H$\beta$ to H$\eta$ are covered by the G430L grating and H$\alpha$ is covered by the G750L grating. In our analysis we opt to exclude H$\alpha$ allowing us to use only the G430L tables with no need to splice together models at different resolutions. The Rel Pixel column shows the exact relative spacing of the grating in pixels, this is multiplied by the the appropriate grating pixel dispersion in order to obtain the spacing in Angstroms. The G430L grating is split between 3200\,\AA\ and 5500\,\AA\ LSF tables therefore we averaged them to give a full coverage overview. 

Convolving our models directly with the LSF proved computationally expensive and not feasible for use on large numbers of stars. Furthermore, severe edge effects arise at the end points of the LSF grid resulting in warping of the model spectra. We instead decided to fit the LSF tabulated data and create a function which closely reproduces the relative pixel spacing values of the grating. We used the sum of three Gaussian profiles and produced a function which matches the actual LSF data to within 1 per cent. This custom LSF function was then used in the convolution to degrade our models to the resolution of the STIS G430L grating.

The Balmer line best-fitting model to the observed spectrophotometry for the 13 warm and four extra flux standard candidates are shown in the right panels of Figures~\ref{fig:SED_Balmer_warm} (continued in Figure~\ref{fig:SED_Balmer_warm_appendix}) and \ref{fig:SED_Balmer_extra}, respectively. The best-fitting \Teff\ and $\log g$ are displayed in Table~\ref{tab:fit_results}. The uncertainties of the fit parameters are dependent on one another and are computed from the covariance matrix of the model atmosphere fit scaled by reduced $\chi^2$ to account for the goodness of fit, therefore they are statistical in nature and do not account for systematic uncertainty.

\subsection{Synthetic magnitudes and photometric fitting}
\label{sec:Synthetic_magnitudes_and_photometric_fitting}

Synthetic magnitudes were computed for the \textit{Gaia} DR3 $G$, $G_{\rm BP}$ and $G_{\rm RP}$ bands, 2MASS $J$, $H$ and $K_S$ bands, and WISE $W$1 and $W$2 bands using our best-fitting STIS and WFC3 \Teff\ and $\log g$ parameters for the 17 flux standard candidates, in addition to GD\,153 and GD\,71 (Table~\ref{tab:fit_results}). Synthetic apparent magnitudes, $m$, for \textit{Gaia} DR3 photometry were computed by integrating under the model curves for each filter bandpass using
\begin{equation}
    m = -2.5 \log \left( \frac{\int T(\lambda) f_{\lambda}(\lambda,\Teff,\log{g}) \lambda d\lambda}{\int T(\lambda) \lambda d\lambda} \right) + c, 
    \label{ref:Gaia_phot_fit_eq}
\end{equation}
{\noindent}where $\lambda$ is the wavelength, $T(\lambda)$ is the filter transmission function for any \textit{Gaia} DR3 filter, obtained from the Spanish Virtual Observatory (SVO) Filter Profile Service \citep{Rodrigo2012}, $f_{\lambda}(\lambda,\Teff,\log{g})$ is the model flux at the distance of the white dwarf from Equation~\ref{flux_scaling}, and $c$ is the zeropoint magnitude for the given filter. 

For 2MASS and WISE, synthetic magnitudes were computed using 
\begin{equation}
    m = -2.5 \log \left( \frac{\int T(\lambda) f_{\lambda}(\lambda,\Teff,\log{g}) d\lambda}{\int T(\lambda) d\lambda} \right) + c,
\end{equation}
{\noindent}where filter transmission functions were also obtained from SVO.

All synthetic and observed magnitudes are in the Vega magnitude system and are displayed in Table~\ref{tab:obs_syn_photometry}. The Vega SEDs from CALSPEC used for synthetic magnitudes were a modified version of alpha$\_$lyr$\_$mod$\_$002.fits\footnote{\url{https://gea.esac.esa.int/archive/documentation/GDR3/Data_processing/chap_cu5pho/cu5pho_sec_photProc/cu5pho_ssec_photCal.html\#SSS3.P2}} in the UV and optical range \citep[to $\approx 1$\,$\mu$m, i.e. \textit{Gaia} filters;][]{Busso2022} and alpha$\_$lyr$\_$stis$\_$011.fits beyond 1\,$\mu$m (i.e. 2MASS and WISE filters). 

We corrected the synthetic magnitudes for reddening by evaluating the extinction $A_{\lambda}$ using the G23 model \citep{Gordon2023} at the effective wavelengths of each filter obtained from SVO. We used $R(V) = 3.1$ and the best-fitting $E(B-V)$ values from our SED fits in Section~\ref{sec:SED_fitting} given in Table~\ref{tab:fit_results}. 

In the NASA/IPAC (Infrared Processing and Analysis Center) Infrared Science Archive, WD\,2117$+$539 has no observed WISE data and WD\,1544$-$377 has no observed 2MASS nor WISE data. WD\,1919$+$145 has no observed 2MASS $J$ magnitude error meaning the reported $J$ magnitude is a 95 per cent confidence upper limit, thus we exclude it from this work.

\begin{table*}
    \centering
    \caption{Observed and synthetic photometry in the \textit{Gaia} $G$, $G_{\rm BP}$ and $G_{\rm RP}$ bands, 2MASS $J$, $H$ and $K_S$ bands, and WISE $W$1 and $W$2 bands. The first row for each object is the observed photometry and error, then the second row is the synthetic magnitude computed from our best-fitting STIS and WFC3 \Teff\ and $\log g$ parameters. Values in parentheses are 1$\sigma$ errors in mmag. WD\,2117$+$539 has no WISE data, WD\,1544$-$377 has no 2MASS nor WISE data, and WD\,1919$+$145 only has an upper limit for the 2MASS $J$ magnitude, so those fields are filled with dashes.}
    \begin{tabular}{ccccccccc}
    \hline
    \hline
    Object & \multicolumn{3}{c}{\textit{Gaia}} & \multicolumn{3}{c}{2MASS} & \multicolumn{2}{c}{WISE} \\
    & $G$ & $G_{\rm BP}$ & $G_{\rm RP}$ & $J$ & $H$ & $K_S$ & $W1$ & $W2$ \\
    & [mag] & [mag] & [mag] & [mag] & [mag] & [mag] & [mag] & [mag]\\
    \hline
    \hline
    \multicolumn{9}{c}{\textbf{Hot primary standards}}\\
    \hline
    GD\,153 & 13.311(3) & 13.151(3) & 13.632(4) & 14.012(25) & 14.209(37) & 14.308(62)  & 14.374(27) & 14.506(52) \\
    & 13.288& 13.118& 13.616& 14.048& 14.162& 14.271& 14.353& 14.393\\
    GD\,71 & 13.000(3) & 12.853(3) & 13.305(4) & 13.728(25) & 13.901(35) & 14.115(65)  & 14.012(29) & 14.103(51) \\
    & 12.979& 12.820& 13.287& 13.701& 13.810& 13.921& 14.005& 14.052\\
    \hline
    \multicolumn{9}{c}{\textbf{Warm flux standard candidates}}\\
    \hline
    WD\,0148$+$467 & 12.496(3) & 12.469(3) & 12.588(4) & 12.768(24) & 12.826(32) & 12.846(30) & 12.906(24) & 12.952(27) \\
    & 12.499& 12.456& 12.588& 12.771& 12.801& 12.864& 12.893& 12.904\\
    WD\,0227$+$050 & 12.821(3) & 12.745(3) & 12.996(4) & 13.282(26) & 13.367(33) & 13.425(36) & 13.435(24) & 13.468(33) \\
    & 12.827& 12.737& 13.003& 13.276& 13.336& 13.416& 13.466& 13.485\\
    WD\,0809$+$177 & 13.429(3) & 13.376(3) & 13.552(4) & 13.762(25) & 13.841(36) & 13.941(51) & 13.970(26) & 14.069(47) \\
    & 13.425& 13.361& 13.553& 13.778& 13.822& 13.892& 13.930& 13.944\\
    WD\,1105$-$340 & 13.700(3) & 13.686(3) & 13.771(4) & 13.954(28) & 13.982(39) & 14.054(66) & 13.951(25) & 13.898(37) \\
    & 13.661& 13.624& 13.739& 13.913& 13.939& 14.001& 14.029& 14.040\\
    WD\,1105$-$048 & 13.091(3) & 13.045(3) & 13.214(4) & 13.405(26) & 13.445(30) & 13.544(56) & 13.570(25) & 13.574(36) \\
    & 13.096& 13.037& 13.217& 13.429& 13.471& 13.537& 13.573& 13.585\\
    WD\,1327$-$083 & 12.355(3) & 12.323(3) & 12.458(4) & 12.621(37) & 12.677(41) & 12.736(48) & 13.328(27) & 13.165(29) \\
    & 12.351& 12.303& 12.452& 12.644& 12.679& 12.743& 12.774& 12.785\\
    WD\,1713$+$695 & 13.328(3) & 13.282(3) & 13.450(4) & 13.618(23) & 13.720(29) & 13.739(55) & 13.811(24) & 13.843(27) \\
    & 13.317& 13.257& 13.441& 13.659& 13.701& 13.770& 13.806& 13.820\\
    WD\,1911$+$536 & 13.247(3) & 13.189(3) & 13.391(4) & 13.616(29) & 13.734(46) & 13.824(47) & 13.795(24) & 13.885(29) \\
    & 13.251& 13.179& 13.395& 13.642& 13.690& 13.766& 13.810& 13.826\\
    WD\,1919$+$145 & 13.019(3) & 12.980(3) & 13.126(4) & -& 13.452(55) & 13.546(66) & 12.412(47) & 12.546(80) \\
    & 13.016& 12.964& 13.122& 13.326& 13.362& 13.429& 13.463& 13.476\\
    WD\,2039$-$682 & 13.339(3) & 13.288(3) & 13.469(4) & 13.729(26) & 13.806(39) & 13.800(50) & 13.879(26) & 13.934(38) \\
    & 13.330& 13.265& 13.463& 13.703& 13.746& 13.822& 13.864& 13.880\\
    WD\,2117$+$539 & 12.393(3) & 12.355(3) & 12.496(4) & 12.681(21) & 12.785(23) & 12.85(38) & - & - \\
    & 12.361& 12.306& 12.475& 12.680& 12.719& 12.785& 12.819& 12.831\\
    WD\,2126$+$734 & 12.887(3) & 12.836(3) & 12.989(5) & 13.096(31) & 13.164(38) & 13.166(44) & 13.075(23) & 13.110(25) \\
    & 12.879& 12.819& 13.001& 13.214& 13.256& 13.323& 13.359& 13.372\\
    WD\,2149$+$021 & 12.777(3) & 12.713(3) & 12.930(4) & 13.203(24) & 13.286(37) & 13.397(37) & 13.385(25) & 13.379(32)\\
    & 12.783& 12.706& 12.936& 13.185& 13.238& 13.313& 13.357& 13.373\\
    \hline
    \multicolumn{9}{c}{\textbf{Extra flux standard candidates}}\\
    \hline
    WD\,0352$+$096 & 14.548(3) & 14.520(3) & 14.635(4) & 14.831(41) & 14.866(59) & 15.061(99) & 14.938(35) & 15.116(81) \\
    & 14.544& 14.500& 14.634& 14.827& 14.856& 14.923& 14.956& 14.969\\
    WD\,1202$-$232 & 12.738(3) & 12.843(3) & 12.548(4) & 12.402(24) & 12.301(27) & 12.342(26) & 12.318(25) & 12.343(24) \\
    & 12.725& 12.814& 12.542& 12.435& 12.331& 12.357& 12.366& 12.368\\
    WD\,1544$-$377 & 13.001(3) & 13.036(3) & 12.931(4) & - & - & - & - & -\\
    & 12.993& 13.028& 12.917& 12.944& 12.907& 12.951& 12.965& 12.973\\
    WD\,2341$+$322 & 12.967(3) & 12.964(3) & 13.007(4) & 13.171(29) & 13.195(37) & 13.179(28) & 13.243(25) & 13.291(29) \\
    & 12.951& 12.929& 12.996& 13.142& 13.159& 13.218& 13.242& 13.254\\
    \end{tabular}
    \label{tab:obs_syn_photometry}
\end{table*}

\begin{figure}
\centering
	\includegraphics[width=\columnwidth]{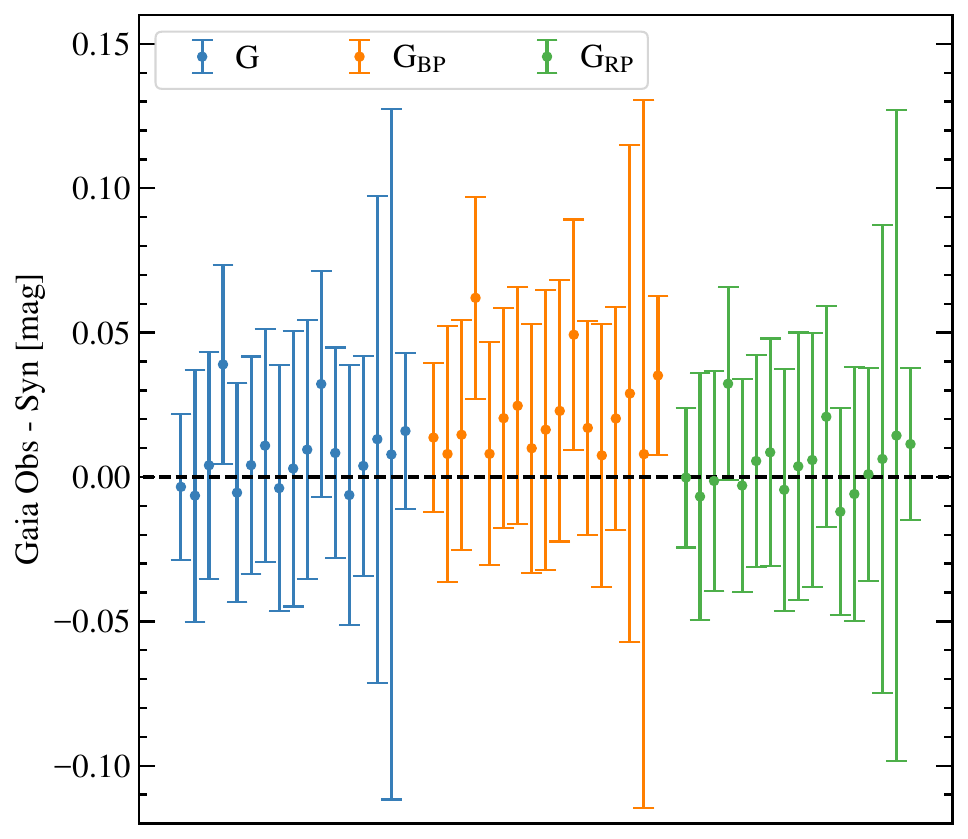}
    \caption{Observed minus synthetic photometry in the \textit{Gaia} $G$, $G_{\rm BP}$ and $G_{\rm RP}$ bands. The $x$-axis coordinates within each band are uniformly spaced, ordered by RA with the warm flux standard candidates first, followed by the extra flux standard candidates. Observed (with errors) and synthetic magnitudes are given in Table~\ref{tab:obs_syn_photometry}. Error bars represent combined 1$\sigma$ observed \textit{Gaia} errors and synthetic magnitude errors, where synthetic errors are calculated from the \Teff\ and $\log g$ errors on our STIS and WFC3 SED fits (Table~\ref{tab:fit_results}). The black dashed line indicates zero difference. Plot is based off fig.~10 in \citet{Axelrod2023}.}
    \label{fig:Gaia_obs_minus_syn_phot}
\end{figure}

\begin{figure}
\centering
	\includegraphics[width=\columnwidth]{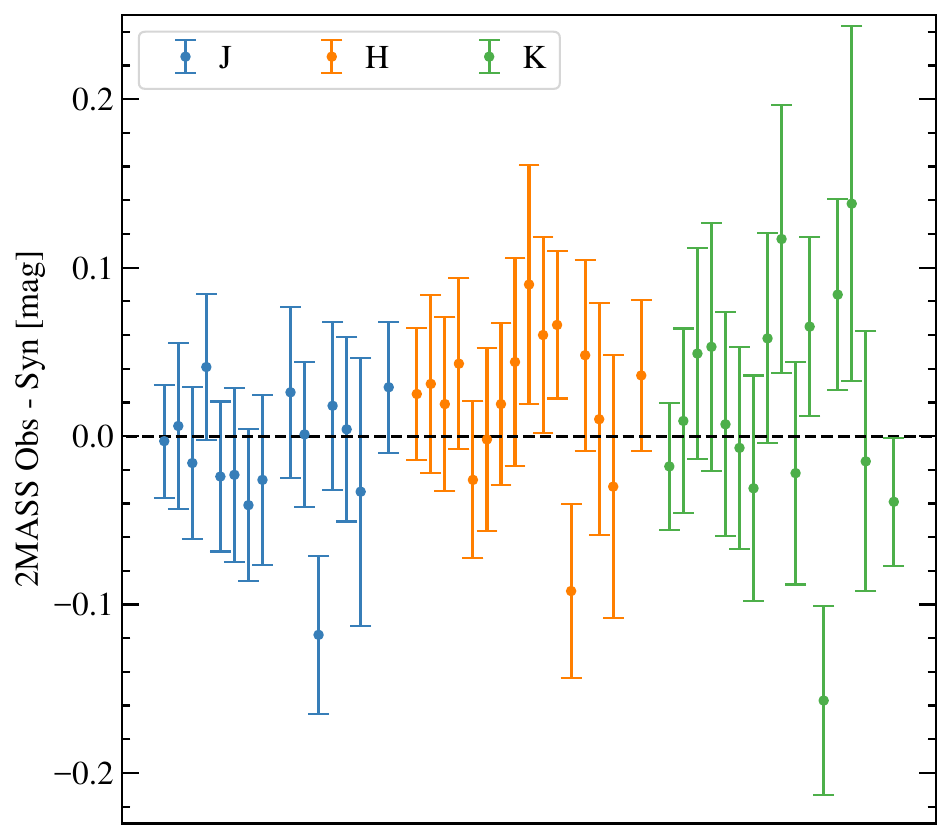}
    \caption{Same as Figure~\ref{fig:Gaia_obs_minus_syn_phot} but for 2MASS $J$, $H$ and $K_S$ bands. WD\,1919$+$145 only has a $J$ magnitude upper limit and WD\,1544$-$377 has no observed 2MASS data so both white dwarfs are not included in this plot.}
    \label{fig:2MASS_obs_minus_syn_phot}
\end{figure}

\begin{figure}
\centering
	\includegraphics[width=0.7\columnwidth]{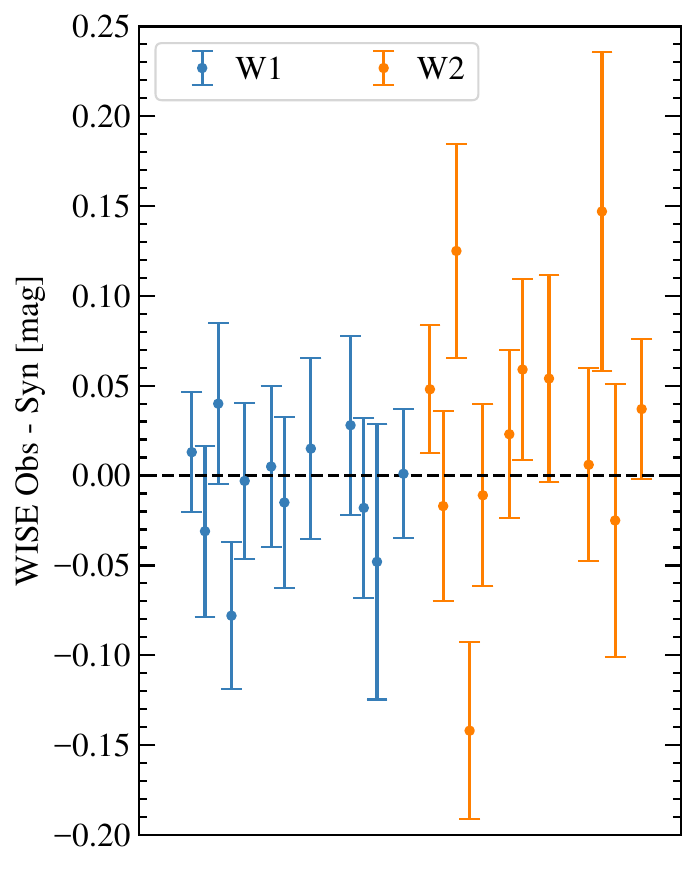}
    \caption{Same as Figure~\ref{fig:Gaia_obs_minus_syn_phot} but for WISE $W$1 and $W$2 bands. WD\,2117$+$539 and WD\,1544$-$377 have no observed WISE data so are not included in this plot. WD\,1327$-$083, WD\,1919$+$145 and WD\,2126$+$734 are not included in this plot either because their observed and synthetic photometries are $> 3\sigma$ discrepant.}
    \label{fig:WISE_obs_minus_syn_phot}
\end{figure}

The magnitude difference between observed and synthetic photometry for our network are shown in each band for \textit{Gaia}, 2MASS and WISE in Figures~\ref{fig:Gaia_obs_minus_syn_phot},~\ref{fig:2MASS_obs_minus_syn_phot}~and~\ref{fig:WISE_obs_minus_syn_phot}, respectively, for those stars with observed photometry. WD\,1327$-$083, WD\,1919$+$145 and WD\,2126$+$734 are not included in Figure~\ref{fig:WISE_obs_minus_syn_phot} because their observed and synthetic photometries are $> 3\sigma$ discrepant (see Section~\ref{sec:photometric_analysis}). The error bars in Figures~\ref{fig:Gaia_obs_minus_syn_phot}~-~\ref{fig:WISE_obs_minus_syn_phot} represent the combined observed and synthetic magnitude errors, where synthetic errors were calculated from the minimum statistical \Teff\ and $\log g$ errors on our STIS and WFC3 SED fits (Table~\ref{tab:fit_results}).

Finally, we photometrically fit the observed \textit{Gaia} DR3 $G$, $G_{\rm BP}$ and $G_{\rm RP}$ band photometry for our network and the two hot standards using Equation~\ref{ref:Gaia_phot_fit_eq} and the same DA model grids as in our SED and Balmer line fitting to obtain \textit{Gaia} atmospheric parameters. Table~\ref{tab:phot_fit_params} displays the best-fitting \Teff\ and $\log g$ parameters obtained from our photometric fits. The differences between \textit{Gaia} photometric fit parameters and those obtained from our \textit{HST} STIS and WFC3 SED fits are shown in Table~\ref{tab:phot_fit_params} as $\sigma_{\Teff}$ and $\sigma_{\log g}$. 

\begin{table}
    \centering
    \caption{Atmospheric parameters of the 17 DA white dwarfs proposed as flux standards from a photometric fit of the \textit{Gaia} $G$, $G_{\rm BP}$ and $G_{\rm RP}$ bands. $\sigma_{\Teff}$ and $\sigma_{\log g}$ correspond to how many $\sigma$ apart the best-fitting \Teff\ and $\log g$ parameters are from the \textit{HST} STIS and WFC3 SED fit (Table~\ref{tab:fit_results}) and \textit{Gaia} photometric fit.}
    \begin{tabular}{ccccc}
    \hline
    \hline
    Object &  \Teff & $\log g$  & $\sigma_{\Teff}$ & $\sigma_{\log g}$ \\
    &  [K] & [dex]  & &\\
    \hline
    \hline
    \multicolumn{5}{c}{\textbf{Hot primary standards}} \\
    \hline
         GD\,153 &  35870 $\pm$ 1119 &  7.723 $\pm$ 0.031 & 3.531&2.553
\\
         GD\,71 &  31526 $\pm$ 1092 &  7.760 $\pm$ 0.048 & 1.609&1.216\\
    \hline
    \multicolumn{5}{c}{\textbf{Warm flux standard candidates}} \\
    \hline
         WD\,0148$+$467 & 14110 $\pm$ 417 & 7.995 $\pm$ 0.037 & 0.834& 0.760
\\
         WD\,0227$+$050 & 18520 $\pm$ 583 & 7.858 $\pm$ 0.040 & 1.080& 1.034
\\
         WD\,0809$+$177 & 15790 $\pm$ 214 & 8.038 $\pm$ 0.017 & 2.071& 1.761
\\
         WD\,1105$-$340 & 13326 $\pm$ 420 & 8.079 $\pm$ 0.035 & 1.726& 0.617
\\
         WD\,1105$-$048 & 15469 $\pm$ 448 & 7.911 $\pm$ 0.036 & 0.743& 0.741
\\
         WD\,1327$-$083 & 15664 $\pm$ 444 & 7.919 $\pm$ 0.036 & 1.189& 2.260
\\
         WD\,1713$+$695 & 15495 $\pm$ 408 & 7.992 $\pm$0.032 & 1.277& 0.961
\\
         WD\,1911$+$536 & 16923 $\pm$ 468 & 8.283 $\pm$ 0.031 & 0.991& 0.948
\\
         WD\,1919$+$145 & 14979 $\pm$ 345 & 8.149 $\pm$ 0.025 & 1.002& 0.791
\\
         WD\,2039$-$682 & 16453 $\pm$ 390 & 8.467 $\pm$ 0.024 & 1.437& 1.052
\\
         WD\,2117$+$539 & 14679 $\pm$ 293 & 7.898 $\pm$ 0.026 & 2.656& 1.690
\\
         WD\,2126$+$734 & 15076 $\pm$ 83 & 7.893 $\pm$ 0.007 & 4.976& 4.454
\\
         WD\,2149$+$021 & 17187 $\pm$ 457 & 7.971 $\pm$ 0.034 & 1.015& 0.994\\
    \hline
    \multicolumn{5}{c}{\textbf{Extra flux standard candidates}} \\
    \hline
         WD\,0352$+$096 & 14308 $\pm$ 355 & 8.291 $\pm$ 0.028 & 1.282& 1.019
\\
         WD\,1202$-$232 & 8655 $\pm$ 87 & 7.981 $\pm$ 0.025 & 1.505& 1.256
\\
         WD\,1544$-$377 & 10377 $\pm$ 33 & 8.021 $\pm$ 0.007 & 0.591& 0.654
\\
         WD\,2341$+$322 & 12358 $\pm$ 293 & 7.996 $\pm$ 0.031 & 1.517& 0.927\\
    \end{tabular}
    \label{tab:phot_fit_params}
\end{table}

\section{Discussion}
\label{sec:Discussion}

As the objective of this work is to fit the \textit{HST} STIS and WFC3 spectrophotometry of 17 white dwarf flux standard candidates as accurately as possible, we tested different options in our fits to determine their implications. 

For our Balmer line fits, we chose to fit from H$\beta$ to H$\eta$. The decision to omit H$\alpha$ was motivated from the fact that H$\alpha$ and the higher order Balmer lines are observed separately with two distinct STIS grating configurations. The G750L grating covers the range $5249 - 10\,270$\,\AA\ with a dispersion of 4.92\,\AA/pix, thus observes H$\alpha$, whereas the higher order Balmer lines are all observed using the G430L grating which covers the range $2900 - 5700$\,\AA\ with a dispersion of 2.73\,\AA/pix. The difference in dispersions shows the STIS sampling for H$\alpha$ is coarser than the other Balmer lines, with a lower resolution by a factor of $\approx 2$. Therefore including H$\alpha$ in the same fit as the higher order Balmer lines requires a well understood instrumental response, and uncertainties on that response may outweigh any additional constraints the extra Balmer line gives on the stellar parameters. 

To obtain the most accurate results from Balmer line fits, models have to be convolved to the resolution of the instrument, i.e. the LSF, as this gives a true account of the light distribution along the primary dispersion axis. Convolving with a Gaussian to a static resolution or full width half maximum (FWHM) is a valid approximation but using the instrumental resolution is the more accurate approach. STIS has four gratings which have slightly different instrumental profiles over their wavelength coverage. We selected the G430L grating LSF as that covered H$\beta$ to H$\eta$ (see Section~\ref{sec:Balmer_fitting} for details). The LSF function we created is accurate within $<1$ per cent to the actual LSF however this small uncertainty will have propagated into the best-fitting Balmer line parameters shown in Table~\ref{tab:fit_results}.

When using the DA $M$-$R$ relation from \citet{Bedard2020} we used thick hydrogen layers with $q_\mathrm{H} = M_\mathrm{H}/M_\mathrm{WD} = 10^{-4}$ as this is the standard procedure for DA white dwarfs based on theories of post-asymptotic-giant-branch (post-AGB) evolution and the maximum hydrogen mass for residual nuclear burning \citep{Iben1984, Althaus2010b}. The other option is to use thin hydrogen layers with $q_\mathrm{H} = 10^{-10}$, which is normally employed for helium-dominated atmosphere white dwarfs (DB, DC, DQ, DZ). However, using thick vs thin H-layers for DA and non-DA white dwarfs is not a strict rule - studies have shown evidence for thin H-layers in DA white dwarfs \citep{Fontaine1987, Fontaine1997, MillerBertolami2017, Bond2017, Cunningham2020}. For the 17 white dwarfs in this study, using thin H-layers altered the \Teff\ best-fitting parameters by $< 50$\,K and $< 0.03$\,dex, resulting in the SED and Balmer line results being slightly more discrepant. Therefore, we conclude that thick H-layers is the more accurate assumption for this white dwarf network.

\subsection{Observed vs model fluxes}
\label{sec:fluxes}

Our 3D DA LTE models successfully predict the observed \textit{HST} STIS and WFC3 spectrophotometry fluxes within 3 per cent over most of the range $1450 - 16\,000$\,\AA\ (Figures~\ref{fig:SED_Balmer_warm},~\ref{fig:SED_Balmer_warm_appendix}~and~\ref{fig:SED_Balmer_extra}). The standard deviation of the flux residuals from the entire SED fit, in addition to the fits of only the UV and combined optical+NIR regions, are shown in Table~\ref{tab:std_flux_residuals}. The median standard deviation flux residual of the entire fit wavelength range is 1.34 per cent for the warm flux standard candidates, 3.25 per cent for the extra candidates, and 1.41 per cent for all 17 candidates. 

The two coolest white dwarfs, WD\,1202$-$232 and WD\,1544$-$377, have slight discrepancies between the observed and model fluxes between $\approx 1700 - 2700$\,\AA, causing their residuals to be $\approx 4.5$ per cent. There are known modelling uncertainties for cool white dwarfs ($\Teff \lesssim 10\,000$\,K) in the near-UV, as the total opacity is affected by the far red wing of the Lyman-$\alpha$ line where H and H$_2$ collisions significantly broaden the absorption line profile at those densities \citep{KowalskiSaumon2006,Saumon2022}. While potential issues with this opacity have been discussed in the context of UV and optical data of white dwarfs cooler than $\approx$ 6000\,K \citep{Saumon2014,Caron2023,OBrien2024}, we speculate that our precise STIS UV data may have uncovered a related, milder issue at warmer temperatures. If we fit the SEDs of the two coolest white dwarfs in the network starting at 2300\,\AA\ then the flux residuals reduce to 1.26 per cent for WD\,1202$-$232 and 2.88 per cent for WD\,1544$-$377. The observed WFC3 data for WD\,1544$-$377 is slightly irregular (Figure~\ref{fig:SED_Balmer_extra}c), with the issues likely caused by its wide binary companion (Section~\ref{sec:flags}) contaminating the detected IR flux. 

Fitting the UV and combined optical+NIR regions of the spectra separately yield median standard deviation flux residuals for the warm candidates of 1.43 per cent and 1.12 per cent, respectively, showing better agreement over the redder wavelengths. The same trend is seen when fitting only the four extra candidates (UV = 2.21 per cent and optical+NIR = 1.40 per cent), and all 17 candidates together (UV = 1.47 per cent and optical+NIR = 1.15 per cent). The proposed network of white dwarfs with $\Teff < 20\,000$\,K can therefore accurately reproduce the CALSPEC flux scale in the optical+NIR regime to within 1.5 per cent. 

We also fit the \textit{HST} STIS and WFC3 spectrophotometry of the current hot standards GD\,153 and GD\,71, with our models successfully predicting the observed fluxes within 0.5 per cent over most of the range $1450 - 16\,000$\,\AA. The standard deviation flux residuals are given in the first two rows of Table~\ref{tab:std_flux_residuals}.

Our \textit{HST} spectrophotometry consists of six separate observations from the four STIS gratings and two WFC3 gratings, which have been merged together. Small glitches in the SED can occur at the grating merge points (edges) causing residuals between observed and model fluxes to be $> 3$ per cent. Users of this white dwarf network as flux calibrators can remove these glitches if desired, but they have not been removed in this work. 

Although we do not fit STIS spectra in the wavelength range $1140-1450$\,\AA, discrepancies between model and observed SEDs can be found from a visual inspection of Figures\,\ref{fig:SED_Balmer_warm}~and~\ref{fig:SED_Balmer_warm_appendix}. We note that the STIS flux calibration from the three primary hot white dwarf standards is uncertain at the $\approx 3$ per cent level in this wavelength regime \citep[see fig.\,4 of][]{Bohlin2020}, a significant contrast to the sub-percent precision at longer wavelengths. This is largely attributed to differences in the predicted spectra from the TLUSTY and TMAP model atmosphere codes, and to a lesser degree on uncertainties in \Teff\ and $\log g$ determinations which propagate to model flux uncertainties. We note that \citet{Sahu2023} found a similar optical vs Lyman-$\alpha$ parameter discrepancy using \textit{HST} Cosmic Origins Spectrograph (COS) observations of warm DA white dwarfs using instead the model atmospheres of \citep{Koester2010}. Therefore these results could be an indication that \textit{HST} primary flux calibration needs to be revisited in the UV. In the present work, additional uncertainties in this wavelength range may be related to missing physics in Lyman-$\alpha$ H$^+_2$ satellite opacities, in particular the treatment of electron broadening and N-body collisions \citep{Allard2009,Gomez2024}. 

\subsection{Atmospheric parameters}
\label{sec:atmospheric_parameters}

Independent \Teff\ and $\log g$ best-fitting parameters are obtained from fitting the \textit{HST} STIS and WFC3 spectrophotometry SEDs and Balmer lines of the flux standard candidates. Statistical errors for both fitting routines are computed from the covariance matrix of the model atmosphere fit scaled by reduced $\chi^2$. However, using statistical errors alone do not account for all uncertainties in the fits thus are underestimates. We calculated the systematic error on \Teff\ and $\log g$ on SED fits as 1.1 per cent and 0.01\,dex, respectively (Section~\ref{sec:SED_fitting}), which gives more realistic uncertainties (Table~\ref{tab:fit_results}). 

It is more challenging to define a systematic uncertainty on Balmer line fits, with sensible options being to compare fits of the same star using different models or fit multiple observations or different lines from the same star with the same model. However, there is only one publicly available model for Stark line broadening \citep{Tremblay2009} and a limited number of Balmer lines to perform fitting. Reducing the number of lines in the fit drastically reduces precision and can introduce degeneracies between \Teff\ and $\log g$ \citep{Tremblay2009}. Therefore we arbitrarily assign a scale factor of $\times 3$ on all Balmer line fit statistical errors, as this brings the Balmer line best-fitting \Teff\ to within $3\sigma$ of the SED best-fitting \Teff\ for 16/17 of the flux standard candidates (13/17 within $2\sigma$ and 8/17 within $1\sigma$). This is a reasonable scale factor to account for uncertainties on Stark broadening profiles and non-ideal gas effects \citep{Cho2022}. The best-fitting $\log g$ parameters agree within $1\sigma$ for 16/17 flux candidates - applying the $\times 3$ scale factor to the $\log g$ uncertainties does not change this result. In this work we only quote the purely statistical uncertainties from Balmer line fitting and not the scaled-up values. However for the purpose of comparing results we suggest users multiply by three the values provided in Table~\ref{tab:fit_results}. Overall the independent \Teff\ and $\log g$ best-fitting parameters from our SED and Balmer line fits have a very tight linear correlation with no systematic deviations, and agree within 2.7 per cent in \Teff\ and 0.013\,dex in $\log g$ as a median for the 17 flux standard candidates. 

The SED and Balmer line best-fitting \Teff\ and $\log g$ parameters for the two hot primary standards, GD\,153 and GD\,71, agree within $\approx 1\sigma$ (Table~\ref{tab:fit_results}). Compared to the parameters for these two standards in Table~1 of \citet{Bohlin2020}, our SED parameters are within $1\sigma$ for GD\,153 and within $1\sigma$ and $2\sigma$ for \Teff\ and $\log g$, respectively, for GD\,71. Similarly, compared to the parameters in Table~2 of \citet{Narayan2019} our SED parameters are within $1\sigma$ and $2\sigma$ for \Teff\ and $\log g$, respectively, for GD\,153, and within $1\sigma$ for GD\,71. The strong agreement between best-fitting parameters in this work and the literature demonstrates the predictive power of the models adopted for our white dwarf flux candidate network.

\subsection{Photometric analysis}
\label{sec:photometric_analysis}

The 17 white dwarfs proposed as flux standards in this work have optical photometry from \textit{Gaia} $G$, $G_{\rm BP}$ and $G_{\rm RP}$ bands, in addition to IR photometry from 2MASS $J$, $H$ and $K_S$ bands and WISE $W$1 and $W$2 bands, with some exceptions. WD\,2117$+$539 has no reported WISE data, potentially because it is near a bright star Gaia DR3 2176116584362421504 which could have contaminated the IR flux. WD\,1544$-$377 has no 2MASS nor WISE data because it was likely not possible to accurately observe it due to its wide and bright binary companion (Section~\ref{sec:flags}).

Our photometric analysis methods are detailed in Section~\ref{sec:Synthetic_magnitudes_and_photometric_fitting} and the synthetic magnitudes computed are shown in Table~\ref{tab:obs_syn_photometry}, alongside observed magnitudes and uncertainties. The difference between observed and synthetic magnitudes for \textit{Gaia}, 2MASS and WISE photometry are plotted in Figures~\ref{fig:Gaia_obs_minus_syn_phot}~-~\ref{fig:WISE_obs_minus_syn_phot} which illustrate the broad agreement across surveys and individual bands. For the \textit{Gaia} photometric bands, there is $1\sigma$ agreement between observed and synthetic photometry for 16 of the white dwarfs in the $G$ band, 14 white dwarfs in the $G_{\rm BP}$ band and all 17 white dwarfs in the $G_{\rm RP}$ band. All 17 white dwarfs agree within $2\sigma$ for the $G$ and $G_{\rm BP}$ bands. The 16 white dwarfs with NIR 2MASS data all agree within $3\sigma$ across all bands, with $1\sigma$ agreement evident for 14, 12 and 10 white dwarfs across the $J$, $H$ and $K_S$ bands respectively. 15 white dwarfs in the network have IR WISE photometric data, of which 12 agree within $3\sigma$ with our synthetic magnitudes across both bands, 11 agree within $1\sigma$ in the $W$1 band and 7 agree within $1\sigma$ in the $W$2 band. 

The WISE observed and synthetic photometries for WD\,1327$-$083, WD\,1919$+$145 and WD\,2126$+$734 are between $6 - 16\sigma$ discrepant. The reason for this discrepancy is likely due to contamination of the observed photometry. WD\,1327$-$083 is in close proximity to a star; WD\,2126$+$734 is in a crowded field and is in a WD+WD binary system (Section~\ref{sec:flags}); and WD\,1919$+$145 is in the Galactic plane with many nearby bright sources in the IR. Nevertheless, these stars are still suitable flux calibrators for spaced-based instruments or at moderate spatial resolution. The relatively high proper motions of our standards means their positions will vary with time and any of them could become temporarily blended with another star in the future.

\citet{Holberg2006} made a similar residual comparison between observed and synthetic 2MASS $J$, $H$ and $K_S$ band photometries to that made in Figure~\ref{fig:2MASS_obs_minus_syn_phot}, but for a larger sample of DA white dwarfs and without an upper \Teff\ constraint. The residual dispersions of $\lesssim 0.2$\,mag across the bands measured in \citet{Holberg2006} is similar to that calculated in this work.

In addition to synthetic photometry, we computed best-fitting atmospheric parameters from photometric fits of the \textit{Gaia} $G$, $G_{\rm BP}$ and $G_{\rm RP}$ observed photometry and parallaxes with our 3D LTE models (Section~\ref{sec:Synthetic_magnitudes_and_photometric_fitting}) which are displayed in Table~\ref{tab:phot_fit_params}. There is good agreement between our best-fitting photometric \Teff\ and $\log g$ parameters and SED parameters with 16 white dwarfs being within $3\sigma$, 14 being within $2\sigma$ for \Teff\ and 15 being within $2\sigma$ for $\log g$. The photometric and SED parameters for WD\,2126$+$734 are just over $3\sigma$ discrepant. We compared our photometric \textit{Gaia} atmospheric parameters with those computed in \citet{GF2021} and found all 17 objects from our network agree within $3\sigma$.

The same photometric analysis on the hot standards GD\,153 and GD\,71 was performed, with the results shown in Tables~\ref{tab:obs_syn_photometry}~and~\ref{tab:phot_fit_params}. The observed and synthetic magnitudes for GD\,153 agree within $1\sigma$ for all three bands in \textit{Gaia} and 2MASS and the WISE $W$1 band, whereas there is a $2\sigma$ agreement for $W$2. For GD\,71, the observed and synthetic magnitudes agree within $1\sigma$ for all three bands in \textit{Gaia}, two bands in WISE and the 2MASS $J$ band, whereas the 2MASS $H$ and $K_S$ bands agree within $2\sigma$ and $3\sigma$, respectively. The best-fitting atmospheric parameters obtained from \textit{Gaia} DR3 photometric fits agree with our STIS and WFC3 SED best-fitting parameters (Table~\ref{tab:fit_results}) within $4\sigma$ and $3\sigma$ for \Teff\ and $\log g$, respectively, for GD\,153, and within $2\sigma$ for GD\,71. 

A small systematic offset is apparent between observed and predicted \textit{Gaia} $G_{\rm BP}$ magnitudes in Fig.\,\ref{fig:Gaia_obs_minus_syn_phot}, which suggests that \textit{Gaia} and \textit{HST} are not on the same relative or absolute flux scales \citep{Maiz2018}. The offset is similar to that found in previous studies that have compared secondary \textit{HST} flux standards with \textit{Gaia} photometry \citep{Narayan2019,Axelrod2023}. Furthermore, the offset is consistent with a systematic difference in \Teff\ and mass found between Balmer line and photometric \textit{Gaia} parameters for DA white dwarfs \citep{Tremblay2019b,Bergeron2019,Cukanovaite2021}. 

\subsection{Flags on individual white dwarfs in our network}
\label{sec:flags}

We searched the literature for our network of 17 white dwarfs to find any instances of metal detections/limits, variability, magnetic field detections/limits or binarity. The Transiting Exoplanet Survey Satellite \citep[\textit{TESS};][]{TESS2014} Input Catalog (TIC) numbers are given for each white dwarf, so readers can check current \textit{TESS} Sectors for variability when they are choosing calibrators. 

Some white dwarfs are flagged with small metal abundances/limits or magnetic fields $\lesssim 150$\,kG, but these would typically have a negligible effect on SEDs so are not removed from our network. One exception is the indirect effect of atmospheric metal pollution from planetary debris on the IR SED; while none of the white dwarfs in our network show a clear \textit{WISE} excess attributable to a debris disk, metal polluted white dwarf calibrators are at greater risk of having so far undetected IR excess from debris. The white dwarfs in the \citet{El-Badry2021} \textit{Gaia} eDR3 catalog of spatially resolved binary stars are flagged, with the projected separation in AU from their companion and the chance alignment ratio, which approximately represents the probability that a binary candidate is in a chance alignment. We present our findings below. 

\textbf{WD\,0148$+$467} is not observed to vary in \textit{TESS} (TIC 415880209). 

\textbf{WD\,0227$+$050} is not observed to vary in \textit{TESS} (TIC 422888592). 

\textbf{WD\,0809$+$177} is not observed to vary in \textit{TESS} (TIC 27634101).

\textbf{WD\,1105$-$340} has potentially contaminated \textit{TESS} data (TIC 23226265). \citet{Landstreet2019} detected a probable dipolar magnetic field on the order of 150\,kG and \citet{OBrien2024} classified this star as DAH. Since it has \Teff\ $\approx$ 14\,000\,K and negligible convective energy transfer, the small magnetic field is not expected to significantly impact the atmospheric structure and predicted SED \citep{Tremblay2015b}. We also note that the continuum and Balmer line fits are in good agreement despite possible Zeeman line splitting.

\textbf{WD\,1105$-$048} is not observed to vary in \textit{TESS} (TIC 53211451). \citet{Valyavin2006} stated this star has a rotation period longer than 3\,h. \citet{AznarCuadrado2004} discovered an average magnetic field of $-$2.1\,kG with polarimetry data from the FOcal Reducer and low dispersion Spectrograph \citep[FORS1;][]{Appenzeller1998}, where different field values from two observations were obtained, potentially indicating stellar rotation. \citet{Valyavin2006} found a variable longitudinal magnetic field from $\approx 0 - 8$\,kG. \citet{Koester2009b} found no detectable magnetic field nor peculiarities in UVES/VLT (Ultraviolet and Visual Echelle Spectrograph/Very Large Telescope) spectra obtained for the ESO SNe Ia Progenitor Survey (SPY). However, this star was flagged as a magnetic white dwarf in \citet{Bagnulo2018} with FORS2 and the Intermediate-dispersion Spectrograph and Imaging System (ISIS) on the William Herschel Telescope (WHT), with a suspected variable weak field between <$B_z$>~$\approx 0 - 2.1$\,kG and unknown period. This star has the spectral type DAH+dM in \citet{Bagnulo2022} and DAH in \citet{OBrien2024}. At $\Teff \approx 15\,800$\,K, it is too warm for magnetic fields to have an effect on atmospheric structure and predicted SED. WD\,1105$-$048 is in a resolved WD+MS binary system with an M dwarf companion LP\,672-2 (Gaia DR3 3788190605663811840) at a projected separation of 6944\,AU and chance alignment of 1.08E-04 \citep{Oswalt1988, AznarCuadrado2004, Holberg2016, Toonen2017, El-Badry2021}. 

\textbf{WD\,1327$-$083} has no \textit{TESS} data (TIC 422888592). \citet{Bagnulo2018} observed this star once with ISIS and once with FORS2, measuring <$B_z$>~$ = -0.9 \pm 0.4$\,kG and <$B_z$>~$ = -0.3 \pm 0.2$\,kG, respectively. Given the non-significant detection, this white dwarf has the spectral type DA \citep{Bagnulo2021, Bagnulo2022}. This star is in a resolved WD+MS binary system with an M dwarf companion LHS\,353 (Gaia DR3 3630015546177181952) at a projected separation of 8085\,AU and chance alignment of 8.66E-09 \citep{Perryman1997, FarihiBecklinZuckerman2005, Holberg2016, Toonen2017, El-Badry2021}. 

\textbf{WD\,1713$+$695} is not observed to vary in \textit{TESS} (TIC 219863212).

\textbf{WD\,1911$+$536} is not observed to vary in \textit{TESS} (TIC 298900716). 

\textbf{WD\,1919$+$145} has potentially contaminated \textit{TESS} data (TIC 338305380) due to it being in the Galactic plane. \citet{Zuckerman2003} state this star has $\log\textnormal{(Ca/H)} < -8.838$ according to High Resolution Echelle Spectrometer (HIRES) echelle spectra from the Keck telescope.

\textbf{WD\,2039$-$682} is not observed to vary in \textit{TESS} (TIC 372109346). \citet{Koester1998} identified a broadened H$\alpha$ core in this white dwarf, which could be explained by rotation or a magnetic field of $\approx 50$\,kG, although a distinction was not possible without circular polarization measurements. No magnetic field has been confirmed in the literature.

\textbf{WD\,2117$+$539} is not observed to vary in \textit{TESS} (TIC 314771701).

\textbf{WD\,2126$+$734} is not observed to vary in \textit{TESS} (TIC 323139945). This star is in a WD+WD binary system with WD\,2126$+$734B (Gaia DR3 2274076301516712704) at a projected separation of 42\,AU and chance alignment of 6.59E-08 \citep{Zuckerman1997, FarihiBecklinZuckerman2005, Holberg2016, El-Badry2021, Heintz2024, OBrien2024}. Despite WD\,2126$+$734 and its companion being physically close together, they have an angular separation of 1.74" because this is a very nearby white dwarf (22.2\,pc; Table~\ref{tab:astrometry}) so they are easily resolved by \textit{HST}, \textit{JWST} and \textit{Gaia}. The observations made by 2MASS and WISE are likely blended due to these telescopes having resolutions $\gtrsim 4$". 

\textbf{WD\,2149$+$021} is not observed to vary in \textit{TESS} (TIC 405122007). A $\log\textnormal{(Ca/H)}$ abundance between $-$8.0 and $-$7.6 has been measured in this white dwarf leading to it being classified as a DAZ in some literature \citep{Koester2005, KoesterWilken2006, Berger2005, Kilic2006, Farihi2009, Kawka2011, OBrien2024}. While this small amount of metal pollution is not expected to impact the predicted SED which is dominated by hydrogen opacities at all continuum wavelengths according to our models, it has more risk of having a so far unseen debris disk.

\textbf{WD\,0352$+$096} is not observed to vary in \textit{TESS} (TIC 415339071). 

\textbf{WD\,1202$-$232} is not observed to vary in \textit{TESS} (TIC 398243520). A $\log\textnormal{(Ca/H)}$ abundance between $-$9.8 and $-$9.7 and $\log\textnormal{(Fe/H)}$ abundance between $-$8.0 and $-$7.85 has been measured in this white dwarf leading to it being classified as a DAZ in some literature \citep{Zuckerman2003, Koester2005, KoesterWilken2006, Farihi2009, Kawka2011, Kawka2012, Subasavage2017, Caron2023, OBrien2024}.

\textbf{WD\,1544$-$377} is not in \textit{TESS}. \citet{Zuckerman2003} state this star has $\log\textnormal{(Ca/H)} < -10.282$ according to HIRES echelle spectra from the Keck telescope. This white dwarf is in a WD+MS binary system with the high proper motion star HD\,140901 (Gaia DR3 6009538585839374336) at a projected separation of 220\,AU and angular separation of 14.6", with a chance alignment of 1.2E-05 \citep{Zuckerman2003, Holberg2016, El-Badry2021, OBrien2024}. The \textit{Gaia} DR3 $G$ magnitude of the companion is 5.834\,mag. Note that the binary companion is incorrectly listed in \citet{Holberg2008}. 

\textbf{WD\,2341$+$322} is not observed to vary in \textit{TESS} (TIC 288144896). \citet{Zuckerman2003} state this star has $\log\textnormal{(Ca/H)} < -9.531$ according to HIRES echelle spectra from the Keck telescope. This white dwarf is not in the \citet{El-Badry2021} wide binary catalog nor is there any evidence of binarity in \textit{Gaia}, but it is listed as a common proper motion companion in a non-interacting binary system in previous literature \citep{McCook1999, Zuckerman2003, Holberg2016}.

\subsection{White dwarfs as IR calibrators}
\label{sec:IR_calibration}

White dwarfs are widely used for flux calibration in UV and optical regimes, along with A- and G-stars. The current three primary white dwarf standards achieve an accuracy better than 1 per cent at optical wavelengths and provide modelled SEDs which extend calibration from the UV to NIR \citep{Bohlin2019, Bohlin2020}. Using predicted SEDs from different model atmosphere codes, \citet{Bohlin2020} suggest that white dwarf flux calibration precision may be as good as 1 per cent up to 10\,$\mu$m. \citet{Bohlin2011} and \citet{GF2020} initially explored the accuracy of extending white dwarf flux calibration into the NIR and found modelled SEDs to be consistent with the \textit{HST} flux scale within a few per cent. Recent work advancing \textit{HST} primary calibration show promising agreement between the Sirius spectrum through the Spitzer Space Telescope Infrared Array Camera (IRAC) bands to $\approx 5$\,$\mu$m \citep{Bohlin2022, Rieke2023}.

\begin{figure*}
\centering
	\includegraphics[width=2\columnwidth]{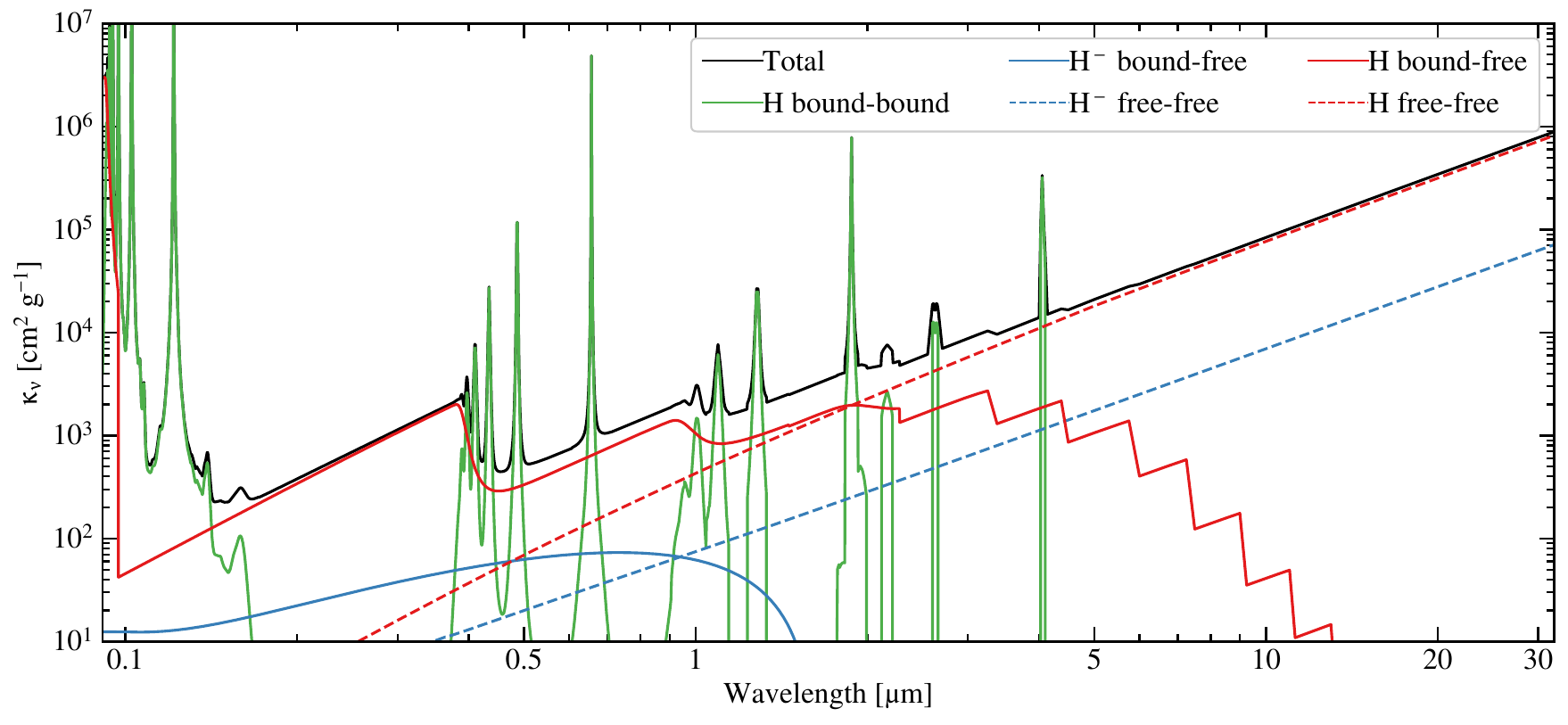}
    \caption{Dominant contributions to the radiative opacity of DA white dwarfs as a function of wavelength. The opacities shown are for a $\Teff = 12\,000$\,K white dwarf. A $\log g$ of 8.0\,dex is assumed and the opacities are evaluated for the conditions at the photosphere. Based off the bottom panel of Fig. 17 in \citet{Saumon2022}.}
    \label{fig:rad_opacity_contributions}
\end{figure*}

This work expands upon previous studies and proposes 17 white dwarfs with $\Teff < 20\,000$\,K as IR calibrators. This network of white dwarfs is well-suited to provide an independent estimate for uniform calibration to the MIR as they have relatively featureless SEDs and lack variability. Also, IR opacities consists in well understood hydrogen free-free, bound-free and bound-bound transitions \citep{Saumon2022}, as shown by Figure~\ref{fig:rad_opacity_contributions} which is based on the same model atmospheres and spectra \citep{Tremblay2013} as those used to fit our observed white dwarf network and predict their full SEDs. Considering this, and since the SEDs of these white dwarfs peak at UV, optical and NIR wavelengths, we predict that these white dwarfs can be reliably used as flux standards for IR observations. This needs to be tested with \textit{JWST} spectrophotometric observations, since \textit{WISE} observations are limited in wavelength coverage and precision ($\approx$ 5 per cent level). Such observations would also determine if any of our calibrators are inappropriate for the IR, such as from having debris disks or planets \citep{Mullally2024}.

The \textit{JWST} absolute flux calibration programme is designed on \textit{HST} absolute calibration therefore our network can provide another estimate for \textit{JWST} calibration, set limits on the precision and provide legacy for its IR instruments \citep{Gordon2022}. We have shown that cool white dwarf stars are valuable NIR and potentially MIR calibrator additions and a composite approach to calibration with white dwarfs, A- and G-stars should be used to obtain the best possible results.

\section{Conclusions}
\label{sec:Conclusions}

We have analysed \textit{HST} STIS and WFC3 spectrophotometry of 17 DA white dwarfs with $\Teff < 20\,000$\,K to investigate their reliability as flux calibrators, specifically in the IR regime. The stars in our network have significantly lower \Teff\ values than current hot DA primary standards \citep{Bohlin2020}, in addition to recent networks of hot but faint DA stars \citep{Narayan2016, Narayan2019, Axelrod2023}, which has multiple advantages: their SEDs peak closer to the optical/NIR regime; have a larger sky density; NLTE effects and UV metal line blanketing have a negligible effect on continuum fluxes. Convection in cooler white dwarfs is not a significant issue either as our state-of-the-art 3D DA LTE model atmospheres account for convective effects without free parameters.

The results from this study show our model atmospheres successfully predict the observed fluxes of all 17 white dwarfs within 3 per cent over most of the wavelength range $1450 - 16\,000$\,\AA. The median standard deviation flux residual for all 17 candidates over the entire fit wavelength range is 1.41 per cent, with the coolest white dwarf residuals being $\approx 3$ per cent. The residuals when fitting the UV and optical+NIR regions separately for the 17 candidates are 1.47 per cent and 1.15 per cent, respectively. Therefore, this network of white dwarfs cooler than current primary standards are fully consistent with the \textit{HST}/CALSPEC flux scale over the UV, optical and NIR regimes.

Spectrophotometric fits of the observed \textit{HST} STIS and WFC3 SEDs and Balmer lines of the 17 flux standard candidates yield independent atmospheric parameters. The best-fitting \Teff\ for 16 stars agree within $3\sigma$ and the best-fitting $\log g$ for all 17 stars agree within $2\sigma$. Photometric fits of the observed \textit{Gaia} DR3 $G$, $G_{\rm BP}$ and $G_{\rm RP}$ photometry results in $3\sigma$ agreement for all 17 stars with the best-fitting SED \Teff\ and $\log g$ parameters. Therefore we find excellent agreement between independent atmospheric parameters for our network. 

We also computed synthetic magnitudes for \textit{Gaia} $G$, $G_{\rm BP}$ and $G_{\rm RP}$ bands, 2MASS $J$, $H$ and $K_S$ bands and WISE $W$1 and $W$2 bands so we could compare them with observed photometry. All 17 white dwarfs in our network agree within $2\sigma$ for the three optical \textit{Gaia} photometric bands. For the IR photometric bands in 2MASS and WISE, all the white dwarfs in our network with 2MASS data agree within $3\sigma$ and 12/15 with WISE data agree within $3\sigma$.

To conclude, our network of 17 DA white dwarfs with $\Teff < 20\,000$\,K represents a valuable addition to available flux standards. These stars provide reliable independent estimates for secondary IR SEDs and the legacy for current and future IR instruments onboard spacecraft such as the \textit{JWST}, \textit{Euclid} and \textit{RST}, in addition to ground based observatories such as the ELT (e.g. MICADO). The synthetic spectra of our network are available with the electronic distribution of this article and on CALSPEC\footnote{The models for WD\,1202$-$232 and WD\,1544$-$377 are not included in CALSPEC due to potential modelling issues and/or contamination in the IR.}. Observers can select the most suitable flux standards for calibrating their observations based on specific wavelength coverage and target requirements.

\section*{Acknowledgements}

This research received funding from the European
Research Council under the European Union’s Horizon 2020 research and
innovation programme number 101002408 (MOS100PC).
This research is based on observations made with the NASA/ESA \textit{Hubble Space Telescope} obtained from the Space Telescope Science Institute, which is operated by the Association of Universities for Research in Astronomy, Inc., under NASA contract NAS 5–26555. Support for Program number (16249) was provided through a grant from the STScI under NASA contract NAS5- 26555. These observations are associated with programs GP-14213, GP-15485, GP-16249.
This work has made use of data from the European Space Agency (ESA) mission
{\it Gaia} (\url{https://www.cosmos.esa.int/gaia}), processed by the {\it Gaia}
Data Processing and Analysis Consortium (DPAC,
\url{https://www.cosmos.esa.int/web/gaia/dpac/consortium}). Funding for the DPAC
has been provided by national institutions, in particular the institutions
participating in the {\it Gaia} Multilateral Agreement.
This publication makes use of data products from the Two Micron All Sky Survey, which is a joint project of the University of Massachusetts and the Infrared Processing and Analysis Center/California Institute of Technology, funded by the National Aeronautics and Space Administration and the National Science Foundation.
This publication makes use of data products from the Wide-field Infrared Survey Explorer, which is a joint project of the University of California, Los Angeles, and the Jet Propulsion Laboratory/California Institute of Technology, funded by the National Aeronautics and Space Administration.

\section*{Data Availability}
All data underlying this paper are publicly available from the relevant survey archives. The model atmospheres used in this paper are available upon reasonable request of the author. The model SEDs for the network of 17 cooler white dwarf flux standards can be found in the electronic distribution of this article and on CALSPEC\footnote{The models for WD\,1202$-$232 and WD\,1544$-$377 are not included in CALSPEC due to potential modelling issues and/or contamination in the IR.} at \url{https://www.stsci.edu/hst/instrumentation/reference-data-for-calibration-and-tools/astronomical-catalogs/calspec}.



\bibliographystyle{mnras}
\bibliography{cal} 

\begin{thebibliography}{}
\makeatletter
\relax
\def\mn@urlcharsother{\let\do\@makeother \do\$\do\&\do\#\do\^\do\_\do\%\do\~}
\def\mn@doi{\begingroup\mn@urlcharsother \@ifnextchar [ {\mn@doi@}
  {\mn@doi@[]}}
\def\mn@doi@[#1]#2{\def\@tempa{#1}\ifx\@tempa\@empty \href
  {http://dx.doi.org/#2} {doi:#2}\else \href {http://dx.doi.org/#2} {#1}\fi
  \endgroup}
\def\mn@eprint#1#2{\mn@eprint@#1:#2::\@nil}
\def\mn@eprint@arXiv#1{\href {http://arxiv.org/abs/#1} {{\tt arXiv:#1}}}
\def\mn@eprint@dblp#1{\href {http://dblp.uni-trier.de/rec/bibtex/#1.xml}
  {dblp:#1}}
\def\mn@eprint@#1:#2:#3:#4\@nil{\def\@tempa {#1}\def\@tempb {#2}\def\@tempc
  {#3}\ifx \@tempc \@empty \let \@tempc \@tempb \let \@tempb \@tempa \fi \ifx
  \@tempb \@empty \def\@tempb {arXiv}\fi \@ifundefined
  {mn@eprint@\@tempb}{\@tempb:\@tempc}{\expandafter \expandafter \csname
  mn@eprint@\@tempb\endcsname \expandafter{\@tempc}}}

\bibitem[\protect\citeauthoryear{{Allard} \& {Kielkopf}}{{Allard} \&
  {Kielkopf}}{2009}]{Allard2009}
{Allard} N.~F.,  {Kielkopf} J.~F.,  2009, \mn@doi [\aap]
  {10.1051/0004-6361:200810294}, \href
  {https://ui.adsabs.harvard.edu/abs/2009A&A...493.1155A} {493, 1155}

\bibitem[\protect\citeauthoryear{{Althaus}, {C{\'o}rsico}, {Bischoff-Kim},
  {Romero}, {Renedo}, {Garc{\'\i}a-Berro}  \& {Miller Bertolami}}{{Althaus}
  et~al.}{2010}]{Althaus2010b}
{Althaus} L.~G.,  {C{\'o}rsico} A.~H.,  {Bischoff-Kim} A.,  {Romero} A.~D.,
  {Renedo} I.,  {Garc{\'\i}a-Berro} E.,   {Miller Bertolami} M.~M.,  2010,
  \mn@doi [\apj] {10.1088/0004-637X/717/2/897}, \href
  {https://ui.adsabs.harvard.edu/abs/2010ApJ...717..897A} {717, 897}

\bibitem[\protect\citeauthoryear{{Appenzeller} et~al.,}{{Appenzeller}
  et~al.}{1998}]{Appenzeller1998}
{Appenzeller} I.,  et~al., 1998, The Messenger, \href
  {https://ui.adsabs.harvard.edu/abs/1998Msngr..94....1A} {94, 1}

\bibitem[\protect\citeauthoryear{{Axelrod} et~al.,}{{Axelrod}
  et~al.}{2023}]{Axelrod2023}
{Axelrod} T.,  et~al., 2023, \mn@doi [\apj] {10.3847/1538-4357/acd333}, \href
  {https://ui.adsabs.harvard.edu/abs/2023ApJ...951...78A} {951, 78}

\bibitem[\protect\citeauthoryear{{Aznar Cuadrado}, {Jordan}, {Napiwotzki},
  {Schmid}, {Solanki}  \& {Mathys}}{{Aznar Cuadrado}
  et~al.}{2004}]{AznarCuadrado2004}
{Aznar Cuadrado} R.,  {Jordan} S.,  {Napiwotzki} R.,  {Schmid} H.~M.,
  {Solanki} S.~K.,   {Mathys} G.,  2004, \mn@doi [\aap]
  {10.1051/0004-6361:20040355}, \href
  {https://ui.adsabs.harvard.edu/abs/2004A&A...423.1081A} {423, 1081}

\bibitem[\protect\citeauthoryear{{Bagnulo} \& {Landstreet}}{{Bagnulo} \&
  {Landstreet}}{2018}]{Bagnulo2018}
{Bagnulo} S.,  {Landstreet} J.~D.,  2018, \mn@doi [\aap]
  {10.1051/0004-6361/201833235}, \href
  {https://ui.adsabs.harvard.edu/abs/2018A&A...618A.113B} {618, A113}

\bibitem[\protect\citeauthoryear{{Bagnulo} \& {Landstreet}}{{Bagnulo} \&
  {Landstreet}}{2021}]{Bagnulo2021}
{Bagnulo} S.,  {Landstreet} J.~D.,  2021, \mn@doi [\mnras]
  {10.1093/mnras/stab2046}, \href
  {https://ui.adsabs.harvard.edu/abs/2021MNRAS.507.5902B} {507, 5902}

\bibitem[\protect\citeauthoryear{{Bagnulo} \& {Landstreet}}{{Bagnulo} \&
  {Landstreet}}{2022}]{Bagnulo2022}
{Bagnulo} S.,  {Landstreet} J.~D.,  2022, \mn@doi [\apjl]
  {10.3847/2041-8213/ac84d3}, \href
  {https://ui.adsabs.harvard.edu/abs/2022ApJ...935L..12B} {935, L12}

\bibitem[\protect\citeauthoryear{{B{\'e}dard}, {Bergeron}, {Brassard}  \&
  {Fontaine}}{{B{\'e}dard} et~al.}{2020}]{Bedard2020}
{B{\'e}dard} A.,  {Bergeron} P.,  {Brassard} P.,   {Fontaine} G.,  2020,
  \mn@doi [\apj] {10.3847/1538-4357/abafbe}, \href
  {https://ui.adsabs.harvard.edu/abs/2020ApJ...901...93B} {901, 93}

\bibitem[\protect\citeauthoryear{{Berger}, {Koester}, {Napiwotzki}, {Reid}  \&
  {Zuckerman}}{{Berger} et~al.}{2005}]{Berger2005}
{Berger} L.,  {Koester} D.,  {Napiwotzki} R.,  {Reid} I.~N.,   {Zuckerman} B.,
  2005, \mn@doi [\aap] {10.1051/0004-6361:20053340}, \href
  {https://ui.adsabs.harvard.edu/abs/2005A&A...444..565B} {444, 565}

\bibitem[\protect\citeauthoryear{{Bergeron}, {Dufour}, {Fontaine}, {Coutu},
  {Blouin}, {Genest-Beaulieu}, {B{\'e}dard}  \& {Rolland}}{{Bergeron}
  et~al.}{2019}]{Bergeron2019}
{Bergeron} P.,  {Dufour} P.,  {Fontaine} G.,  {Coutu} S.,  {Blouin} S.,
  {Genest-Beaulieu} C.,  {B{\'e}dard} A.,   {Rolland} B.,  2019, \mn@doi [\apj]
  {10.3847/1538-4357/ab153a}, \href
  {https://ui.adsabs.harvard.edu/abs/2019ApJ...876...67B} {876, 67}

\bibitem[\protect\citeauthoryear{{Bohlin}}{{Bohlin}}{2014}]{Bohlin2014}
{Bohlin} R.~C.,  2014, \mn@doi [\aj] {10.1088/0004-6256/147/6/127}, \href
  {https://ui.adsabs.harvard.edu/abs/2014AJ....147..127B} {147, 127}

\bibitem[\protect\citeauthoryear{{Bohlin} et~al.,}{{Bohlin}
  et~al.}{2011}]{Bohlin2011}
{Bohlin} R.~C.,  et~al., 2011, \mn@doi [\aj] {10.1088/0004-6256/141/5/173},
  \href {https://ui.adsabs.harvard.edu/abs/2011AJ....141..173B} {141, 173}

\bibitem[\protect\citeauthoryear{{Bohlin}, {Gordon}  \& {Tremblay}}{{Bohlin}
  et~al.}{2014}]{BohlinGordonTremblay2014}
{Bohlin} R.~C.,  {Gordon} K.~D.,   {Tremblay} P.~E.,  2014, \mn@doi [\pasp]
  {10.1086/677655}, \href
  {https://ui.adsabs.harvard.edu/abs/2014PASP..126..711B} {126, 711}

\bibitem[\protect\citeauthoryear{{Bohlin}, {Deustua}  \& {de Rosa}}{{Bohlin}
  et~al.}{2019}]{Bohlin2019}
{Bohlin} R.~C.,  {Deustua} S.~E.,   {de Rosa} G.,  2019, \mn@doi [\aj]
  {10.3847/1538-3881/ab480c}, \href
  {https://ui.adsabs.harvard.edu/abs/2019AJ....158..211B} {158, 211}

\bibitem[\protect\citeauthoryear{{Bohlin}, {Hubeny}  \& {Rauch}}{{Bohlin}
  et~al.}{2020}]{Bohlin2020}
{Bohlin} R.~C.,  {Hubeny} I.,   {Rauch} T.,  2020, \mn@doi [\aj]
  {10.3847/1538-3881/ab94b4}, \href
  {https://ui.adsabs.harvard.edu/abs/2020AJ....160...21B} {160, 21}

\bibitem[\protect\citeauthoryear{{Bohlin}, {Krick}, {Gordon}  \&
  {Hubeny}}{{Bohlin} et~al.}{2022}]{Bohlin2022}
{Bohlin} R.~C.,  {Krick} J.~E.,  {Gordon} K.~D.,   {Hubeny} I.,  2022, \mn@doi
  [\aj] {10.3847/1538-3881/ac6fe1}, \href
  {https://ui.adsabs.harvard.edu/abs/2022AJ....164...10B} {164, 10}

\bibitem[\protect\citeauthoryear{{Bond}, {Bergeron}  \& {B{\'e}dard}}{{Bond}
  et~al.}{2017}]{Bond2017}
{Bond} H.~E.,  {Bergeron} P.,   {B{\'e}dard} A.,  2017, \mn@doi [\apj]
  {10.3847/1538-4357/aa8a63}, \href
  {https://ui.adsabs.harvard.edu/abs/2017ApJ...848...16B} {848, 16}

\bibitem[\protect\citeauthoryear{{Brout} et~al.,}{{Brout}
  et~al.}{2022}]{Brout2022}
{Brout} D.,  et~al., 2022, \mn@doi [\apj] {10.3847/1538-4357/ac8bcc}, \href
  {https://ui.adsabs.harvard.edu/abs/2022ApJ...938..111B} {938, 111}

\bibitem[\protect\citeauthoryear{{Busso} et~al.,}{{Busso}
  et~al.}{2022}]{Busso2022}
{Busso} G.,  et~al., 2022, {Gaia DR3 documentation Chapter 5: Photometric
  data}, Gaia DR3 documentation, European Space Agency; Gaia Data Processing
  and Analysis Consortium.

\bibitem[\protect\citeauthoryear{{Byrd}, {Schnabel}  \& {Shultz}}{{Byrd}
  et~al.}{1987}]{Byrd1987}
{Byrd} R.~H.,  {Schnabel} R.~B.,   {Shultz} G.~A.,  1987, \mn@doi [SIAM Journal
  on Numerical Analysis] {10.1137/0724076}, \href
  {https://ui.adsabs.harvard.edu/abs/1987SJNA...24.1152B} {24, 1152}

\bibitem[\protect\citeauthoryear{{Calamida} et~al.,}{{Calamida}
  et~al.}{2022}]{Calamida2022}
{Calamida} A.,  et~al., 2022, \mn@doi [\apj] {10.3847/1538-4357/ac96f4}, \href
  {https://ui.adsabs.harvard.edu/abs/2022ApJ...940...19C} {940, 19}

\bibitem[\protect\citeauthoryear{{Caron}, {Bergeron}, {Blouin}  \&
  {Leggett}}{{Caron} et~al.}{2023}]{Caron2023}
{Caron} A.,  {Bergeron} P.,  {Blouin} S.,   {Leggett} S.~K.,  2023, \mn@doi
  [\mnras] {10.1093/mnras/stac3733}, \href
  {https://ui.adsabs.harvard.edu/abs/2023MNRAS.519.4529C} {519, 4529}

\bibitem[\protect\citeauthoryear{{Cho}, {Gomez}, {Montgomery}, {Dunlap}, {Fitz
  Axen}, {Hobbs}, {Hubeny}  \& {Winget}}{{Cho} et~al.}{2022}]{Cho2022}
{Cho} P.~B.,  {Gomez} T.~A.,  {Montgomery} M.~H.,  {Dunlap} B.~H.,  {Fitz Axen}
  M.,  {Hobbs} B.,  {Hubeny} I.,   {Winget} D.~E.,  2022, \mn@doi [\apj]
  {10.3847/1538-4357/ac4df3}, \href
  {https://ui.adsabs.harvard.edu/abs/2022ApJ...927...70C} {927, 70}

\bibitem[\protect\citeauthoryear{{Cukanovaite}, {Tremblay}, {Bergeron},
  {Freytag}, {Ludwig}  \& {Steffen}}{{Cukanovaite}
  et~al.}{2021}]{Cukanovaite2021}
{Cukanovaite} E.,  {Tremblay} P.-E.,  {Bergeron} P.,  {Freytag} B.,  {Ludwig}
  H.-G.,   {Steffen} M.,  2021, \mn@doi [\mnras] {10.1093/mnras/staa3684},
  \href {https://ui.adsabs.harvard.edu/abs/2021MNRAS.501.5274C} {501, 5274}

\bibitem[\protect\citeauthoryear{{Cunningham}, {Tremblay}, {Gentile Fusillo},
  {Hollands}  \& {Cukanovaite}}{{Cunningham} et~al.}{2020}]{Cunningham2020}
{Cunningham} T.,  {Tremblay} P.-E.,  {Gentile Fusillo} N.~P.,  {Hollands} M.,
  {Cukanovaite} E.,  2020, \mn@doi [\mnras] {10.1093/mnras/stz3638}, \href
  {https://ui.adsabs.harvard.edu/abs/2020MNRAS.492.3540C} {492, 3540}

\bibitem[\protect\citeauthoryear{{El-Badry}, {Rix}  \& {Heintz}}{{El-Badry}
  et~al.}{2021}]{El-Badry2021}
{El-Badry} K.,  {Rix} H.-W.,   {Heintz} T.~M.,  2021, \mn@doi [\mnras]
  {10.1093/mnras/stab323}, \href
  {https://ui.adsabs.harvard.edu/abs/2021MNRAS.506.2269E} {506, 2269}

\bibitem[\protect\citeauthoryear{{Farihi}, {Becklin}  \& {Zuckerman}}{{Farihi}
  et~al.}{2005}]{FarihiBecklinZuckerman2005}
{Farihi} J.,  {Becklin} E.~E.,   {Zuckerman} B.,  2005, \mn@doi [\apjs]
  {10.1086/444362}, \href
  {https://ui.adsabs.harvard.edu/abs/2005ApJS..161..394F} {161, 394}

\bibitem[\protect\citeauthoryear{{Farihi}, {Jura}  \& {Zuckerman}}{{Farihi}
  et~al.}{2009}]{Farihi2009}
{Farihi} J.,  {Jura} M.,   {Zuckerman} B.,  2009, \mn@doi [\apj]
  {10.1088/0004-637X/694/2/805}, \href
  {https://ui.adsabs.harvard.edu/abs/2009ApJ...694..805F} {694, 805}

\bibitem[\protect\citeauthoryear{{Fitzpatrick} \& {Massa}}{{Fitzpatrick} \&
  {Massa}}{1999}]{Fitzpatrick1999}
{Fitzpatrick} E.~L.,  {Massa} D.,  1999, \mn@doi [\apj] {10.1086/307944}, \href
  {https://ui.adsabs.harvard.edu/abs/1999ApJ...525.1011F} {525, 1011}

\bibitem[\protect\citeauthoryear{{Fontaine} \& {Wesemael}}{{Fontaine} \&
  {Wesemael}}{1987}]{Fontaine1987}
{Fontaine} G.,  {Wesemael} F.,  1987, in {Philip} A.~G.~D.,  {Hayes} D.~S.,
  {Liebert} J.~W.,  eds, IAU Colloq. 95: Second Conference on Faint Blue Stars.
  pp 319--326

\bibitem[\protect\citeauthoryear{{Fontaine} \& {Wesemael}}{{Fontaine} \&
  {Wesemael}}{1997}]{Fontaine1997}
{Fontaine} G.,  {Wesemael} F.,  1997, in {Isern} J.,  {Hernanz} M.,
  {Garcia-Berro} E.,  eds,  Astrophysics and Space Science Library Vol. 214,
  White dwarfs. p.~173, \mn@doi{10.1007/978-94-011-5542-7_26}

\bibitem[\protect\citeauthoryear{{Gaia Collaboration} et~al.,}{{Gaia
  Collaboration} et~al.}{2023}]{Gaia2023}
{Gaia Collaboration} et~al., 2023, \mn@doi [\aap]
  {10.1051/0004-6361/202243940}, \href
  {https://ui.adsabs.harvard.edu/abs/2023A&A...674A...1G} {674, A1}

\bibitem[\protect\citeauthoryear{{Gentile Fusillo}, {Tremblay}, {Bohlin},
  {Deustua}  \& {Kalirai}}{{Gentile Fusillo} et~al.}{2020}]{GF2020}
{Gentile Fusillo} N.~P.,  {Tremblay} P.-E.,  {Bohlin} R.~C.,  {Deustua} S.~E.,
   {Kalirai} J.~S.,  2020, \mn@doi [\mnras] {10.1093/mnras/stz2984}, \href
  {https://ui.adsabs.harvard.edu/abs/2020MNRAS.491.3613G} {491, 3613}

\bibitem[\protect\citeauthoryear{{Gentile Fusillo} et~al.,}{{Gentile Fusillo}
  et~al.}{2021}]{GF2021}
{Gentile Fusillo} N.~P.,  et~al., 2021, \mn@doi [\mnras]
  {10.1093/mnras/stab2672}, \href
  {https://ui.adsabs.harvard.edu/abs/2021MNRAS.508.3877G} {508, 3877}

\bibitem[\protect\citeauthoryear{{Gianninas}, {Bergeron}, {Dupuis}  \&
  {Ruiz}}{{Gianninas} et~al.}{2010}]{Gianninas2010}
{Gianninas} A.,  {Bergeron} P.,  {Dupuis} J.,   {Ruiz} M.~T.,  2010, \mn@doi
  [\apj] {10.1088/0004-637X/720/1/581}, \href
  {https://ui.adsabs.harvard.edu/abs/2010ApJ...720..581G} {720, 581}

\bibitem[\protect\citeauthoryear{{Gomez}, {Stambulchik}  \& {White}}{{Gomez}
  et~al.}{2024}]{Gomez2024}
{Gomez} T.~A.,  {Stambulchik} E.,   {White} J.,  2024, \mn@doi [\pra]
  {10.1103/PhysRevA.109.052804}, \href
  {https://ui.adsabs.harvard.edu/abs/2024PhRvA.109e2804G} {109, 052804}

\bibitem[\protect\citeauthoryear{{Gordon} et~al.,}{{Gordon}
  et~al.}{2022}]{Gordon2022}
{Gordon} K.~D.,  et~al., 2022, \mn@doi [\aj] {10.3847/1538-3881/ac66dc}, \href
  {https://ui.adsabs.harvard.edu/abs/2022AJ....163..267G} {163, 267}

\bibitem[\protect\citeauthoryear{{Gordon}, {Clayton}, {Decleir}, {Fitzpatrick},
  {Massa}, {Misselt}  \& {Tollerud}}{{Gordon} et~al.}{2023}]{Gordon2023}
{Gordon} K.~D.,  {Clayton} G.~C.,  {Decleir} M.,  {Fitzpatrick} E.~L.,  {Massa}
  D.,  {Misselt} K.~A.,   {Tollerud} E.~J.,  2023, \mn@doi [\apj]
  {10.3847/1538-4357/accb59}, \href
  {https://ui.adsabs.harvard.edu/abs/2023ApJ...950...86G} {950, 86}

\bibitem[\protect\citeauthoryear{{Greenstein} \& {Oke}}{{Greenstein} \&
  {Oke}}{1979}]{Greenstein1979}
{Greenstein} J.~L.,  {Oke} J.~B.,  1979, \mn@doi [\apjl] {10.1086/182948},
  \href {https://ui.adsabs.harvard.edu/abs/1979ApJ...229L.141G} {229, L141}

\bibitem[\protect\citeauthoryear{{Heintz}, {Hermes}, {Tremblay}, {Ould Rouis},
  {Reding}, {Kaiser}  \& {van Saders}}{{Heintz} et~al.}{2024}]{Heintz2024}
{Heintz} T.~M.,  {Hermes} J.~J.,  {Tremblay} P.~E.,  {Ould Rouis} L.~B.,
  {Reding} J.~S.,  {Kaiser} B.~C.,   {van Saders} J.~L.,  2024, \mn@doi [\apj]
  {10.3847/1538-4357/ad479b}, \href
  {https://ui.adsabs.harvard.edu/abs/2024ApJ...969...68H} {969, 68}

\bibitem[\protect\citeauthoryear{{Holberg} \& {Bergeron}}{{Holberg} \&
  {Bergeron}}{2006}]{Holberg2006}
{Holberg} J.~B.,  {Bergeron} P.,  2006, \mn@doi [\aj] {10.1086/505938}, \href
  {https://ui.adsabs.harvard.edu/abs/2006AJ....132.1221H} {132, 1221}

\bibitem[\protect\citeauthoryear{{Holberg}, {Wesemael}, {Wegner}  \&
  {Bruhweiler}}{{Holberg} et~al.}{1985}]{Holberg1985}
{Holberg} J.~B.,  {Wesemael} F.,  {Wegner} G.,   {Bruhweiler} F.~C.,  1985,
  \mn@doi [\apj] {10.1086/163237}, \href
  {https://ui.adsabs.harvard.edu/abs/1985ApJ...293..294H} {293, 294}

\bibitem[\protect\citeauthoryear{{Holberg}, {Sion}, {Oswalt}, {McCook}, {Foran}
   \& {Subasavage}}{{Holberg} et~al.}{2008}]{Holberg2008}
{Holberg} J.~B.,  {Sion} E.~M.,  {Oswalt} T.,  {McCook} G.~P.,  {Foran} S.,
  {Subasavage} J.~P.,  2008, \mn@doi [\aj] {10.1088/0004-6256/135/4/1225},
  \href {https://ui.adsabs.harvard.edu/abs/2008AJ....135.1225H} {135, 1225}

\bibitem[\protect\citeauthoryear{{Holberg}, {Oswalt}, {Sion}  \&
  {McCook}}{{Holberg} et~al.}{2016}]{Holberg2016}
{Holberg} J.~B.,  {Oswalt} T.~D.,  {Sion} E.~M.,   {McCook} G.~P.,  2016,
  \mn@doi [\mnras] {10.1093/mnras/stw1357}, \href
  {https://ui.adsabs.harvard.edu/abs/2016MNRAS.462.2295H} {462, 2295}

\bibitem[\protect\citeauthoryear{{Hounsell} et~al.,}{{Hounsell}
  et~al.}{2018}]{Hounsell2018}
{Hounsell} R.,  et~al., 2018, \mn@doi [\apj] {10.3847/1538-4357/aac08b}, \href
  {https://ui.adsabs.harvard.edu/abs/2018ApJ...867...23H} {867, 23}

\bibitem[\protect\citeauthoryear{{Iben} \& {Tutukov}}{{Iben} \&
  {Tutukov}}{1984}]{Iben1984}
{Iben} I. J.,  {Tutukov} A.~V.,  1984, \mn@doi [\apjs] {10.1086/190932}, \href
  {https://ui.adsabs.harvard.edu/abs/1984ApJS...54..335I} {54, 335}

\bibitem[\protect\citeauthoryear{{Johnson}}{{Johnson}}{1965}]{Johnson1965}
{Johnson} H.~L.,  1965, \mn@doi [\apj] {10.1086/148186}, \href
  {https://ui.adsabs.harvard.edu/abs/1965ApJ...141..923J} {141, 923}

\bibitem[\protect\citeauthoryear{{Kawka} \& {Vennes}}{{Kawka} \&
  {Vennes}}{2012}]{Kawka2012}
{Kawka} A.,  {Vennes} S.,  2012, \mn@doi [\mnras]
  {10.1111/j.1365-2966.2012.21574.x}, \href
  {https://ui.adsabs.harvard.edu/abs/2012MNRAS.425.1394K} {425, 1394}

\bibitem[\protect\citeauthoryear{{Kawka}, {Vennes}, {Dinnbier}, {Cibulkov{\'a}}
   \& {N{\'e}meth}}{{Kawka} et~al.}{2011}]{Kawka2011}
{Kawka} A.,  {Vennes} S.,  {Dinnbier} F.,  {Cibulkov{\'a}} H.,   {N{\'e}meth}
  P.,  2011, in {Schuh} S.,  {Drechsel} H.,   {Heber} U.,  eds,  American
  Institute of Physics Conference Series Vol. 1331, Planetary Systems Beyond
  the Main Sequence. AIP, pp 238--245 (\mn@eprint {arXiv} {1012.2639}),
  \mn@doi{10.1063/1.3556206}

\bibitem[\protect\citeauthoryear{{Kilic}, {von Hippel}, {Leggett}  \&
  {Winget}}{{Kilic} et~al.}{2006}]{Kilic2006}
{Kilic} M.,  {von Hippel} T.,  {Leggett} S.~K.,   {Winget} D.~E.,  2006,
  \mn@doi [\apj] {10.1086/504682}, \href
  {https://ui.adsabs.harvard.edu/abs/2006ApJ...646..474K} {646, 474}

\bibitem[\protect\citeauthoryear{{Koester}}{{Koester}}{2010}]{Koester2010}
{Koester} D.,  2010, \memsai, \href
  {https://ui.adsabs.harvard.edu/abs/2010MmSAI..81..921K} {81, 921}

\bibitem[\protect\citeauthoryear{{Koester} \& {Wilken}}{{Koester} \&
  {Wilken}}{2006}]{KoesterWilken2006}
{Koester} D.,  {Wilken} D.,  2006, \mn@doi [\aap] {10.1051/0004-6361:20064843},
  \href {https://ui.adsabs.harvard.edu/abs/2006A&A...453.1051K} {453, 1051}

\bibitem[\protect\citeauthoryear{{Koester}, {Weidemann}, {Zeidler-K.~T.}  \&
  {Vauclair}}{{Koester} et~al.}{1985}]{Koester1985}
{Koester} D.,  {Weidemann} V.,  {Zeidler-K.~T.} E.~M.,   {Vauclair} G.,  1985,
  \aap, \href {https://ui.adsabs.harvard.edu/abs/1985A&A...142L...5K} {142, L5}

\bibitem[\protect\citeauthoryear{{Koester}, {Dreizler}, {Weidemann}  \&
  {Allard}}{{Koester} et~al.}{1998}]{Koester1998}
{Koester} D.,  {Dreizler} S.,  {Weidemann} V.,   {Allard} N.~F.,  1998, \aap,
  \href {https://ui.adsabs.harvard.edu/abs/1998A&A...338..612K} {338, 612}

\bibitem[\protect\citeauthoryear{{Koester}, {Rollenhagen}, {Napiwotzki},
  {Voss}, {Christlieb}, {Homeier}  \& {Reimers}}{{Koester}
  et~al.}{2005}]{Koester2005}
{Koester} D.,  {Rollenhagen} K.,  {Napiwotzki} R.,  {Voss} B.,  {Christlieb}
  N.,  {Homeier} D.,   {Reimers} D.,  2005, \mn@doi [\aap]
  {10.1051/0004-6361:20041927}, \href
  {https://ui.adsabs.harvard.edu/abs/2005A&A...432.1025K} {432, 1025}

\bibitem[\protect\citeauthoryear{{Koester}, {Voss}, {Napiwotzki}, {Christlieb},
  {Homeier}, {Lisker}, {Reimers}  \& {Heber}}{{Koester}
  et~al.}{2009}]{Koester2009b}
{Koester} D.,  {Voss} B.,  {Napiwotzki} R.,  {Christlieb} N.,  {Homeier} D.,
  {Lisker} T.,  {Reimers} D.,   {Heber} U.,  2009, \mn@doi [\aap]
  {10.1051/0004-6361/200912531}, \href
  {https://ui.adsabs.harvard.edu/abs/2009A&A...505..441K} {505, 441}

\bibitem[\protect\citeauthoryear{{Kowalski} \& {Saumon}}{{Kowalski} \&
  {Saumon}}{2006}]{KowalskiSaumon2006}
{Kowalski} P.~M.,  {Saumon} D.,  2006, \mn@doi [\apjl] {10.1086/509723}, \href
  {https://ui.adsabs.harvard.edu/abs/2006ApJ...651L.137K} {651, L137}

\bibitem[\protect\citeauthoryear{{Landstreet} \& {Bagnulo}}{{Landstreet} \&
  {Bagnulo}}{2019}]{Landstreet2019}
{Landstreet} J.~D.,  {Bagnulo} S.,  2019, \mn@doi [\aap]
  {10.1051/0004-6361/201834638}, \href
  {https://ui.adsabs.harvard.edu/abs/2019A&A...623A..46L} {623, A46}

\bibitem[\protect\citeauthoryear{{Ma{\'\i}z Apell{\'a}niz} \&
  {Weiler}}{{Ma{\'\i}z Apell{\'a}niz} \& {Weiler}}{2018}]{Maiz2018}
{Ma{\'\i}z Apell{\'a}niz} J.,  {Weiler} M.,  2018, \mn@doi [\aap]
  {10.1051/0004-6361/201834051}, \href
  {https://ui.adsabs.harvard.edu/abs/2018A&A...619A.180M} {619, A180}

\bibitem[\protect\citeauthoryear{{McCook} \& {Sion}}{{McCook} \&
  {Sion}}{1999}]{McCook1999}
{McCook} G.~P.,  {Sion} E.~M.,  1999, \mn@doi [\apjs] {10.1086/313186}, \href
  {https://ui.adsabs.harvard.edu/abs/1999ApJS..121....1M} {121, 1}

\bibitem[\protect\citeauthoryear{{Megessier}}{{Megessier}}{1995}]{Megessier1995}
{Megessier} C.,  1995, \aap, \href
  {https://ui.adsabs.harvard.edu/abs/1995A&A...296..771M} {296, 771}

\bibitem[\protect\citeauthoryear{{Miller Bertolami}, {Althaus}  \&
  {C{\'o}rsico}}{{Miller Bertolami} et~al.}{2017}]{MillerBertolami2017}
{Miller Bertolami} M.~M.,  {Althaus} L.~G.,   {C{\'o}rsico} A.~H.,  2017, in
  {Tremblay} P.~E.,  {Gaensicke} B.,   {Marsh} T.,  eds,  Astronomical Society
  of the Pacific Conference Series Vol. 509, 20th European White Dwarf
  Workshop. p.~435 (\mn@eprint {arXiv} {1609.08683}),
  \mn@doi{10.48550/arXiv.1609.08683}

\bibitem[\protect\citeauthoryear{{Mullally} et~al.,}{{Mullally}
  et~al.}{2024}]{Mullally2024}
{Mullally} S.~E.,  et~al., 2024, \mn@doi [\apjl] {10.3847/2041-8213/ad2348},
  \href {https://ui.adsabs.harvard.edu/abs/2024ApJ...962L..32M} {962, L32}

\bibitem[\protect\citeauthoryear{{Munday} et~al.,}{{Munday}
  et~al.}{2024}]{Munday2024}
{Munday} J.,  et~al., 2024, \mn@doi [\mnras] {10.1093/mnras/stae1645}, \href
  {https://ui.adsabs.harvard.edu/abs/2024MNRAS.532.2534M} {532, 2534}

\bibitem[\protect\citeauthoryear{{Napiwotzki} et~al.,}{{Napiwotzki}
  et~al.}{2020}]{Napi2020}
{Napiwotzki} R.,  et~al., 2020, \mn@doi [\aap] {10.1051/0004-6361/201629648},
  \href {https://ui.adsabs.harvard.edu/abs/2020A&A...638A.131N} {638, A131}

\bibitem[\protect\citeauthoryear{{Narayan} et~al.,}{{Narayan}
  et~al.}{2016}]{Narayan2016}
{Narayan} G.,  et~al., 2016, \mn@doi [\apj] {10.3847/0004-637X/822/2/67}, \href
  {https://ui.adsabs.harvard.edu/abs/2016ApJ...822...67N} {822, 67}

\bibitem[\protect\citeauthoryear{{Narayan} et~al.,}{{Narayan}
  et~al.}{2019}]{Narayan2019}
{Narayan} G.,  et~al., 2019, \mn@doi [\apjs] {10.3847/1538-4365/ab0557}, \href
  {https://ui.adsabs.harvard.edu/abs/2019ApJS..241...20N} {241, 20}

\bibitem[\protect\citeauthoryear{{Nelan} \& {Wegner}}{{Nelan} \&
  {Wegner}}{1985}]{Nelan1985}
{Nelan} E.~P.,  {Wegner} G.,  1985, \mn@doi [\apjl] {10.1086/184428}, \href
  {https://ui.adsabs.harvard.edu/abs/1985ApJ...289L..31N} {289, L31}

\bibitem[\protect\citeauthoryear{{O'Brien} et~al.,}{{O'Brien}
  et~al.}{2024}]{OBrien2024}
{O'Brien} M.~W.,  et~al., 2024, \mn@doi [\mnras] {10.1093/mnras/stad3773},
  \href {https://ui.adsabs.harvard.edu/abs/2024MNRAS.527.8687O} {527, 8687}

\bibitem[\protect\citeauthoryear{{Oswalt}, {Hintzen}  \& {Luyten}}{{Oswalt}
  et~al.}{1988}]{Oswalt1988}
{Oswalt} T.~D.,  {Hintzen} P.~M.,   {Luyten} W.~J.,  1988, \mn@doi [\apjs]
  {10.1086/191263}, \href
  {https://ui.adsabs.harvard.edu/abs/1988ApJS...66..391O} {66, 391}

\bibitem[\protect\citeauthoryear{{Perryman} et~al.,}{{Perryman}
  et~al.}{1997}]{Perryman1997}
{Perryman} M.~A.~C.,  et~al., 1997, \aap, \href
  {https://ui.adsabs.harvard.edu/abs/1997A&A...323L..49P} {323, L49}

\bibitem[\protect\citeauthoryear{{Rauch}, {Werner}, {Bohlin}  \&
  {Kruk}}{{Rauch} et~al.}{2013}]{Rauch2013}
{Rauch} T.,  {Werner} K.,  {Bohlin} R.,   {Kruk} J.~W.,  2013, \mn@doi [\aap]
  {10.1051/0004-6361/201322336}, \href
  {https://ui.adsabs.harvard.edu/abs/2013A&A...560A.106R} {560, A106}

\bibitem[\protect\citeauthoryear{{Ricker} et~al.,}{{Ricker}
  et~al.}{2014}]{TESS2014}
{Ricker} G.~R.,  et~al., 2014, in {Oschmann} Jacobus~M. J.,  {Clampin} M.,
  {Fazio} G.~G.,   {MacEwen} H.~A.,  eds,  Society of Photo-Optical
  Instrumentation Engineers (SPIE) Conference Series Vol. 9143, Space
  Telescopes and Instrumentation 2014: Optical, Infrared, and Millimeter Wave.
  p. 914320 (\mn@eprint {arXiv} {1406.0151}), \mn@doi{10.1117/12.2063489}

\bibitem[\protect\citeauthoryear{{Rieke}, {Engelke}, {Su}  \&
  {Casagrande}}{{Rieke} et~al.}{2023}]{Rieke2023}
{Rieke} G.~H.,  {Engelke} C.,  {Su} K.,   {Casagrande} L.,  2023, \mn@doi [\aj]
  {10.3847/1538-3881/ac9f1b}, \href
  {https://ui.adsabs.harvard.edu/abs/2023AJ....165...99R} {165, 99}

\bibitem[\protect\citeauthoryear{{Rodrigo}, {Solano}  \& {Bayo}}{{Rodrigo}
  et~al.}{2012}]{Rodrigo2012}
{Rodrigo} C.,  {Solano} E.,   {Bayo} A.,  2012, {SVO Filter Profile Service
  Version 1.0}, IVOA Working Draft 15 October 2012,
  \mn@doi{10.5479/ADS/bib/2012ivoa.rept.1015R}

\bibitem[\protect\citeauthoryear{{Sahu} et~al.,}{{Sahu}
  et~al.}{2023}]{Sahu2023}
{Sahu} S.,  et~al., 2023, \mn@doi [\mnras] {10.1093/mnras/stad2663}, \href
  {https://ui.adsabs.harvard.edu/abs/2023MNRAS.526.5800S} {526, 5800}

\bibitem[\protect\citeauthoryear{{Saumon}, {Holberg}  \& {Kowalski}}{{Saumon}
  et~al.}{2014}]{Saumon2014}
{Saumon} D.,  {Holberg} J.~B.,   {Kowalski} P.~M.,  2014, \mn@doi [\apj]
  {10.1088/0004-637X/790/1/50}, \href
  {https://ui.adsabs.harvard.edu/abs/2014ApJ...790...50S} {790, 50}

\bibitem[\protect\citeauthoryear{{Saumon}, {Blouin}  \& {Tremblay}}{{Saumon}
  et~al.}{2022}]{Saumon2022}
{Saumon} D.,  {Blouin} S.,   {Tremblay} P.-E.,  2022, \mn@doi [\physrep]
  {10.1016/j.physrep.2022.09.001}, \href
  {https://ui.adsabs.harvard.edu/abs/2022PhR...988....1S} {988, 1}

\bibitem[\protect\citeauthoryear{{Schultz} \& {Wiemer}}{{Schultz} \&
  {Wiemer}}{1975}]{Schultz1975}
{Schultz} G.~V.,  {Wiemer} W.,  1975, \aap, \href
  {https://ui.adsabs.harvard.edu/abs/1975A&A....43..133S} {43, 133}

\bibitem[\protect\citeauthoryear{{Scolnic} et~al.,}{{Scolnic}
  et~al.}{2015}]{Scolnic2015}
{Scolnic} D.,  et~al., 2015, \mn@doi [\apj] {10.1088/0004-637X/815/2/117},
  \href {https://ui.adsabs.harvard.edu/abs/2015ApJ...815..117S} {815, 117}

\bibitem[\protect\citeauthoryear{{Scolnic} et~al.,}{{Scolnic}
  et~al.}{2022}]{Scolnic2022}
{Scolnic} D.,  et~al., 2022, \mn@doi [\apj] {10.3847/1538-4357/ac8b7a}, \href
  {https://ui.adsabs.harvard.edu/abs/2022ApJ...938..113S} {938, 113}

\bibitem[\protect\citeauthoryear{{Skrutskie} et~al.,}{{Skrutskie}
  et~al.}{2006}]{Skrutskie2006}
{Skrutskie} M.~F.,  et~al., 2006, \mn@doi [\aj] {10.1086/498708}, \href
  {https://ui.adsabs.harvard.edu/abs/2006AJ....131.1163S} {131, 1163}

\bibitem[\protect\citeauthoryear{{Stubbs} \& {Brown}}{{Stubbs} \&
  {Brown}}{2015}]{Stubbs2015}
{Stubbs} C.~W.,  {Brown} Y.~J.,  2015, \mn@doi [Modern Physics Letters A]
  {10.1142/S021773231530030X}, \href
  {https://ui.adsabs.harvard.edu/abs/2015MPLA...3030030S} {30, 1530030}

\bibitem[\protect\citeauthoryear{{Subasavage} et~al.,}{{Subasavage}
  et~al.}{2017}]{Subasavage2017}
{Subasavage} J.~P.,  et~al., 2017, \mn@doi [\aj] {10.3847/1538-3881/aa76e0},
  \href {https://ui.adsabs.harvard.edu/abs/2017AJ....154...32S} {154, 32}

\bibitem[\protect\citeauthoryear{{Tayar}, {Claytor}, {Huber}  \& {van
  Saders}}{{Tayar} et~al.}{2022}]{Tayar2022}
{Tayar} J.,  {Claytor} Z.~R.,  {Huber} D.,   {van Saders} J.,  2022, \mn@doi
  [\apj] {10.3847/1538-4357/ac4bbc}, \href
  {https://ui.adsabs.harvard.edu/abs/2022ApJ...927...31T} {927, 31}

\bibitem[\protect\citeauthoryear{{Toonen}, {Hollands}, {G{\"a}nsicke}  \&
  {Boekholt}}{{Toonen} et~al.}{2017}]{Toonen2017}
{Toonen} S.,  {Hollands} M.,  {G{\"a}nsicke} B.~T.,   {Boekholt} T.,  2017,
  \mn@doi [\aap] {10.1051/0004-6361/201629978}, \href
  {https://ui.adsabs.harvard.edu/abs/2017A&A...602A..16T} {602, A16}

\bibitem[\protect\citeauthoryear{{Tremblay} \& {Bergeron}}{{Tremblay} \&
  {Bergeron}}{2009}]{Tremblay2009}
{Tremblay} P.~E.,  {Bergeron} P.,  2009, \mn@doi [\apj]
  {10.1088/0004-637X/696/2/1755}, \href
  {https://ui.adsabs.harvard.edu/abs/2009ApJ...696.1755T} {696, 1755}

\bibitem[\protect\citeauthoryear{{Tremblay}, {Bergeron}  \&
  {Gianninas}}{{Tremblay} et~al.}{2011}]{Tremblay2011}
{Tremblay} P.~E.,  {Bergeron} P.,   {Gianninas} A.,  2011, \mn@doi [\apj]
  {10.1088/0004-637X/730/2/128}, \href
  {https://ui.adsabs.harvard.edu/abs/2011ApJ...730..128T} {730, 128}

\bibitem[\protect\citeauthoryear{{Tremblay}, {Ludwig}, {Steffen}  \&
  {Freytag}}{{Tremblay} et~al.}{2013}]{Tremblay2013}
{Tremblay} P.~E.,  {Ludwig} H.~G.,  {Steffen} M.,   {Freytag} B.,  2013,
  \mn@doi [\aap] {10.1051/0004-6361/201322318}, \href
  {https://ui.adsabs.harvard.edu/abs/2013A&A...559A.104T} {559, A104}

\bibitem[\protect\citeauthoryear{{Tremblay}, {Gianninas}, {Kilic}, {Ludwig},
  {Steffen}, {Freytag}  \& {Hermes}}{{Tremblay} et~al.}{2015a}]{Tremblay2015a}
{Tremblay} P.~E.,  {Gianninas} A.,  {Kilic} M.,  {Ludwig} H.~G.,  {Steffen} M.,
   {Freytag} B.,   {Hermes} J.~J.,  2015a, \mn@doi [\apj]
  {10.1088/0004-637X/809/2/148}, \href
  {https://ui.adsabs.harvard.edu/abs/2015ApJ...809..148T} {809, 148}

\bibitem[\protect\citeauthoryear{{Tremblay}, {Fontaine}, {Freytag}, {Steiner},
  {Ludwig}, {Steffen}, {Wedemeyer}  \& {Brassard}}{{Tremblay}
  et~al.}{2015b}]{Tremblay2015b}
{Tremblay} P.~E.,  {Fontaine} G.,  {Freytag} B.,  {Steiner} O.,  {Ludwig}
  H.~G.,  {Steffen} M.,  {Wedemeyer} S.,   {Brassard} P.,  2015b, \mn@doi
  [\apj] {10.1088/0004-637X/812/1/19}, \href
  {https://ui.adsabs.harvard.edu/abs/2015ApJ...812...19T} {812, 19}

\bibitem[\protect\citeauthoryear{{Tremblay} et~al.,}{{Tremblay}
  et~al.}{2017}]{Tremblay2017}
{Tremblay} P.~E.,  et~al., 2017, \mn@doi [\mnras] {10.1093/mnras/stw2854},
  \href {https://ui.adsabs.harvard.edu/abs/2017MNRAS.465.2849T} {465, 2849}

\bibitem[\protect\citeauthoryear{{Tremblay}, {Cukanovaite}, {Gentile Fusillo},
  {Cunningham}  \& {Hollands}}{{Tremblay} et~al.}{2019}]{Tremblay2019b}
{Tremblay} P.~E.,  {Cukanovaite} E.,  {Gentile Fusillo} N.~P.,  {Cunningham}
  T.,   {Hollands} M.~A.,  2019, \mn@doi [\mnras] {10.1093/mnras/sty3067},
  \href {https://ui.adsabs.harvard.edu/abs/2019MNRAS.482.5222T} {482, 5222}

\bibitem[\protect\citeauthoryear{{Valyavin}, {Bagnulo}, {Fabrika},
  {Reisenegger}, {Wade}, {Han}  \& {Monin}}{{Valyavin}
  et~al.}{2006}]{Valyavin2006}
{Valyavin} G.,  {Bagnulo} S.,  {Fabrika} S.,  {Reisenegger} A.,  {Wade} G.~A.,
  {Han} I.,   {Monin} D.,  2006, \mn@doi [\apj] {10.1086/505781}, \href
  {https://ui.adsabs.harvard.edu/abs/2006ApJ...648..559V} {648, 559}

\bibitem[\protect\citeauthoryear{{Wegner}}{{Wegner}}{1982}]{Wegner1982}
{Wegner} G.,  1982, \mn@doi [\apjl] {10.1086/183894}, \href
  {https://ui.adsabs.harvard.edu/abs/1982ApJ...261L..87W} {261, L87}

\bibitem[\protect\citeauthoryear{{Werner}}{{Werner}}{1996}]{Werner1996}
{Werner} K.,  1996, \mn@doi [\apjl] {10.1086/309889}, \href
  {https://ui.adsabs.harvard.edu/abs/1996ApJ...457L..39W} {457, L39}

\bibitem[\protect\citeauthoryear{{Whittet} \& {van Breda}}{{Whittet} \& {van
  Breda}}{1980}]{Whittet1980}
{Whittet} D.~C.~B.,  {van Breda} I.~G.,  1980, \mn@doi [\mnras]
  {10.1093/mnras/192.3.467}, \href
  {https://ui.adsabs.harvard.edu/abs/1980MNRAS.192..467W} {192, 467}

\bibitem[\protect\citeauthoryear{{Wilson} et~al.,}{{Wilson}
  et~al.}{2023}]{Wilson2023}
{Wilson} R.~F.,  et~al., 2023, \mn@doi [\apjs] {10.3847/1538-4365/acf3df},
  \href {https://ui.adsabs.harvard.edu/abs/2023ApJS..269....5W} {269, 5}

\bibitem[\protect\citeauthoryear{{Wright} et~al.,}{{Wright}
  et~al.}{2010}]{WISE2010}
{Wright} E.~L.,  et~al., 2010, \mn@doi [\aj] {10.1088/0004-6256/140/6/1868},
  \href {https://ui.adsabs.harvard.edu/abs/2010AJ....140.1868W} {140, 1868}

\bibitem[\protect\citeauthoryear{{Xu}, {Jura}, {Koester}, {Klein}  \&
  {Zuckerman}}{{Xu} et~al.}{2013}]{Xu2013}
{Xu} S.,  {Jura} M.,  {Koester} D.,  {Klein} B.,   {Zuckerman} B.,  2013,
  \mn@doi [\apjl] {10.1088/2041-8205/766/2/L18}, \href
  {https://ui.adsabs.harvard.edu/abs/2013ApJ...766L..18X} {766, L18}

\bibitem[\protect\citeauthoryear{{Zuckerman}, {Becklin}, {Macintosh}  \&
  {Bida}}{{Zuckerman} et~al.}{1997}]{Zuckerman1997}
{Zuckerman} B.,  {Becklin} E.~E.,  {Macintosh} B.~A.,   {Bida} T.,  1997,
  \mn@doi [\aj] {10.1086/118296}, \href
  {https://ui.adsabs.harvard.edu/abs/1997AJ....113..764Z} {113, 764}

\bibitem[\protect\citeauthoryear{{Zuckerman}, {Koester}, {Reid}  \&
  {H{\"u}nsch}}{{Zuckerman} et~al.}{2003}]{Zuckerman2003}
{Zuckerman} B.,  {Koester} D.,  {Reid} I.~N.,   {H{\"u}nsch} M.,  2003, \mn@doi
  [\apj] {10.1086/377492}, \href
  {https://ui.adsabs.harvard.edu/abs/2003ApJ...596..477Z} {596, 477}

\makeatother
\end{thebibliography}




\appendix

\section{\textit{HST} observation log}

\begin{table*}
    \centering
    \caption{Log of \textit{HST} STIS and WFC3 spectrophotometric observations for the 17 DA white dwarfs proposed as flux standards. The exposure time ($t_{\rm exp}$), number of exposures ($n_{\rm exp}$) and the duration of the observing run is given. Where observations were taken over multiple dates with multiple exposure times, numbers are separated by a forward slash (/). }
    \label{tab:HST_log}
    \begin{tabular}{ccccccc}
        \hline
        \hline
         Object & Telescope/Instrument & Date & Grating & $t_{\rm exp}$ & $n_{\rm exp}$ & Duration\\
         &   & [yyyy-mm-dd] &  & [s] &  & [s]\\
        \hline
        \hline
        \multicolumn{7}{c}{\textbf{Warm flux standard candidates}}\\
        \hline
        \multirow{6}{2cm}{WD\,0148$+$467} & \multirow{4}{2cm}{HST/STIS} & 2021-08-07 & G140L & 1355.0 & 1 & 1355.0 \\
          &  & 2021-08-07 & G230L & 718.0 & 1 & 718.0 \\
          &  & 2021-08-07 & G430L & 410.0 & 1 & 410.0 \\
          &  & 2021-08-07 & G750L & 2000.0 & 1 & 2000.0 \\
          & \multirow{2}{2cm}{HST/WFC3} & 2021-07-30 & G102 & 138.4 & 3 & 415.1 \\
          &  & 2021-07-30 & G141 & 115.5 & 3 & 346.4 \\
        \hline
        \multirow{6}{2cm}{WD\,0227$+$050} & \multirow{4}{2cm}{HST/STIS} & 2021-12-19 & G140L & 953.0 & 1 & 953.0 \\
          &  & 2021-12-19 & G230L & 980.0 & 1 & 980.0 \\
          &  & 2021-12-19 & G430L & 306.0 & 1 & 306.0 \\
          &  & 2021-12-19 & G750L & 1920.0 & 1 & 1920.0 \\
          & \multirow{2}{2cm}{HST/WFC3} & 2021-09-22 & G102 & 253.0 & 3 & 759.0 \\
          &  & 2021-07-30 & G141 & 207.1 & 3 & 621.4 \\
        \hline
        \multirow{6}{2cm}{WD\,0809$+$177} & \multirow{4}{2cm}{HST/STIS} & 2022-02-21 & G140L & 1074.0 & 1 & 1074.0 \\
          &  & 2022-02-21 & G230L & 870.0 & 1 & 870.0 \\
          &  & 2022-02-21 & G430L & 300.0 & 1 & 300.0 \\
          &  & 2022-02-21 & G750L & 1980.0 & 1 & 1980.0 \\
          & \multirow{2}{2cm}{HST/WFC3} & 2021-09-18 & G102 & 432.1 & 3 & 1296.4 \\
          &  & 2021-09-18 & G141 & 336.3 & 3 & 1008.9 \\
        \hline
        \multirow{6}{2cm}{WD\,1105$-$340} & \multirow{4}{2cm}{HST/STIS} & 2022-01-27 & G140L & 1368.0 & 1 & 1368.0 \\
          &  & 2022-01-27 & G230L & 600.0 & 1 & 600.0 \\
          &  & 2022-01-27 & G430L & 324.0 & 1 & 324.0 \\
          &  & 2022-01-27 & G750L & 1980.0 & 1 & 1980.0 \\
          & \multirow{2}{2cm}{HST/WFC3} & 2021-08-03 & G102 & 432.1 & 3 & 1296.4 \\
          &  & 2021-08-03 & G141 & 384.2 & 3 & 1152.7 \\
        \hline
        \multirow{6}{2cm}{WD\,1105$-$048} & \multirow{4}{2cm}{HST/STIS} & 2022-04-02 & G140L & 1143.0 & 1 & 1143.0 \\
          &  & 2022-04-02 & G230L & 780.0 & 1 & 780.0 \\
          &  & 2022-04-02 & G430L & 340.0 & 1 & 340.0 \\
          &  & 2022-04-02 & G750L & 1928.0 & 1 & 1928.0 \\
          & \multirow{2}{2cm}{HST/WFC3} & 2021-12-15 & G102 & 275.9 & 3 & 827.7 \\
          &  & 2021-12-15 & G141 & 230.1 & 3 & 690.2 \\
        \hline
        \multirow{6}{2cm}{WD\,1327$-$083} & \multirow{4}{2cm}{HST/STIS} & 2016-04-15 & G140L & 1244.0 & 1 & 1244.0 \\
          &  & 2016-04-15 & G230L & 605.0 & 1 & 605.0 \\
          &  & 2016-04-15 & G430L & 200.0 & 1 & 200.0 \\
          &  & 2016-04-15 & G750L & 936.0 & 1 & 936.0 \\
          & \multirow{2}{2cm}{HST/WFC3} & 2016-04-15 & G102 & 83.5 & 1 & 83.5 \\
          &  & 2016-04-15 & G141 & 60.1 & 3 & 60.1 \\
        \hline
        \multirow{6}{2cm}{WD\,1713$+$695} & \multirow{4}{2cm}{HST/STIS} & 2022-07-07 & G140L & 1364.0 & 1 & 1364.0 \\
          &  & 2022-07-07 & G230L & 910.0 & 1 & 910.0 \\
          &  & 2022-07-07 & G430L & 342.0 & 1 & 342.0 \\
          &  & 2022-08-29 & G750L & 2475.0 & 1 & 2475.0 \\
          & \multirow{2}{2cm}{HST/WFC3} & 2021-09-20 & G102 & 321.8 & 3 & 965.3 \\
          &  & 2021-09-20 & G141 & 275.9 & 3 & 827.7 \\
        \hline
        \multirow{6}{2cm}{WD\,1911$+$536} & \multirow{4}{2cm}{HST/STIS} & 2021-10-01 & G140L & 1112.0 & 1 & 1112.0 \\
          &  & 2021-10-01 & G230L & 1000.0 & 1 & 1000.0 \\
          &  & 2021-10-01 & G430L & 330.0 & 1 & 330.0 \\
          &  & 2021-10-01 & G750L & 2080.0 & 1 & 2080.0 \\
          & \multirow{2}{2cm}{HST/WFC3} & 2021-07-24 & G102 & 336.3 & 3 & 1008.9 \\
          &  & 2021-07-24 & G141 & 298.8 & 3 & 896.5 \\
        \hline
        \multirow{6}{2cm}{WD\,1919$+$145} & \multirow{4}{2cm}{HST/STIS} & 2022-08-23 & G140L & 1237.0 & 1 & 1237.0 \\
          &  & 2022-08-23 & G230L & 700.0 & 1 & 700.0 \\
          &  & 2021/2022-10/08-06/23 & G430L & 330.0 & 2 & 660.0 \\
          &  & 2022-08-23 & G750L & 1944.0 & 2 & 1944.0 \\
          & \multirow{2}{2cm}{HST/WFC3} & 2021-07/09-21/14 & G102 & 253.0 & 6 & 1517.9 \\
          &  & 2021-07/09-21/14 & G141 & 207.1 & 6 & 1242.9 \\
        \hline
        \multirow{6}{2cm}{WD\,2039$-$682} & \multirow{4}{2cm}{HST/STIS} & 2022-08-01 & G140L & 1220.0 & 1 & 1220.0 \\
          &  & 2022-08-01 & G230L & 1054.0 & 1 & 1054.0 \\
          &  & 2022-08-01 & G430L & 342.0 & 1 & 342.0 \\
          &  & 2022-08-01 & G750L & 2225.0 & 1 & 2225.0 \\
          & \multirow{2}{2cm}{HST/WFC3} & 2021-07-27 & G102 & 336.3 & 3 & 1008.9 \\
          &  & 2021-07-27 & G141 & 298.8 & 3 & 896.5 \\
        \hline
    \end{tabular}
\end{table*}

\begin{table*}
    \centering
    \contcaption{}
    \begin{tabular}{ccccccc}
    \hline
    \hline
     Object & Telescope/Instrument & Date & Grating & $t_{\rm exp}$ & $n_{\rm exp}$ & Duration\\
     &   & [yyyy-mm-dd] &  & [s] &  & [s]\\
    \hline
    \hline
        \multirow{6}{2cm}{WD\,2117$+$539} & \multirow{4}{2cm}{HST/STIS} & 2022-08-27 & G140L & 1437.0 & 1 & 1437.0 \\
          &  & 2022-08-27 & G230L & 700.0 & 1 & 700.0 \\
          &  & 2022-08-27 & G430L & 354.0 & 1 & 354.0 \\
          &  & 2022-08-27 & G750L & 2060.0 & 1 & 2060.0 \\
          & \multirow{2}{2cm}{HST/WFC3} & 2021-07/09-31/15 & G102 & 230.1 & 6 & 1380.4 \\
          &  & 2021-07/09-31/15 & G141 & 184.2 & 6 & 1105.3 \\
        \hline
        \multirow{6}{2cm}{WD\,2126$+$734} & \multirow{4}{2cm}{HST/STIS} & 2021-10-02 & G140L & 1400.0 & 1 & 1400.0 \\
          &  & 2021-10-02 & G230L & 913.0 & 1 & 913.0 \\
          &  & 2021-10-02 & G430L & 406.0 & 1 & 406.0 \\
          &  & 2021-10-02 & G750L & 2200.0 & 1 & 2200.0 \\
          & \multirow{2}{2cm}{HST/WFC3} & 2021-08-03 & G102 & 253.0 & 3 & 759.0 \\
          &  & 2021-08-03 & G141 & 207.1 & 3 & 621.4 \\
        \hline
        \multirow{6}{2cm}{WD\,2149$+$021} & \multirow{4}{2cm}{HST/STIS} & 2021-10-02 & G140L & 965.0 & 1 & 965.0 \\
          &  & 2021-10-02 & G230L & 965.0 & 1 & 965.0 \\
          &  & 2021-10-02 & G430L & 302.0 & 1 & 302.0 \\
          &  & 2021-10-02 & G750L & 1920.0 & 1 & 1920.0 \\
          & \multirow{2}{2cm}{HST/WFC3} & 2021-07-26 & G102 & 207.1 & 3 & 621.4 \\
          &  & 2021-07-26 & G141 & 161.3/184.2 & 3 & 506.8 \\
        \hline
        \multicolumn{7}{c}{\textbf{Extra flux standard candidates}}\\
        \hline
        \multirow{4}{2cm}{WD\,0352$+$096} & \multirow{4}{2cm}{HST/STIS} & 2019-02-09 & G140L & 2379.0 & 1 & 2379.0 \\
          &  & 2019-02-09 & G230L & 1101.0 & 1 & 1101.0 \\
          &  & 2019-02-09 & G430L & 1200.0 & 1 & 1200.0 \\
          &  & 1998-02/03-19 & G750L & 1980.0 & 2 & 3960.0 \\
        \hline
        \multirow{6}{2cm}{WD\,1202$-$232} & \multirow{4}{2cm}{HST/STIS} & 2021-08-05 & G140L & 749.0/761.0 & 2 & 1510.0 \\
          &  & 2016-06-15 & G230L & 2302.2 & 1 & 2302.2 \\
          &  & 2021-08-05 & G430L & 1200.0 & 1 & 1200.0 \\
          &  & 2021-08-05 & G750L & 1600.0 & 1 & 1600.0 \\
          & \multirow{2}{2cm}{HST/WFC3} & 2021-07-25 & G102 & 138.4 & 3 & 415.1 \\
          &  & 2021-07-25 & G141 & 88.0 & 3 & 264.0 \\
        \hline
        \multirow{6}{2cm}{WD\,1544$-$377} & \multirow{4}{2cm}{HST/STIS} & 2021-10-03 & G140L & 1047.0 & 1 & 1047.0 \\
          &  & 2021-10-03 & G230L & 1790.0 & 1 & 1790.0 \\
          &  & 2021-10-03 & G430L & 300.0 & 1 & 300.0 \\
          &  & 2021-10-03 & G750L & 1401.0 & 1 & 1401.0 \\
          & \multirow{2}{2cm}{HST/WFC3} & 2021-07/09-29/27 & G102 & 161.3/177.9 & 3/3 & 1017.7 \\
          &  & 2021-07/09-29/27 & G141 & 115.2/127.9 & 3/3 & 730.2 \\
        \hline
        \multirow{6}{2cm}{WD\,2341$+$322} & \multirow{4}{2cm}{HST/STIS} & 2016-05-27 & G140L & 1370.0 & 1 & 1370.0 \\
          &  & 2016-05-27 & G230L & 650.0 & 1 & 650.0 \\
          &  & 2016-05-27 & G430L & 240.0 & 1 & 240.0 \\
          &  & 2016-05-27 & G750L & 939.0 & 1 & 939.0 \\
          & \multirow{2}{2cm}{HST/WFC3} & 2016-05-27 & G102 & 176.0 & 1 & 176.0 \\
          &  & 2016-05-27 & G141 & 128.5 & 1 & 128.5 \\
          \hline
    \end{tabular}
\end{table*}

\section{Spectrophotometric fits for 11/13 warm white dwarfs}

\begin{figure*}
    \includegraphics[width=1.42\textwidth,height=0.36\textheight,keepaspectratio]{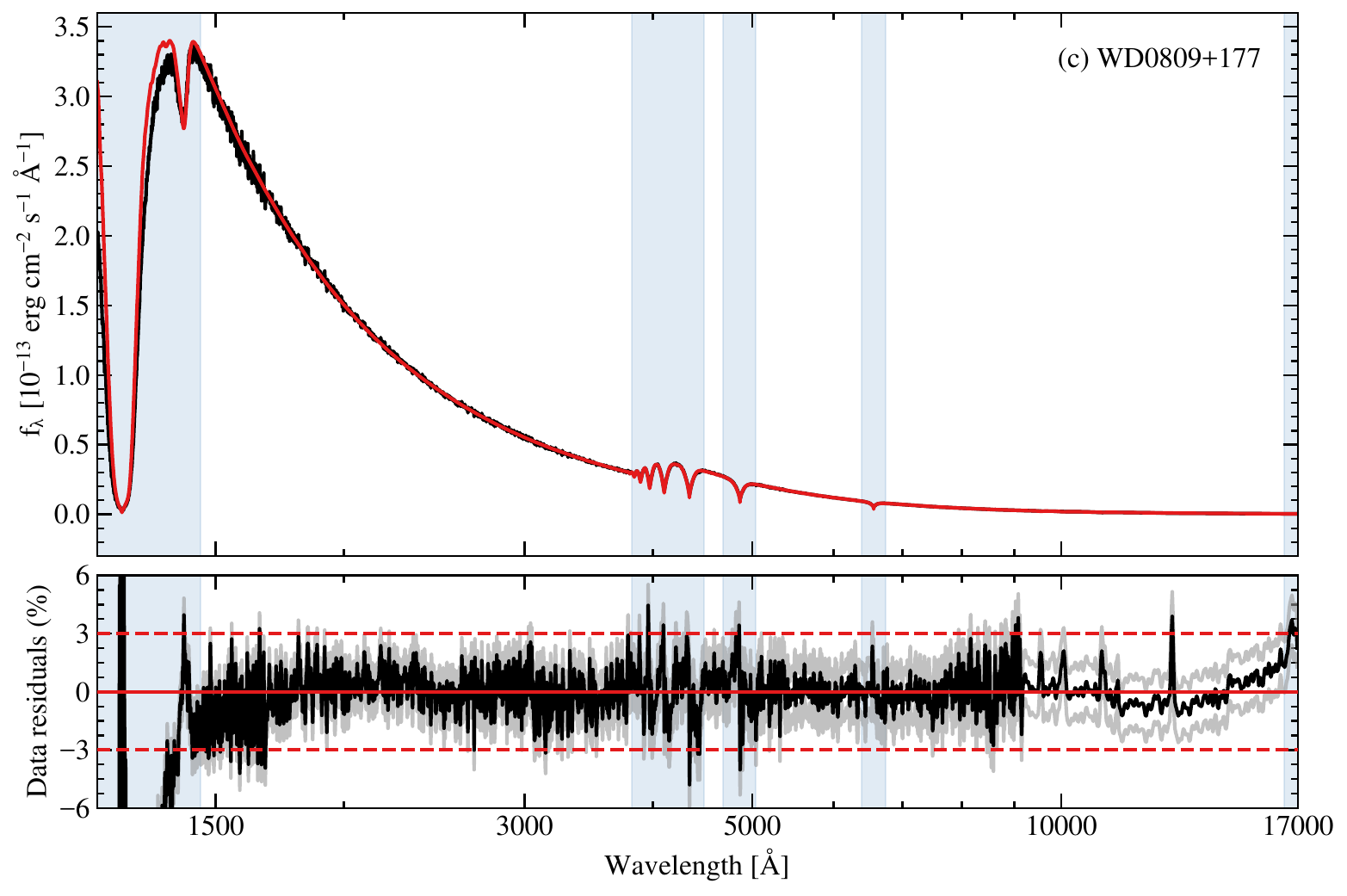}
	\includegraphics[width=0.45\textwidth,height=0.37\textheight,keepaspectratio]{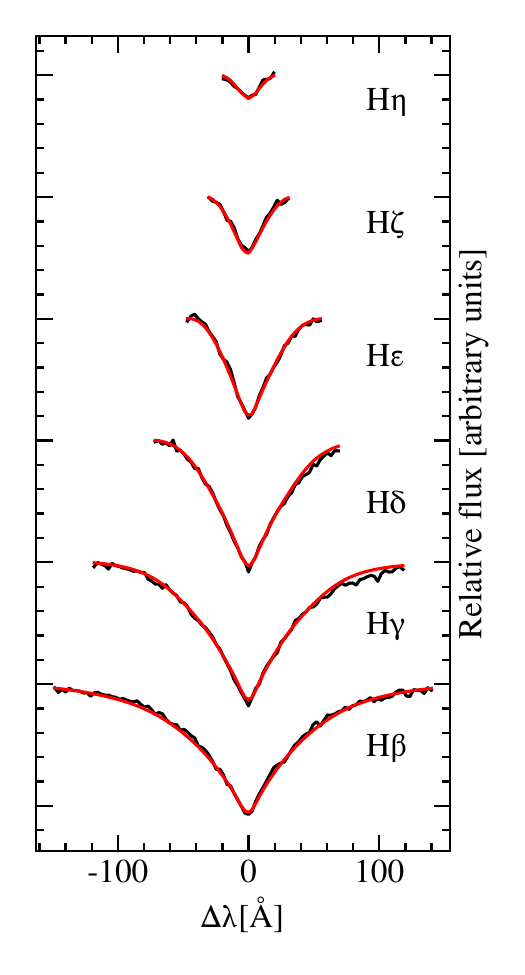}
 	\includegraphics[width=1.42\textwidth,height=0.36\textheight,keepaspectratio]{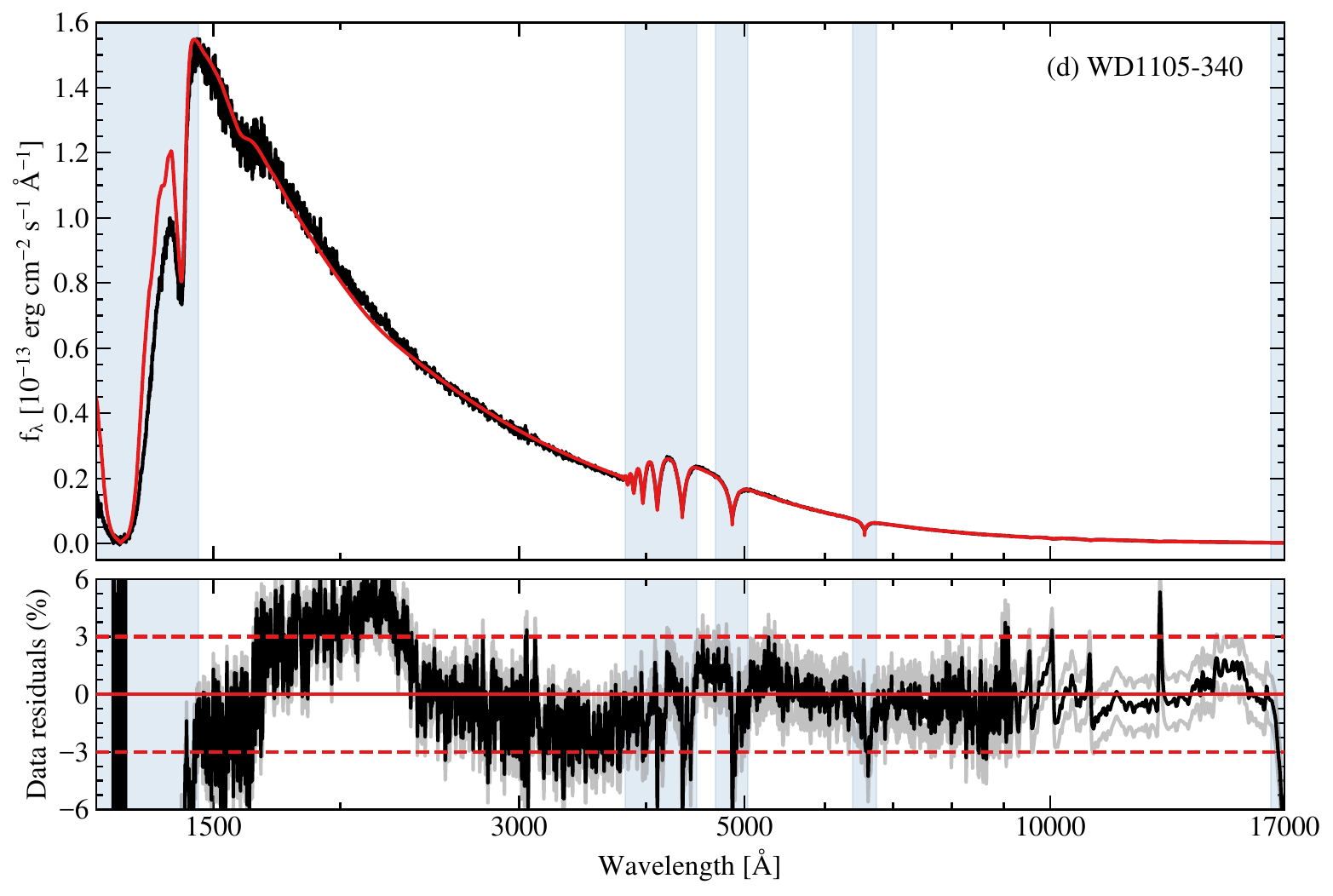}
	\includegraphics[width=0.45\textwidth,height=0.37\textheight,keepaspectratio]{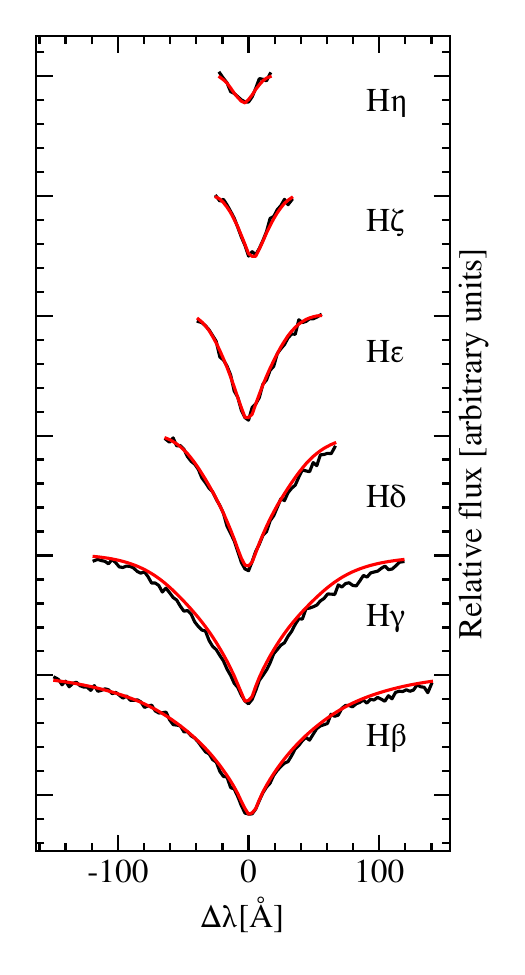}
    \caption{Spectrophotometric fits of the STIS and WFC3 data for 11/13 warm white dwarfs proposed as flux standard candidates, with the WD name of each star given in the corner of the top left panels. \textit{Top left}: SED fit between the observed spectrophotometry (black) and best-fitting model (red). \textit{Bottom left}: Flux residuals from the corresponding SED fit, where the black line is the calculated residual, grey lines indicate residuals $\pm 1\sigma$ using only the statistical errors from the fits, and red lines indicate 0 and $\pm 3$ per cent flux residuals as a guide. The shaded blue regions in the left panels indicate wavelength ranges excluded from the fits. \textit{Right}: Balmer line fits for H$\beta$ to H$\eta$ between the observed spectrophotometry (black) and best-fitting model (red). The line profiles are vertically offset for clarity. The best-fitting parameters for the SED and Balmer line fits are found in Table~\ref{tab:fit_results}.}
\label{fig:SED_Balmer_warm_appendix}
\end{figure*}

\begin{figure*}
\contcaption{}
	\includegraphics[width=1.42\textwidth,height=0.36\textheight,keepaspectratio]{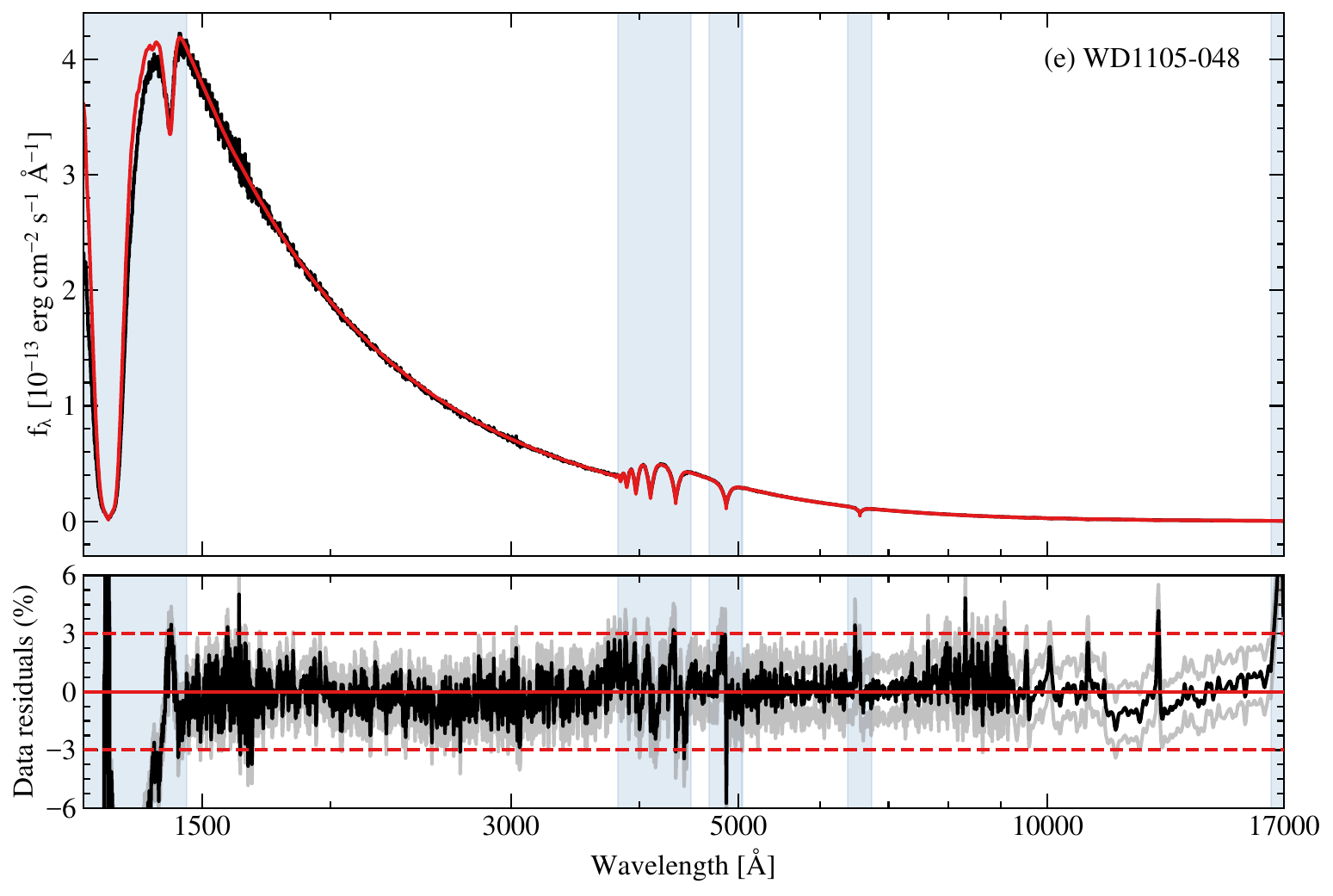}
	\includegraphics[width=0.45\textwidth,height=0.37\textheight,keepaspectratio]{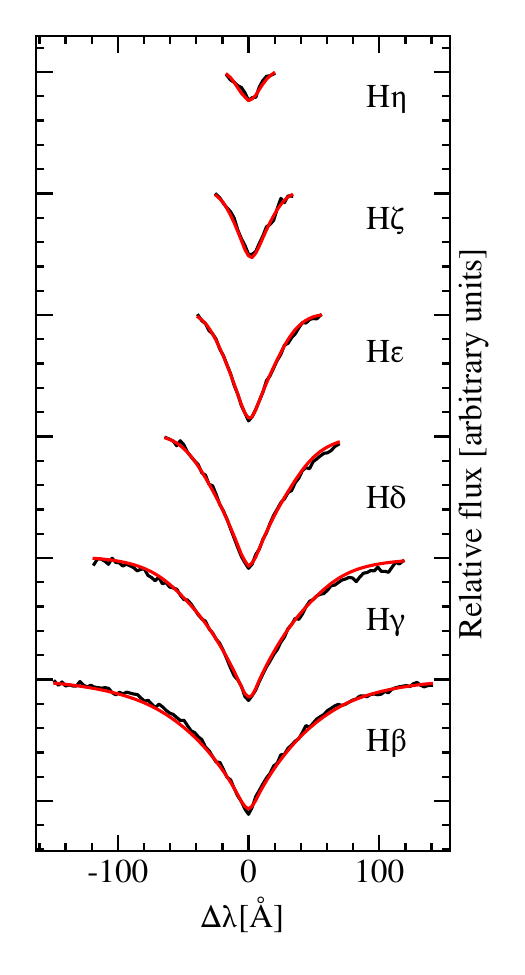}
 	\includegraphics[width=1.42\textwidth,height=0.36\textheight,keepaspectratio]{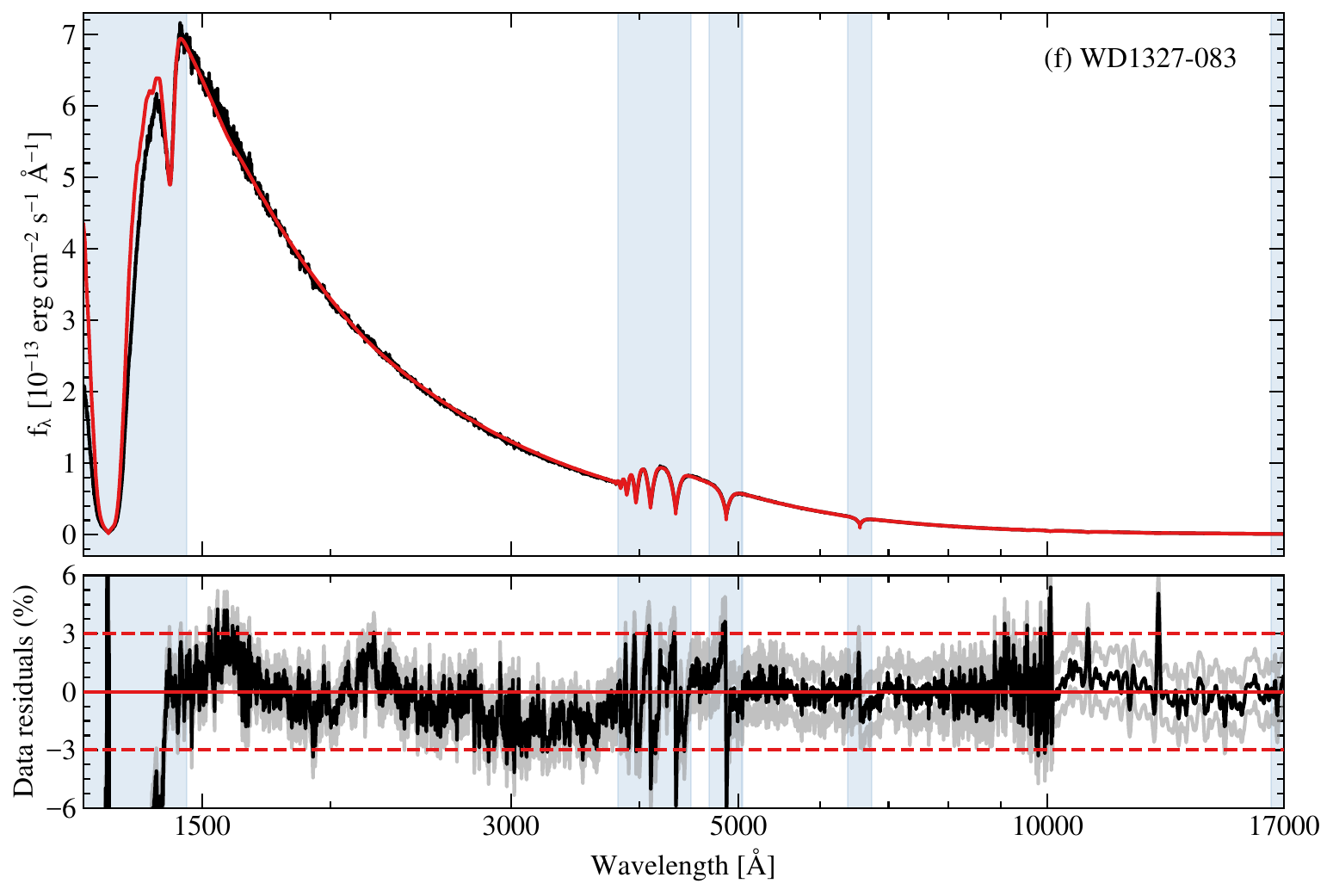}
	\includegraphics[width=0.45\textwidth,height=0.37\textheight,keepaspectratio]{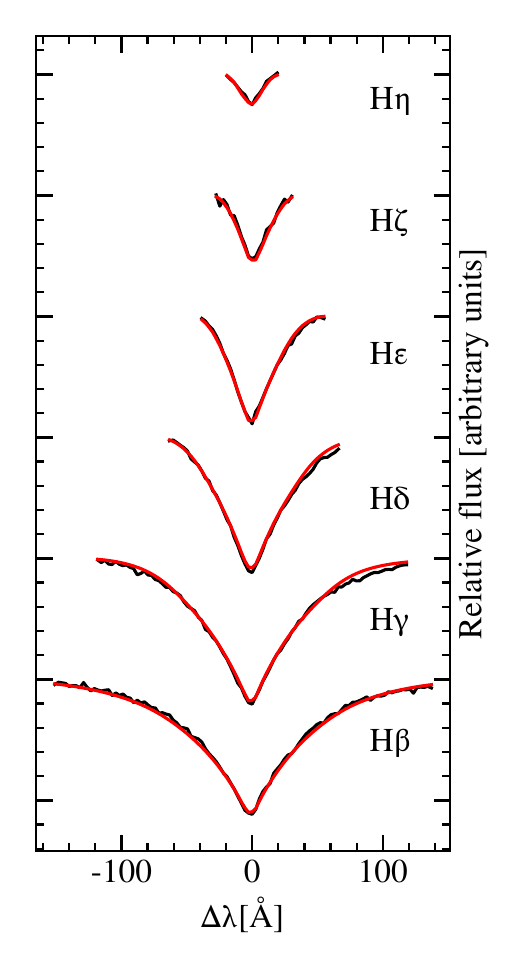}
\end{figure*}

\begin{figure*}
\contcaption{}
	\includegraphics[width=1.42\textwidth,height=0.36\textheight,keepaspectratio]{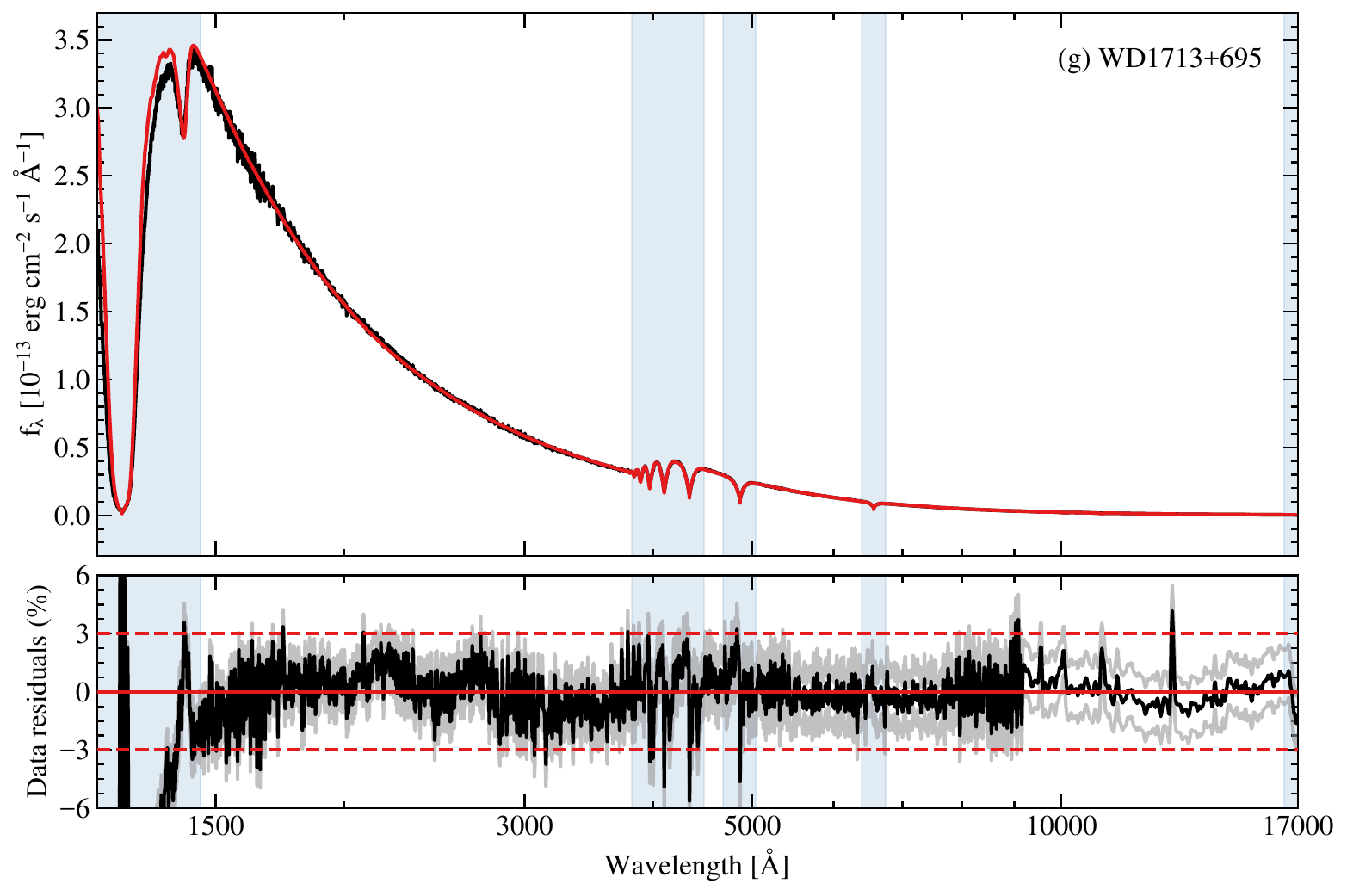}
	\includegraphics[width=0.45\textwidth,height=0.37\textheight,keepaspectratio]{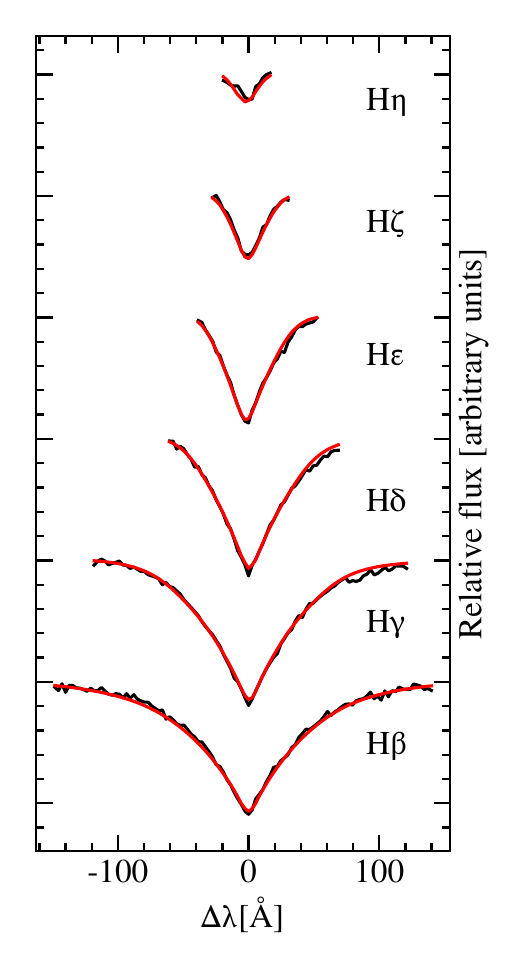}
 	\includegraphics[width=1.42\textwidth,height=0.36\textheight,keepaspectratio]{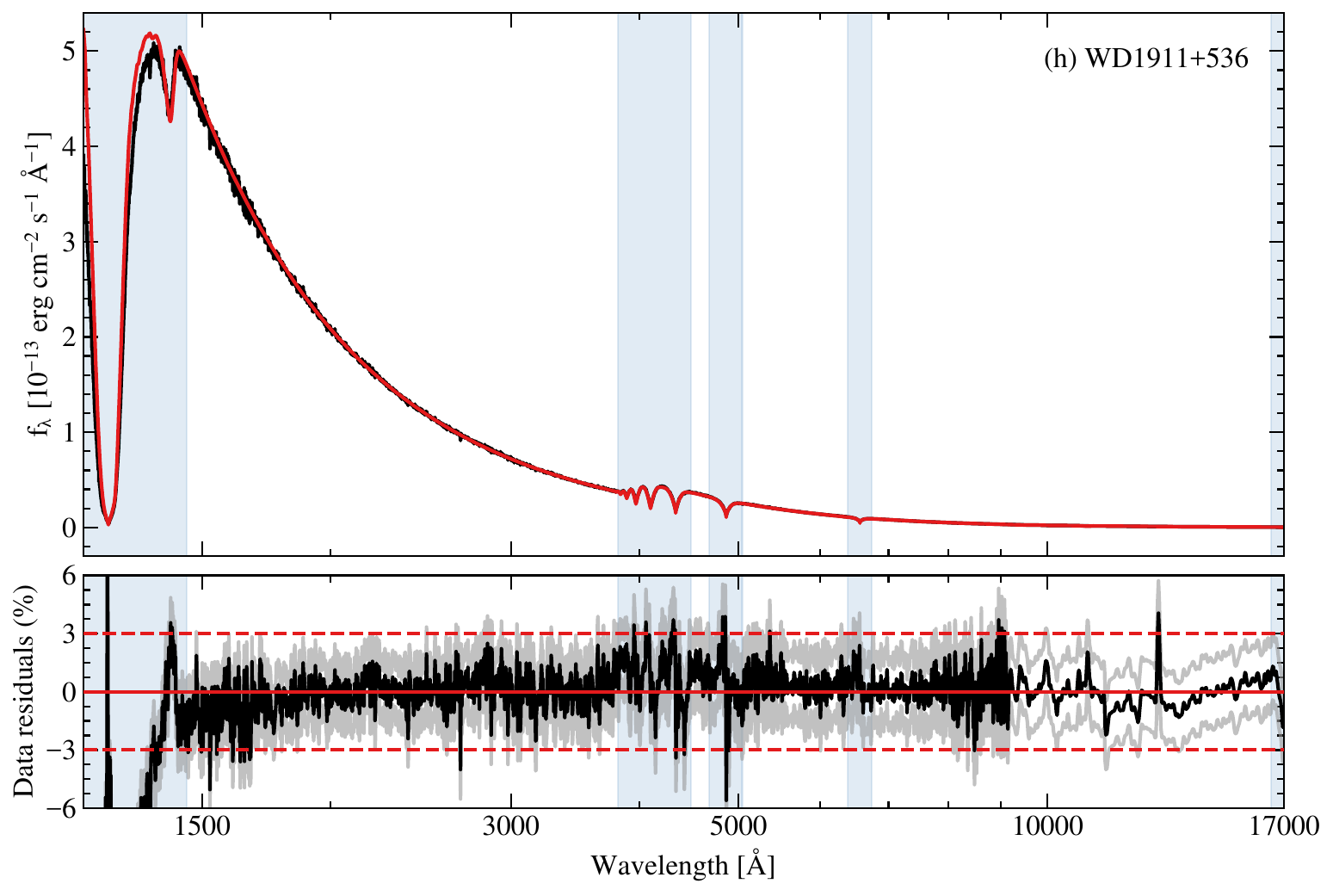}
	\includegraphics[width=0.45\textwidth,height=0.37\textheight,keepaspectratio]{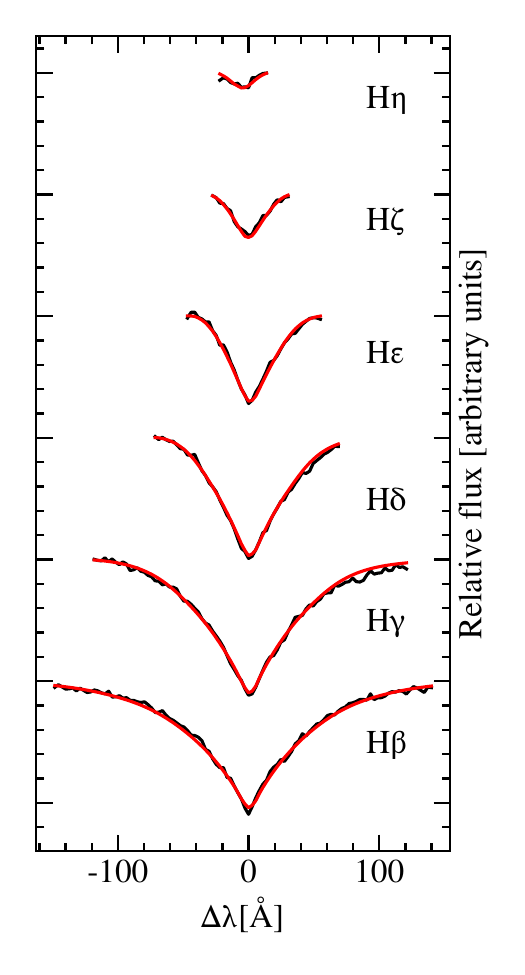}
\end{figure*}

\begin{figure*}
\contcaption{}
	\includegraphics[width=1.42\textwidth,height=0.36\textheight,keepaspectratio]{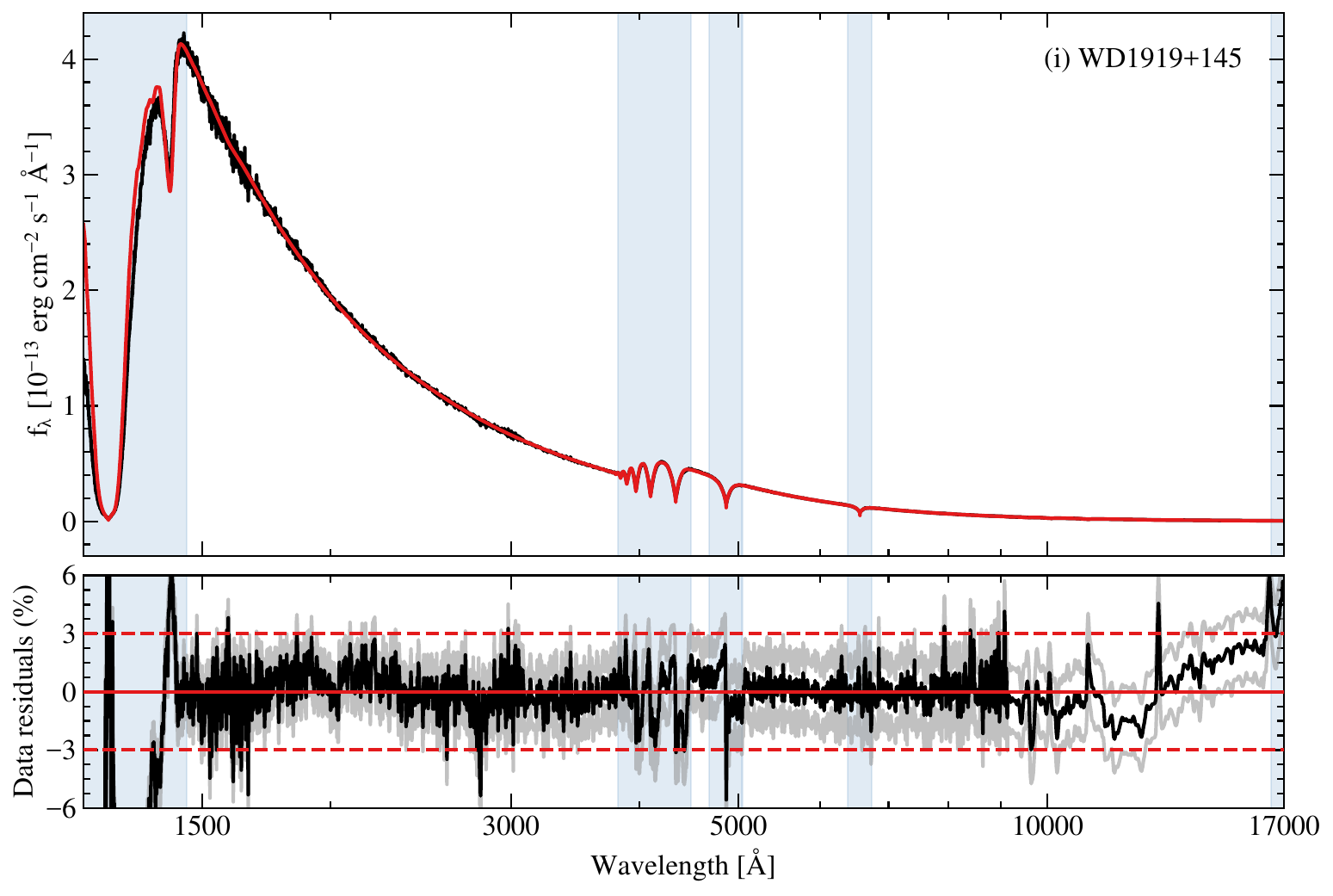}
	\includegraphics[width=0.45\textwidth,height=0.37\textheight,keepaspectratio]{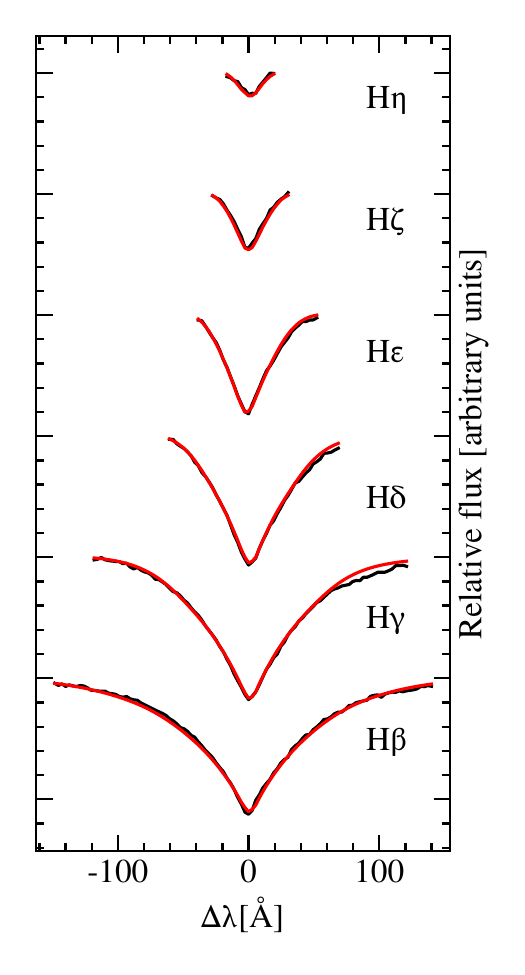}
 	\includegraphics[width=1.42\textwidth,height=0.36\textheight,keepaspectratio]{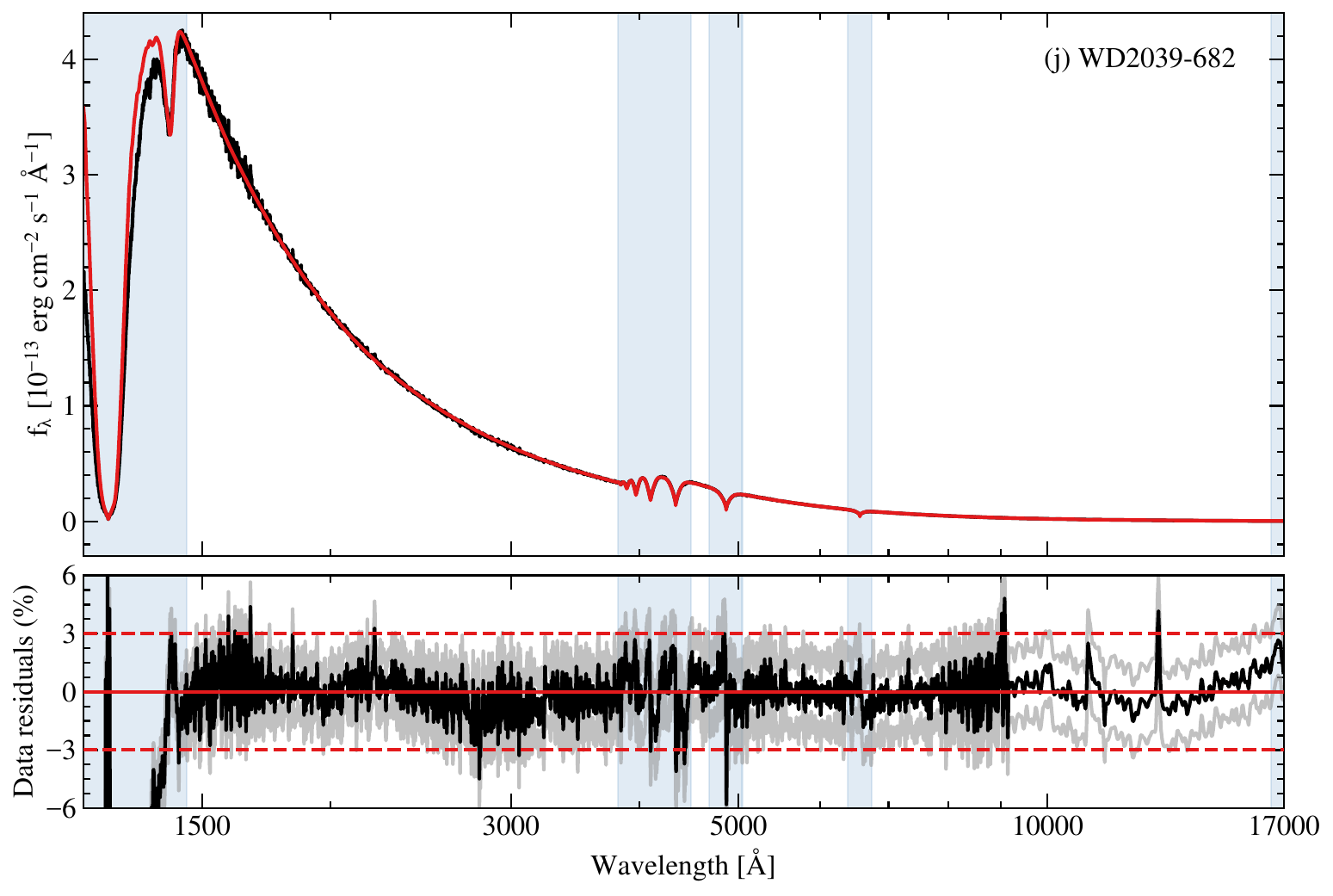}
	\includegraphics[width=0.45\textwidth,height=0.37\textheight,keepaspectratio]{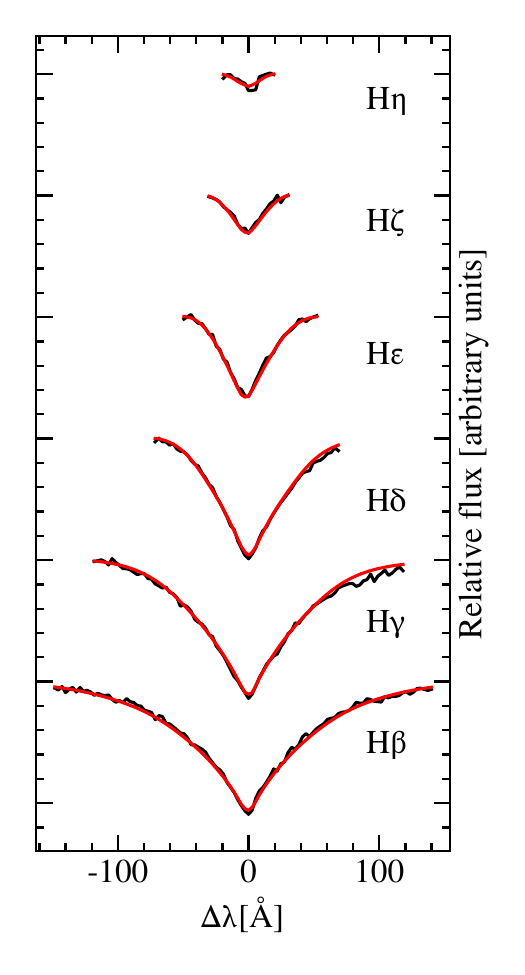}
\end{figure*}

\begin{figure*}
\contcaption{}
	\includegraphics[width=1.42\textwidth,height=0.36\textheight,keepaspectratio]{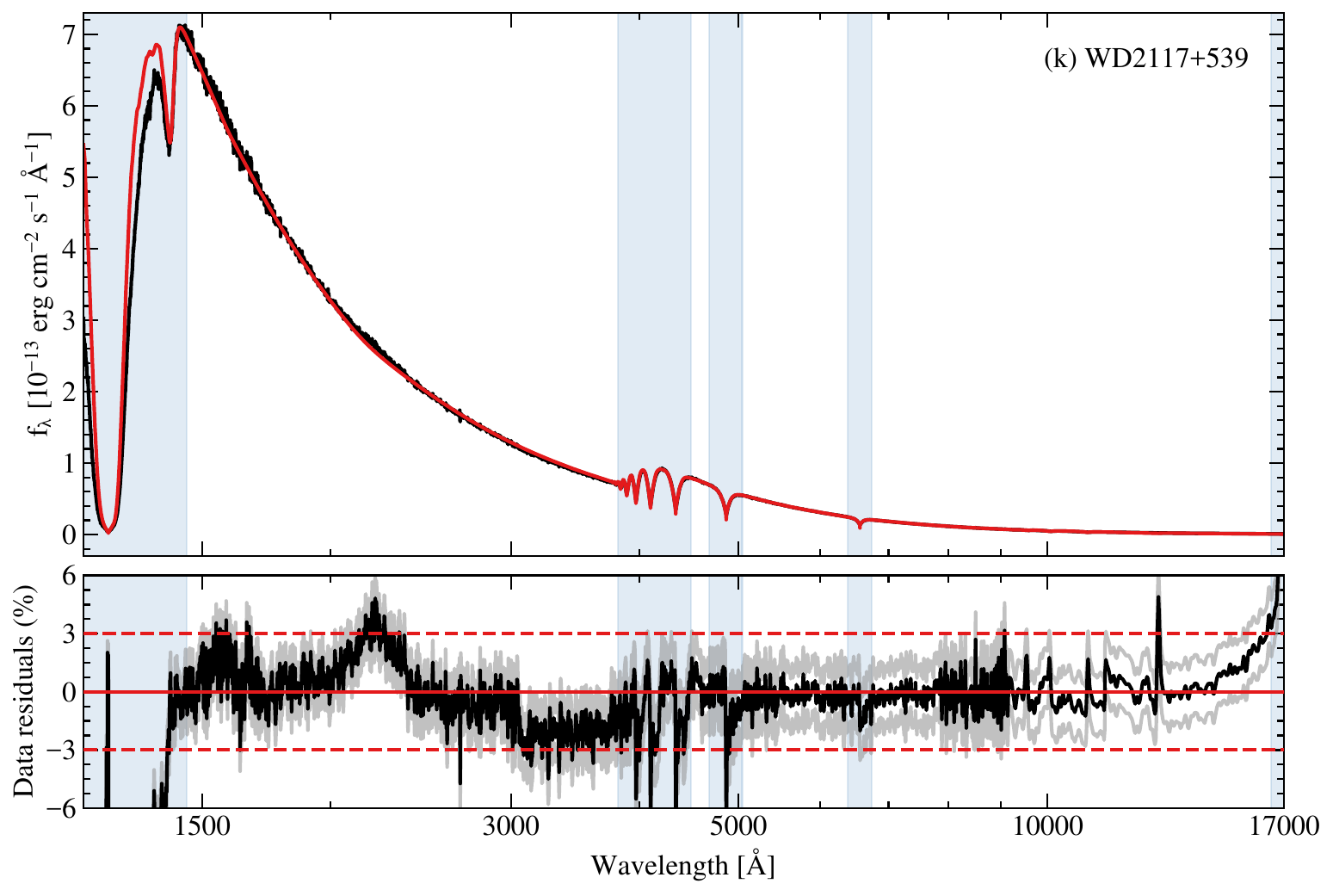}
	\includegraphics[width=0.45\textwidth,height=0.37\textheight,keepaspectratio]{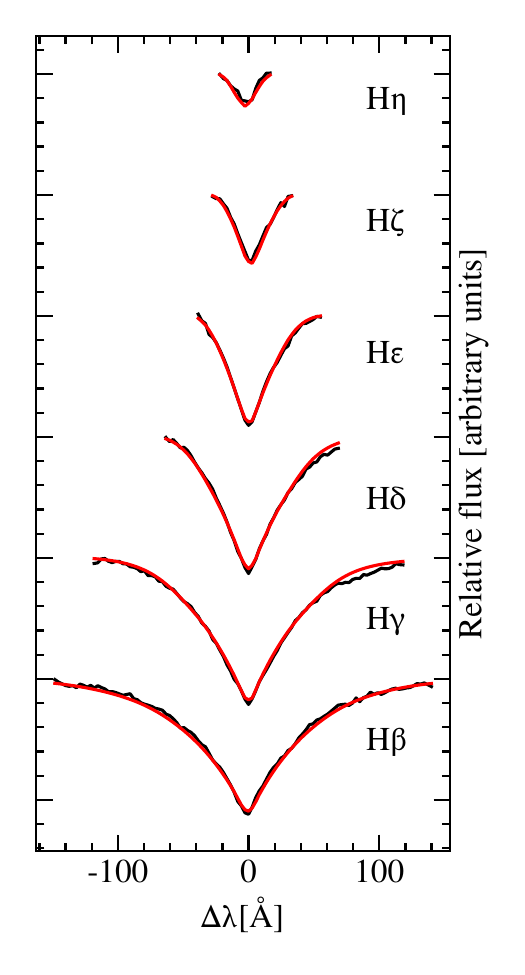}
 	\includegraphics[width=1.42\textwidth,height=0.36\textheight,keepaspectratio]{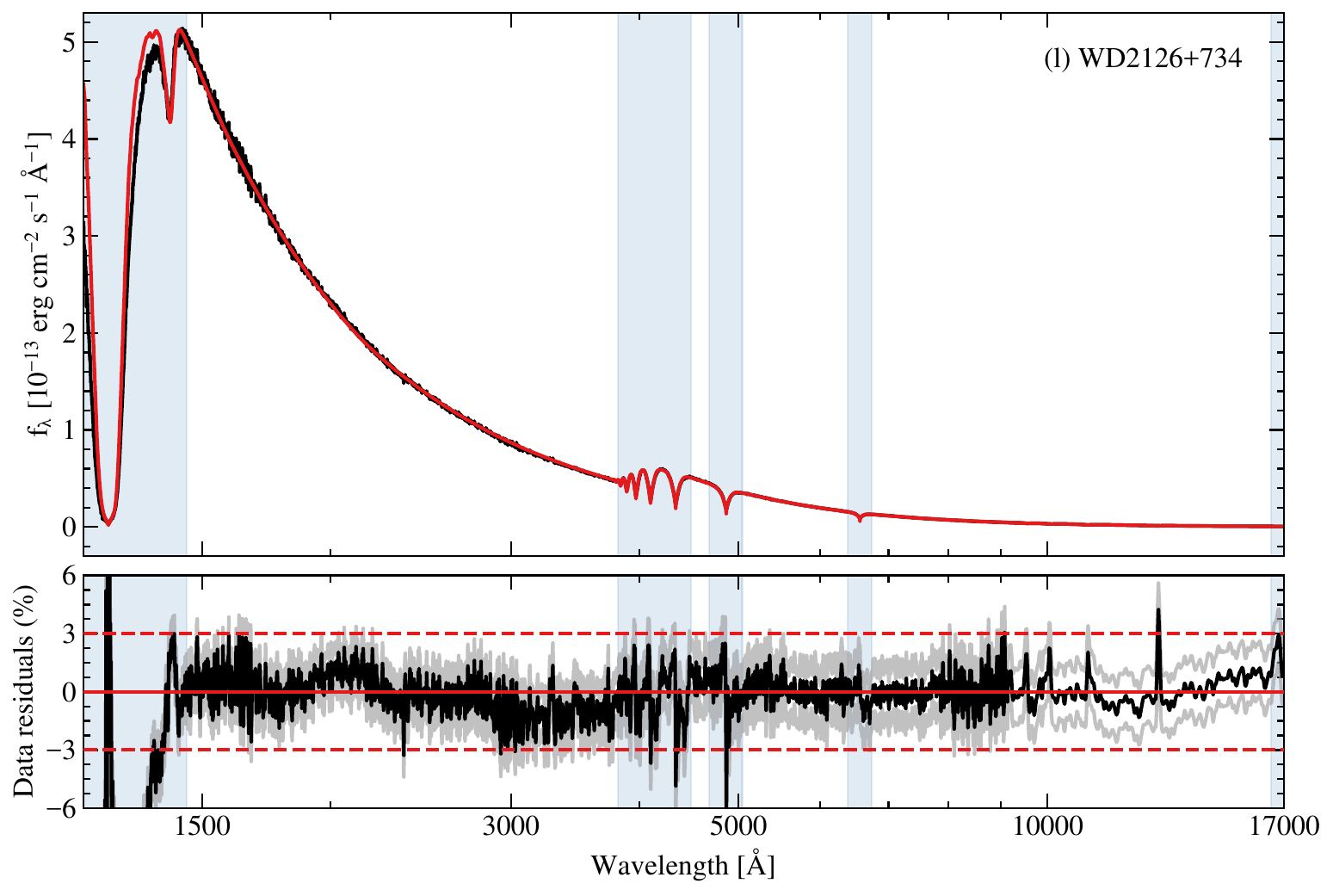}
	\includegraphics[width=0.45\textwidth,height=0.37\textheight,keepaspectratio]{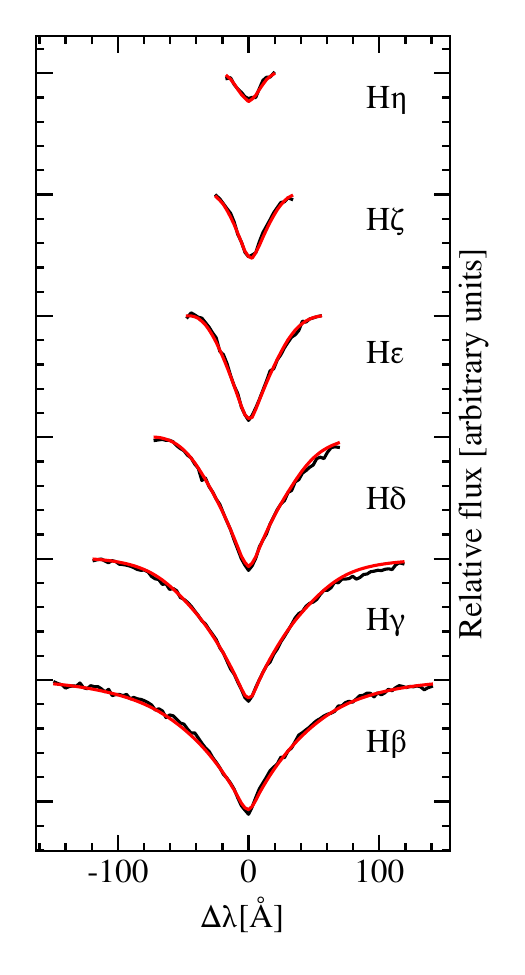}
\end{figure*}

\begin{figure*}
\contcaption{}
	\includegraphics[width=1.42\textwidth,height=0.36\textheight,keepaspectratio]{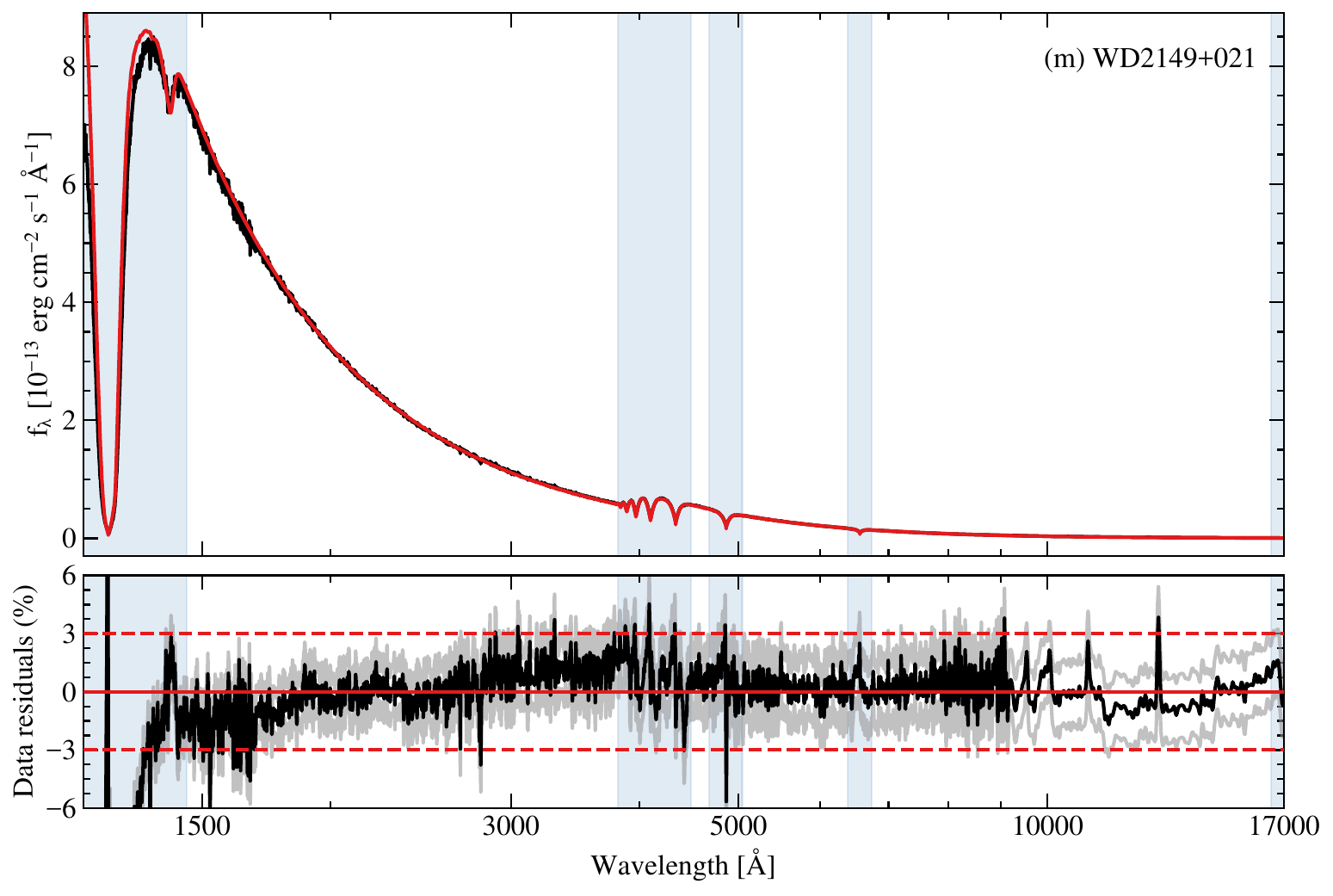}
	\includegraphics[width=0.45\textwidth,height=0.37\textheight,keepaspectratio]{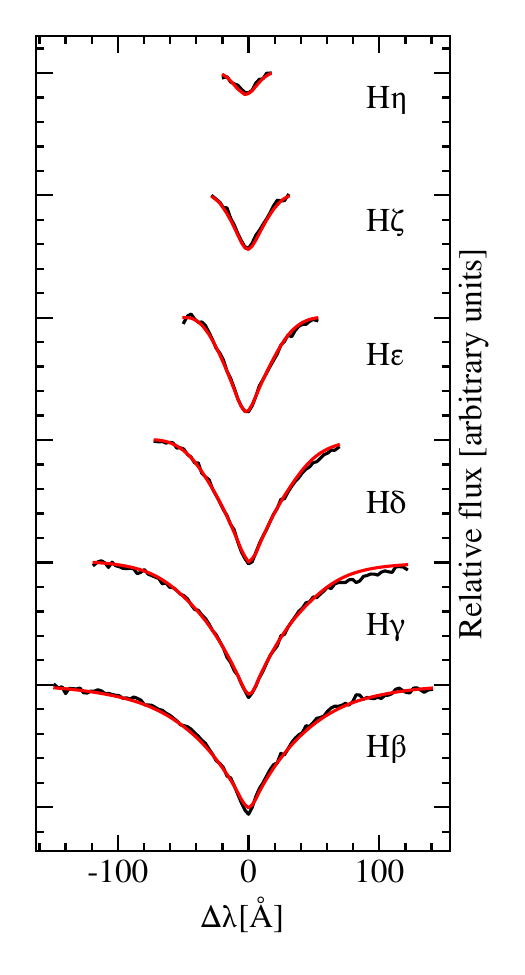}
\end{figure*}


\bsp	
\label{lastpage}
\end{document}